\documentclass[aps,prc,twocolumn,amsmath]{revtex4}
\usepackage[normalem]{ulem}

\usepackage{graphicx}
\usepackage{epsfig}
\usepackage{color}
\usepackage{multirow,ulem}
\usepackage{booktabs}
\usepackage{adjustbox,pdflscape,longtable,rotating}

\usepackage{color,graphicx}
\usepackage{mathptmx}                % math+times fonts
\usepackage{dcolumn}                 % Align table columns on decimal point
\usepackage{bm}                      % bold math
\usepackage[utf8]{inputenc}
\def\roughly#1{\mathrel{\raise.3ex\hbox{$#1$\kern-.75em%
\lower1ex\hbox{$\sim$}}}}

\definecolor{red}{rgb}{0.4,0,0.4}
\definecolor{green}{rgb}{0,0.5,0.0}
\definecolor{navy}{rgb}{0.0,0.0,0.6}
\definecolor{orange}{rgb}{0.8,0.2,0.0}

\begin{document}

\title{Thermal evolution of relativistic hyperonic compact stars
  with calibrated equations of state}

\author{Morgane Fortin}\email{fortin@camk.edu.pl,ganetin@gmail.com}
\affiliation{N. Copernicus Astronomical Center, Polish Academy of Sciences,
  Bartycka 18, 00-716 Warszawa, Poland}

\author{Adriana R. Raduta}\email{araduta@nipne.ro}
\affiliation{National Institute for Physics and Nuclear Engineering (IFIN-HH),
  RO-077125 Bucharest, Romania}

\author{Sidney Avancini}\email{sidney.avancini@ufsc.br}
\affiliation{Departamento de Fisica, Universidade Federal de Santa Catarina,
  88040-900 Florianopolis, Santa Catarina, Brazil}

\author{Constança Provid\^encia}\email{cp@uc.pt}
\affiliation{$^4$CFisUC, Department of Physics, University of Coimbra,
  Portugal}

\begin{abstract}
  A set of unified relativistic mean-field equations of state for hyperonic compact stars
  recently built in
  [M. Fortin, Ad. R. Raduta, S. Avancini, and C. Provid\^encia, Phys. Rev. D {\bf 101}, 034017 (2020)]
  is used to study the thermal evolution of non-magnetized and non-rotating spherically-symmetric
  isolated and accreting neutron stars under different hypothesis concerning
  proton $S$-wave superfluidity.
  These equations of state have been obtained in the following way:
  the slope of the symmetry energy is in agreement with experimental data;
  the coupling constants of $\Lambda$ and $\Xi$-hyperons
  are determined from experimental hypernuclear data;
  uncertainties in the nucleon-$\Sigma$ interaction potential are accounted for;
  current constraints on the lower bound of the maximum neutron star mass are satisfied.
  Within the considered set of equations of state, the presence of hyperons is essential for the
  description of the cooling/heating curves.
  One of the conclusions we reach is that the criterion of best agreement with observational data
  leads to different equations of states and proton $S$-wave superfluidity gaps when applied
  separately for isolated neutron stars and accreting neutron stars in quiescence.
  This means that at least in one situation the traditional simulation framework
  that we employ is not complete and/or the equations of state are inappropriate.
  Another result is that,  considering  equations of state which
  do not allow for nucleonic dUrca   or allow for it only in very massive NS,
  the low luminosity of SAX J1808 requires a repulsive
  $\Sigma$-hyperon potential  in symmetric nuclear matter in the range
  $U_\Sigma^{(N)}\approx 10-30$ MeV. This range of values for $U_\Sigma^{(N)}
  $ is also supported  by the criterion of best agreement with all available data from INS and XRT.
 \end{abstract}

\date{\today}

% \pacs{26.60.-c  % Neutron stars; nuclear physics aspects
% 21.65.Mn, % Equations of state of nuclear matter
% 64.10.+h, % General theory of equations of state and phase equilibria
% }

\maketitle

\section{Introduction}

The composition of neutron star (NS) interiors is a lively subject of research
in both nuclear physics and astrophysics.
Over time the nucleation of various exotic species
(hyperons, Delta resonances, pion and kaon condensates and quarks)
has been considered, in addition to nucleons.
The discovery of several massive pulsars in binary systems with white
dwarfs \cite{Demorest10,Antoniadis13,Cromartie2019}
brought into focus the issue of hyperonisation of NS inner cores.
A large number of studies, most of which accounting for
all experimental constraints from nuclear and hypernuclear physics, have
shown that $\approx 2M_{\odot}$ NS are not incompatible with hyperons
\cite{Weissenborn_PRC_2012,Weissenborn_NPA_2013,Bonanno2012,Bednarek2012,Long2012,Providencia13,Colucci_13,Miyatsu_PRC2013,vandalen_14,Gusakov_MNRAS2014,Oertel:2014qza,Maslov2015,Fortin2016,Tolos2016,Fortin17,Li2018EPJA}.
The way in which the hyperons modify other NS observables,
{\em e.g.}: radius, tidal deformability and moment of inertia,
has been also addressed in \cite{Providencia13,Fortin_PRD_2020}.

The composition of NS cores also determines the neutrino emission
upon which the surface temperature depends.
Neutrino emission is the dominant heat loss mechanism
in the first $10^3$ kyr after the star birth in a supernova explosion
and during the quiescence periods following accretion episodes.
Cooling tracks of isolated NS (INS) as a function of age
\cite{Haensel_AA_1994,Schaab_NPA_1996,Schaab_1998,Yakovlev04,Page_ARNPS_2006,Page_NPA_2006,Tsuruta_2009}
and heating tracks of X-ray transients (XRT) as a function of accretion rate
\cite{Yakovlev_AA_2003,Potekhin_AA_2019}
present qualitatively different features depending on whether direct Urca (dUrca)
process(es) operate or not.
The stars accommodating at least one dUrca process cool much faster than
the stars whose thermal evolution is regulated by the series of the so-called
slow and intermediate cooling processes (modified Urca, bremsstrahlung and, for paired species,
also Cooper pair breaking and formation (PBF)).
dUrca reactions exist for all particles which, under different scenarios,
are expected to nucleate in NS cores.
In the case of baryons and quarks, 
in order for a dUrca process to operate,
the relative abundances of involved baryonic (quark) and leptonic species
should verify the triangle inequality of Fermi momenta,
$p_{F,i}+p_{F,j} \geq p_{F,k}$ \cite{DUY92},
which typically leads to a density threshold ($n_{DU}$).

The threshold density of nucleonic dUrca depends on the relative
neutron and proton abundances in cold catalyzed NS matter.
These are regulated by the density behavior of the symmetry energy, which is
typically expressed in terms of the symmetry energy of symmetric saturated
nuclear matter ($J$) and its slope ($L$) and curvature ($K_{sym}$).
Correlations between the nuclear symmetry energy and the threshold
density of nucleonic dUrca have been sought after in \cite{Fortin16,Providencia19}.
By considering a large number of non-relativistic Skyrme-like and
relativistic mean-field (RMF) models Ref. \cite{Fortin16} concludes that
i) models with high $L$-values manifest a clear (anti)correlation between $L$ and
$n_{DU}$, and
ii) no definite behavior can be identified for models with low  $L$-values.
Ref. \cite{Providencia19} showed that the (anti)correlation between $L$ and
$n_{DU}$ survives also in hyperonic stars and nucleation of hyperons
does not alter significantly the threshold density of nucleonic dUrca.
$\Lambda$, $\Xi^-$, $\Xi^0$ and $\Sigma^-$, which are the only 
hyperons that nucleate in NS matter, may - in principle -
be involved in dUrca reactions, too.
The threshold densities of hyperonic dUrca are determined by the (poorly constrained) hyperon-nucleon interaction potentials
\cite{Li2018EPJA,Fortin_PRD_2020}, in addition to the
isoscalar and isovector behaviors of the nucleonic equation of state (EoS).

The neutrino emission from compact star interiors depends sensitively on
the magnitude and density dependence of the pairing gaps of fermions, too.
The magnitudes of the pairing gaps are controlled by the baryon interactions
and by the many particle correlations accounted for;
their density dependences are determined by the different particle density profiles,
which are regulated by baryonic effective potentials,
with the isovector nucleon potential playing an important role.

The consequences of various dUrca processes and the way in which pairing
  of different baryonic species in the core impact the thermal evolution
  was discussed at length in
  \cite{Haensel_AA_1994,Yakovlev04,Schaab_NPA_1996,Schaab_1998,Tsuruta_2009,BY15,Beloin_PRC_2018,Beloin_PRC_2019}
  and is qualitatively well understood.

Recent progress on the EoS of hyperonic stars and neutrino emissivities in
different channels  \cite{Yakovlev_PhysRep_2001}
motivated  \cite{Raduta17,Grigorian2018,Negreiros18,Raduta19}
to reconsider the cooling of hyperonic isolated stars.
Different strategies have been used by the different groups of authors.
\cite{Raduta17,Raduta19} focused on the maximum suppression of hyperonic dUrca reactions,
by considering $\Lambda$ and $\Xi$ $^1S_0$ pairing gaps provided by the most attractive
bare $\Lambda\Lambda$ and $\Xi\Xi$ interaction potentials.
Based on symmetry arguments, Ref. \cite{Raduta19} has also investigated
the possibility of high order pairing of $\Lambda$ hyperons in the core together
with its impact on cooling.
In both cases, the studies employed a bunch of EoS which
explore present day uncertainties in the nucleonic sector.
Refs. \cite{Grigorian2018,Negreiros18} focused on the possibility to obtain,
for a given EoS, agreement with observed effective surface temperature versus
age data of INS. To this aim, the neutron $^3P_2-^3F_2$ and proton $^1S_0$ pairing 
and, in the case of \cite{Grigorian2018}, also the $\Lambda$ $^1S_0$ pairing
in the core were considered as free parameters.
\cite{Grigorian2018} used the MKVORH$\phi$ \cite{Maslov2015,Maslov2016}
RMF model with hadron effective couplings and masses dependent on the scalar field;
\cite{Negreiros18} used the FSU2R and FSU2H \cite{Tolos17} extensions
of the relativistic mean field model FSU2 \cite{Chen2014}.
Different modelization of the nucleonic sector in \cite{Maslov2015,Maslov2016} and \cite{Tolos17}
makes that the first of these models does not allow for nucleonic dUrca in any stable NS,
while the second does. 
The common conclusion of \cite{Raduta17,Grigorian2018,Negreiros18,Raduta19} is that
thermal states of INS do not rule out hyperon nucleation in NS cores,
contrary to what was previously claimed by \cite{Haensel_AA_1994,Schaab_1998,Tsuruta_2009}.

The aim of the present work is to investigate whether thermal evolution data of NS
may be interpreted such as to get extra information on the EoS,
including nucleation of hyperons, or, alternatively, assess the most probable composition
of one of the NS for which thermal data are available.
To this end evolutionary tracks simulated with a benchmark code and corresponding
to different mass NS built upon {\em realistic} EoS \cite{Fortin_PRD_2020}
are {\em systematically} compared with data corresponding to
{\em middle age isolated NS} and {\em accreting NS in low mass X-ray binaries}.
Uncertainties related to nucleon effective interactions at high densities,
including the isovector channel, are accounted for by allowing for different EoS models;
uncertainties in the $\Sigma$-nucleon interaction are accounted for
by varying the well depth of the $\Sigma$-hyperon at rest in symmetric nuclear matter;
uncertainties related to the proton $^1S_0$ pairing in the core, expected to
regulate the $n \to p+l+\tilde \nu$ and $\Lambda \to p+l+\tilde \nu$ dUrca processes,
are accounted for by considering two {\em extreme} prescriptions
for the dependence of the pairing gap on the Fermi momentum.
With the purpose of limiting the parameter space and the motivation that
hyperon-hyperon interactions are little known hyperonic pairing is
here disregarded.

Let us point out that other approaches have been considered to describe hyperonic matter.
  In particular, in \cite{Jiang2012} the authors have constrained the density dependence of the meson
  masses to the 2$M_\odot$ condition. 
  A G-matrix approach to the EoS was considered in \cite{Takatsuka2002}. In this case it is
  known that the EoS becomes too soft with the introduction of hyperons, as
  shown in \cite{Baldo2000,Vidana2000}, and 2$M_\odot$ are not described.
  However, three-body forces play an essential role in this approach
  and are still not constrained, in particular, the terms
  involving hyperons, see the discussion in \cite{Vidana2010,Logoteta2013,Lonardoni2014}.

To our knowledge the only attempts done so far in order to constrain the NS EoS,
in particular, the threshold density for dUrca, and/or the nucleonic paring gaps 
based on systematic comparison with thermal data have disregarded hyperons and employed 
parameterized EoS and/or pairing gaps \cite{BY15,Beznogov_MNRAS_2015,Beloin_PRC_2018,Beloin_PRC_2019}.
Out of these only \cite{BY15,Beznogov_MNRAS_2015} performed a best fit analyses of
thermal data of both INS and XRT.

The paper is organized as follows.
The set of EoS is presented in Sec.~\ref{sec:eos}.
Properties of NS built on these EoS are discussed in Sec.~\ref{sec:NS}
along with the constraints coming from gravitational wave (GW) detection
and mass-radius measurements.
Observational data are commented in Sec.~\ref{sec:obsdata}.
A short review of cooling/heating theory of NS is offered in Sec.~\ref{sec:cool}.
The results of our simulations are confronted with data in Sec.~\ref{sec:res}.
Finally the conclusions are drawn in Sec.~\ref{sec:concl}. 

\section{Equations of State}
\label{sec:eos}

A set of phenomenological EoS for hyperonic compact stars has been recently
built in Ref. \cite{Fortin_PRD_2020} by tuning the $\Lambda$-nucleon
and $\Xi$-nucleon interactions on experimental data of $\Lambda$ and
$\Xi$-hypernuclei, following a procedure described later on in this
section.
Uncertainties in the $\Sigma$-nucleon interaction,
for which poor experimental constraints exist,
have been accounted for by allowing the well depth of $\Sigma$ at rest in
symmetric saturated nuclear matter $U_{\Sigma}^{(N)}$ to span a wide range of values.

 These EoS  rely on the covariant density functional theory, in which
the nucleons interact among each others by the exchange of
scalar-isoscalar ($\sigma$), vector-isoscalar ($\omega$) and
vector-isovector ($\rho$) mesons.
Interactions between nucleons and hyperons are additionally mediated by the
hidden strangeness vector meson $\phi$. 

For the nucleonic part, 11 models have been considered:
FSU2 \cite{Chen2014}, FSU2H and FSU2R \cite{Tolos17,Negreiros18},
NL3 \cite{nl3}, NL3$\omega\rho$ \cite{Pais16,Horowitz01},
TM1 \cite{tm1}, TM1$\omega\rho$ \cite{Providencia13,Bao2014,Pais16},
TM1-2 and TM1-2$\omega\rho$ \cite{Providencia13},
DD2 \cite{typel10} and DDME2 \cite{ddme2}.
The first nine belong to the category of RMF models with non-linear (NL) terms.
The last two belong to the category of RMF models with density dependent (DD)
coupling constants. For a general review on RMF models, refer to \cite{Dutra2014}.

For each nucleonic EoS, the couplings of the scalar-isoscalar $\sigma$ meson
to the $\Lambda$ and $\Xi$-hyperons are determined from the
fit of experimental binding energies with the predictions
of the single-particle Dirac equations for baryons.
The calibration of the EoS to $\Lambda$-hypernuclei was performed in \cite{Fortin17,Fortin18},
where a vast collection of experimental data corresponding
to both hypernuclei in $s$ and $p$ shells was considered. These results
are used in the present study.
Experimental data exist so far only for two $\Xi$-hypernuclei:
$^{12}_{\Xi^-}$Be \cite{khaustov00} and $^{15}_{\Xi^-}$C \cite{kiso},
with the second affected by uncertainties
related to the state in which one of the daughter nuclei is produced.
We use the coupling constants obtained in \cite{Fortin_PRD_2020}
under the hypothesis according to which $^{15}_{\Xi^-}$C is produced
in the first excited state. Based on consistency with data
corresponding to $^{12}_{\Xi^-}$Be, this seems to be the most plausible scenario,
in agreement with previous conclusion of \cite{Sun2016}.
Uncertainties related to the interaction between nucleons and the
$\Sigma$-hyperon are dealt with
by allowing $U_{\Sigma}^{(N)}$ to explore the domain
$-10 \leq U_{\Sigma}^{(N)} \leq 40$ MeV. We remind that, according to \cite{Gal2016},
$U_{\Sigma}^{(N)} \approx 30 \pm 20$ MeV.
The couplings of the vector mesons to hyperons are expressed in terms of
coupling constants of the meson under consideration to the nucleon and are
derived by assuming SU(6) flavor symmetry.
They are as follows:
$g_{\omega \Lambda}=2/3 g_{\omega N}$, $g_{\rho \Lambda}=0$, $g_{\phi \Lambda}=-\sqrt{2}/3 g_{\omega N}$,
$g_{\omega \Xi}=1/3 g_{\omega N}$, $g_{\rho \Xi}=g_{\rho N}$, $g_{\phi \Xi}=-2\sqrt{2}/3 g_{\omega N}$,
$g_{\omega \Sigma}=2/3 g_{\omega N}$, $g_{\rho \Sigma}=2 g_{\rho N}$,
$g_{\phi \Sigma}=-\sqrt{2}/3 g_{\omega N}$.
The hidden strangeness meson $\phi$ does not interact with nucleons, $g_{\phi N}=0$.
For models with DD coupling constants, we assume that the coupling constants
of the various vector fields to the hyperons have the same density dependence
as for the coupling to the nucleons.

Out of these EoS models we here select those which satisfy the following
conditions:
i) the maximum mass of hyperonic stars exceeds the widely accepted lower limit,
$\approx 2M_{\odot}$,
ii) the slope of the symmetry energy of saturated symmetric matter 
is  in agreement with experimental constraints,
$40 \lesssim L \lesssim 62$ MeV \cite{Lattimer13} or
$30 \lesssim L \lesssim 86$ MeV \cite{Oertel17}.
Only the following EoS satisfy the above conditions:
DD2 \cite{typel10}, DDME2 \cite{ddme2},
NL3$\omega\rho$ \cite{Pais16,Horowitz01} and FSU2H \cite{Tolos17,Negreiros18}.
With $K_{sym}$ values equal to $-93.2$, $-87.1$, $-7.6$ and, respectively,
$86.7$ MeV, only the two
DD models comply, at $1-\sigma$ confidence level,
with the recent constraint $K_{sym}=-102 \pm 71$ MeV \cite{Zimmerman_2020}
extracted from LIGO/Virgo and NICER data, or with  $K_{sym}=-111.8\pm 71.3$ MeV
obtained in \cite{Mondal_PRC_2017} from nuclear properties.
In \cite{Malik_PRC_2018}, within a different analysis imposing constraints from  LIGO/Virgo,
it was proposed $-140< K_{sym}< 16$ MeV, with which the first three models comply.
Others studies obtain wider ranges as \cite{Zhang_EPJA_2018},
where  an upper limit at about 68 MeV is obtained.
Finally, Bayesian analyses of neutron skin thickness data that incorporate prior knowledge
  of the pure neutron matter equation of state from chiral effective field
  theory calculations predict much smaller values, $K_{sym}=-260^{+35}_{-33}$ MeV \cite{Newton_2020}.

\section{Properties of hyperonic stars}
\label{sec:NS}

\begin{figure}[tb]
\includegraphics[width=0.7\linewidth]{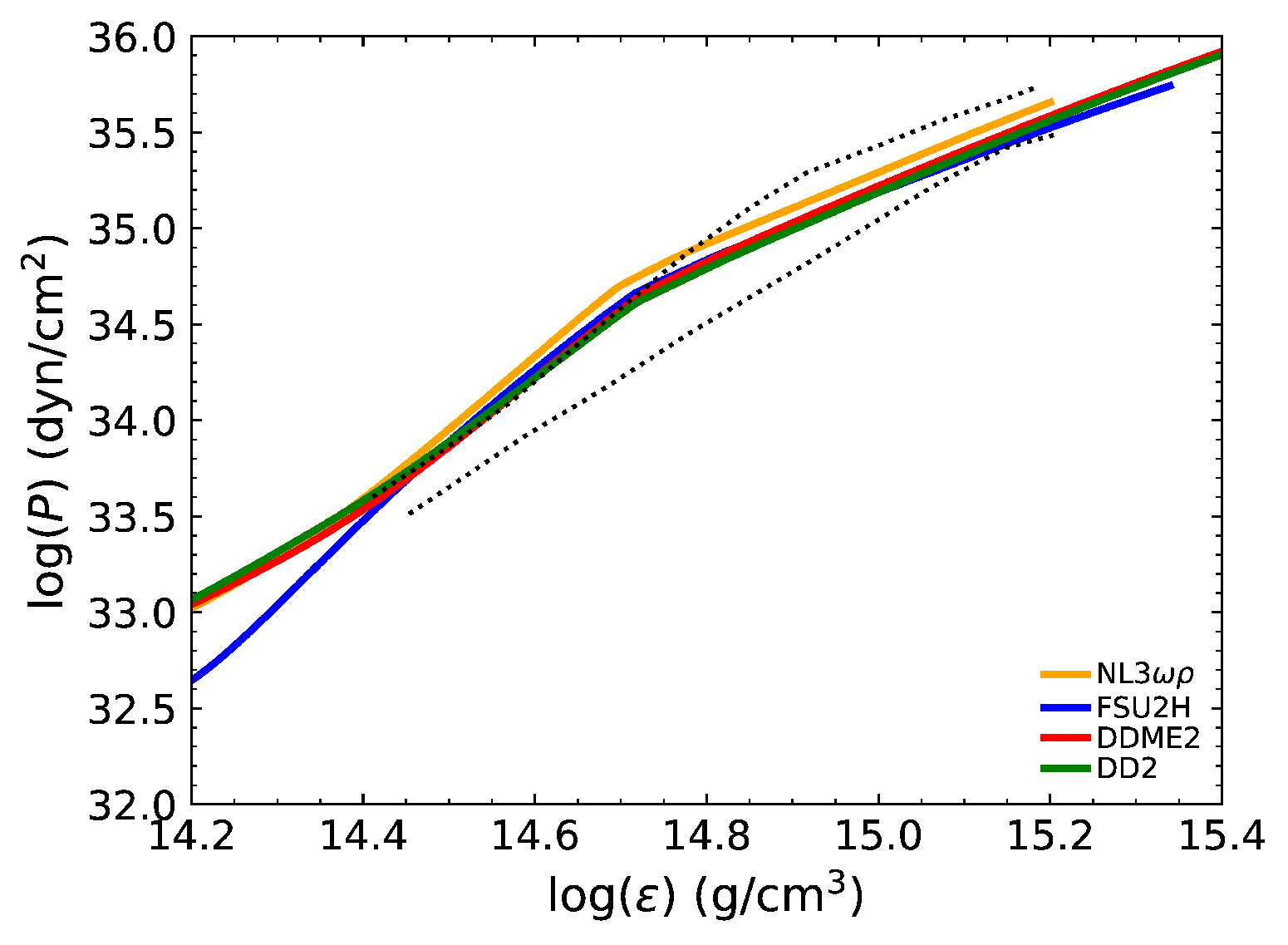}
\includegraphics[width=0.7\linewidth]{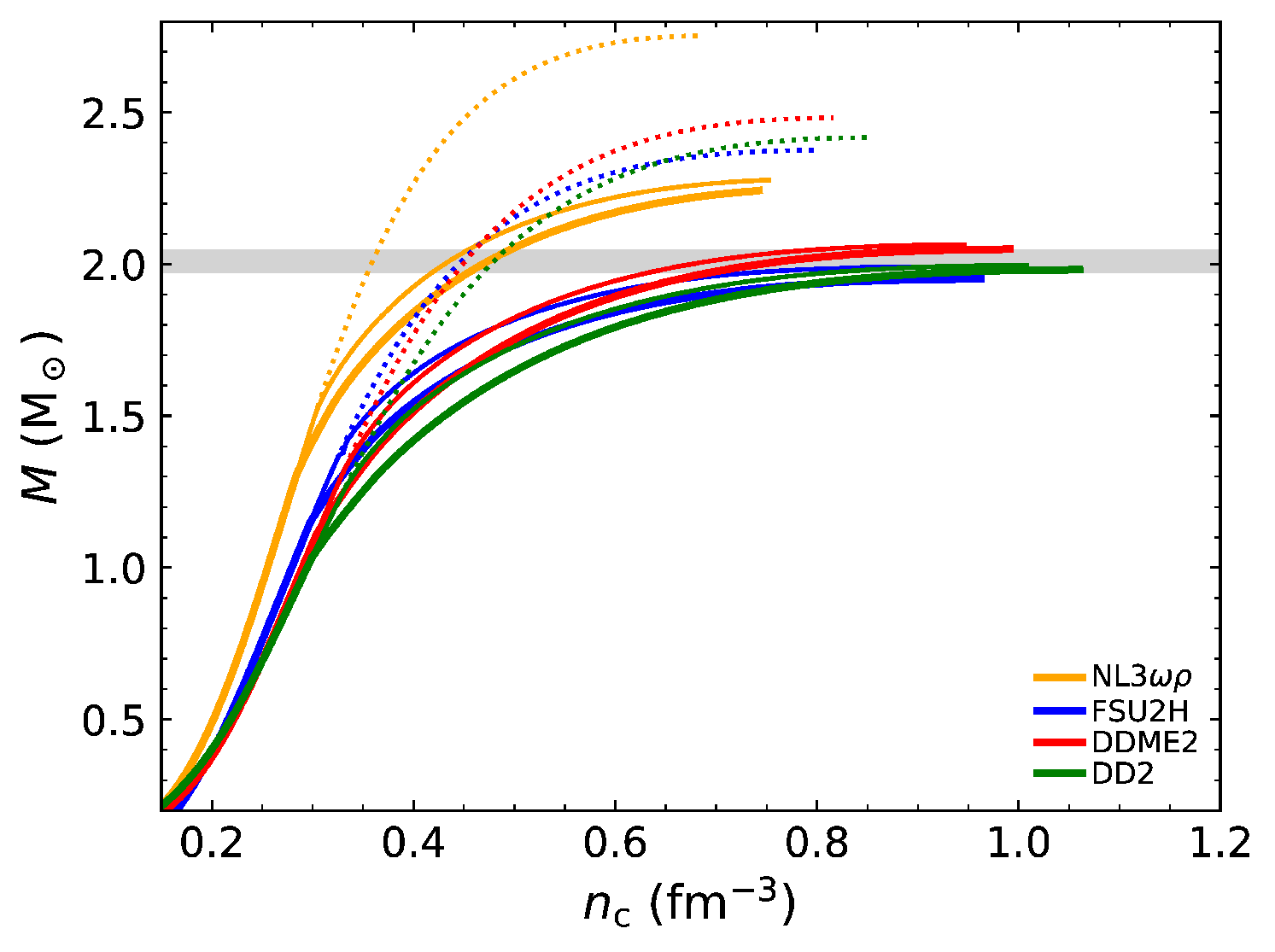}
\includegraphics[width=0.7\linewidth]{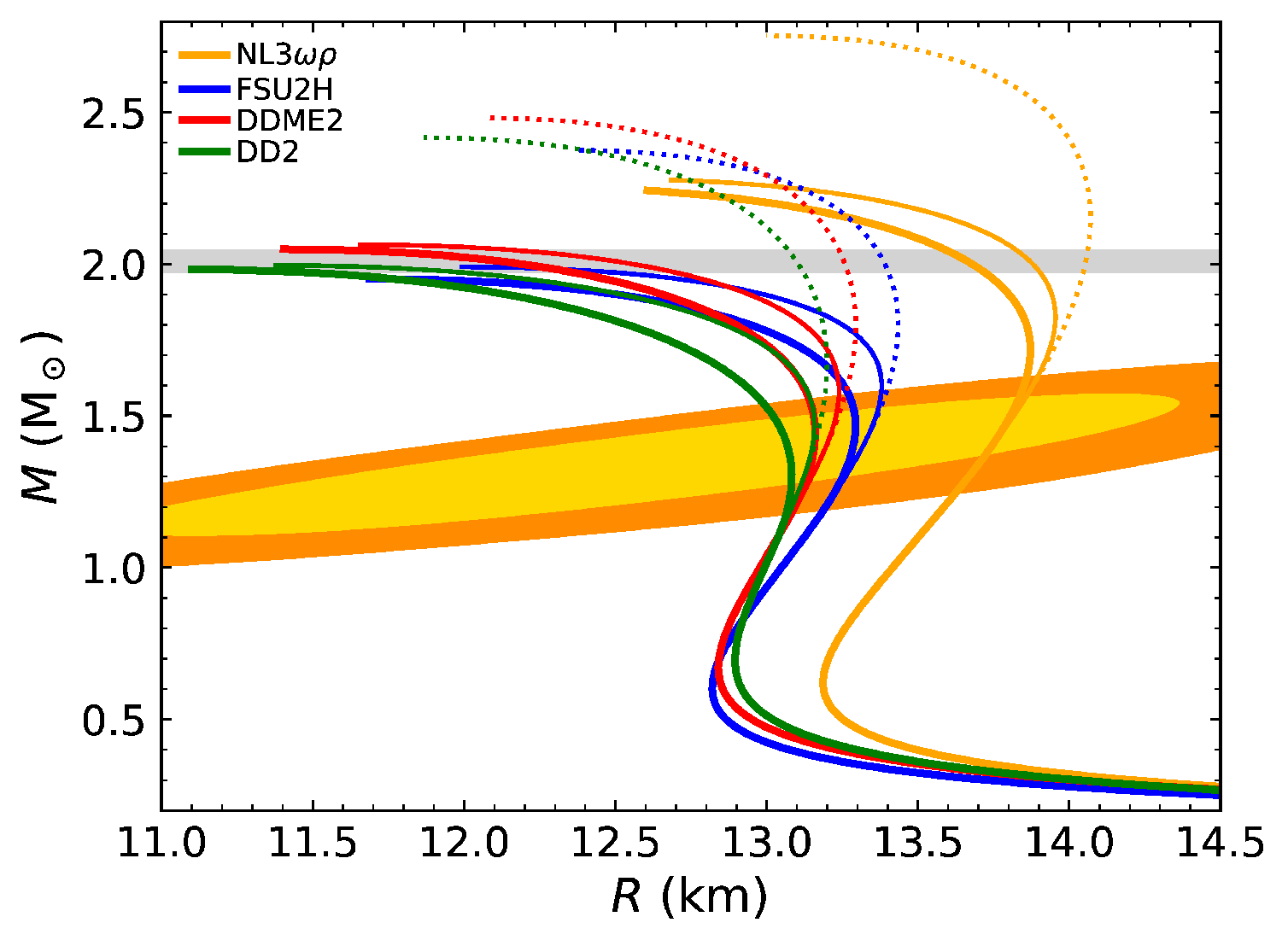}
\includegraphics[width=0.7\linewidth]{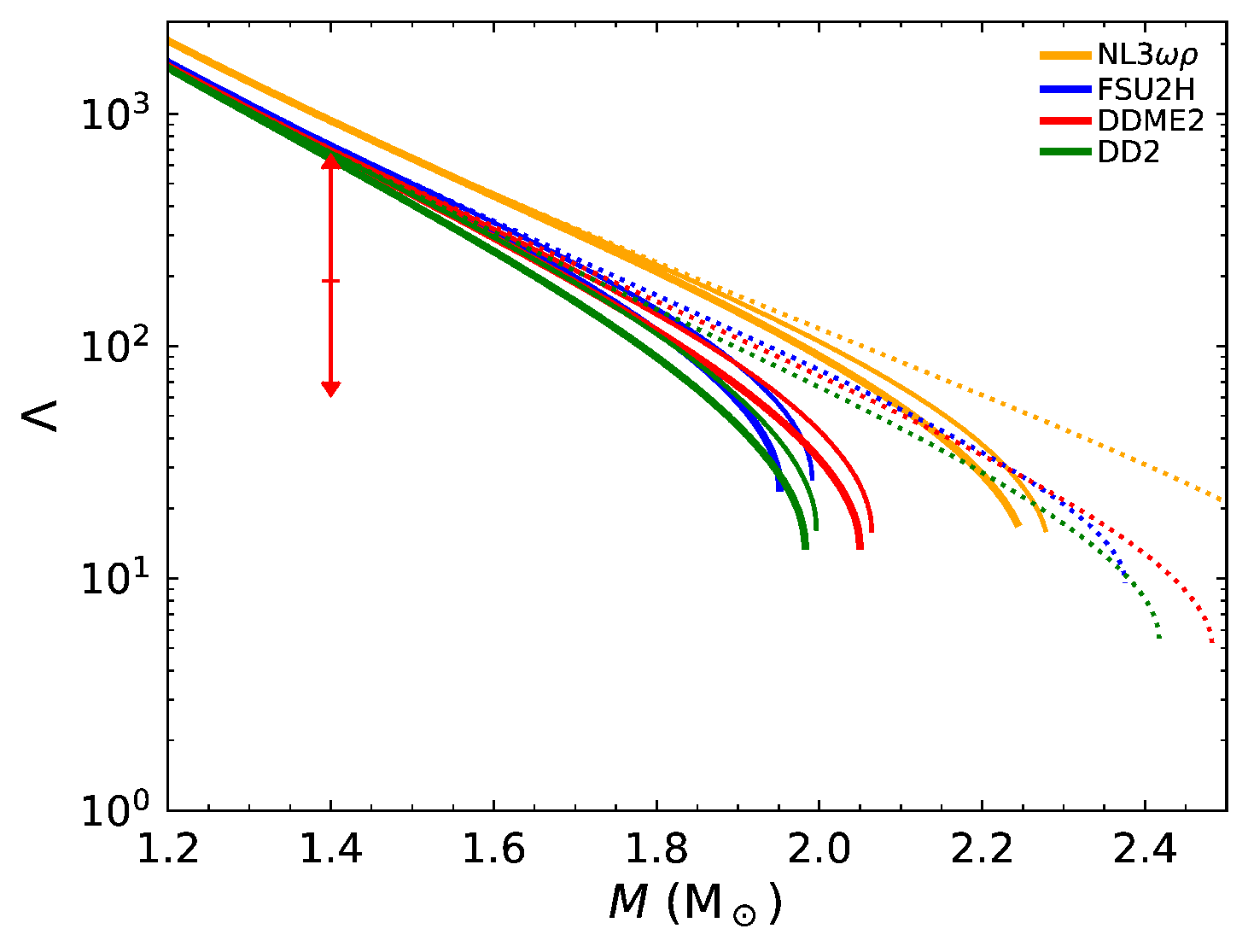}
\caption{Equation of state (first top panel),
  gravitational mass versus central baryonic number density (second panel),
  gravitational mass versus radius (third panel) and
  tidal deformability versus gravitational mass (bottom panel)
  corresponding to the stellar models considered in this work.
  Thin (thick) curves correspond to $U_{\Sigma}^{(N)}=30$ MeV
  ($U_{\Sigma}^{(N)}=-10$ MeV) and the dotted ones to purely nucleonic stars.
  The thin dot-dashed curves in the $P(\varepsilon)$ plot mark the borders of the domain
  obtained in \cite{Raaijmakers_2019} from the analyses of NICER
  and LIGO/Virgo data,
  corresponding to PSR J0030+0451 \cite{Miller_2019,Riley_2019}
  and, respectively, GW170817 \cite{GW170817}.
  The horizontal shaded area in $M-n_c$ and $M-R$ plots shows the observed mass of
  pulsar PSR J0348+0432, $M=2.01 \pm 0.04 M_{\odot}$ \cite{Antoniadis2013}.
  The experimental constraints on mass-radius relation based on NICER measurements
  of the millisecond pulsar PSR J0030+0451 
  are represented in the $M-R$ plot \cite{Riley_2019} (68\% and 95\% confidence contours).
  Finally, the vertical error bar in the $\Lambda-M$ plot illustrates the constraints
  from GW170817, as obtained in \cite{GW170817}. }
\label{fig:eos}
\end{figure}

We now turn to discuss the properties of cold catalyzed NS built upon
the core EoS described in Section~\ref{sec:eos}.
For the crust we use EoS models which are consistent with the core \cite{Fortin16,Providencia19,Fortin_PRD_2020};
for the employed approach, see \cite{Fortin16}. 
In Fig.~\ref{fig:eos} results corresponding to extreme values of $U_{\Sigma}^{(N)}$ are
illustrated with thick and thin lines for $U_\Sigma^{(N)}=-10$ and 30 MeV, respectively.

The top panel of Fig.~\ref{fig:eos} illustrates
the pressure versus energy density in the core.
Also shown are the constraints obtained by \cite{Raaijmakers_2019},
based on NICER and LIGO/Virgo data,
corresponding to PSR J0030+0451 \cite{Miller_2019,Riley_2019}
and, respectively, GW170817 \cite{GW170817}.
It comes out that, with the exception of NL3$\omega\rho$ which provides a slightly
stiffer behavior over the density range $0.13 \lesssim n_B \lesssim 0.28~{\rm fm}^{-3}$,
all our EOS comply with the constraints in
\cite{Raaijmakers_2019}.
For the energy density range considered, FSU2H shows the  softest behavior
at low energy densities and  NL3$\omega\rho$ the  stiffest behavior at
high energy densities.
The different behaviors of the $P-\varepsilon$ curves
lead to different values for the central baryonic number densities
for a given mass, as clearly seen in the second panel of Fig. ~\ref{fig:eos}.
The lowest central baryonic densities correspond, for all NS masses,
to the stiffest EoS NL3$\omega\rho$.
As expected, the dispersion among the various EoS increases with the NS mass.
To give an example, for $U_{\Sigma}^{(N)}$=30 MeV and a mass of $1.8M_\odot$,
$n_{\rm c}^{\rm NL3\omega\rho}=0.34~{\rm fm}^{-3}$,
$n_{\rm c}^{\rm FSU2H}=0.48~{\rm fm}^{-3}$,
$n_{\rm c}^{\rm DDME2}=0.47~{\rm fm}^{-3}$,
$n_{\rm c}^{\rm DD2}=0.54~{\rm fm}^{-3}$.
  The reason why the $n_{\rm c}^{\rm DDME2} < n_{\rm c}^{\rm DD2}$ is attributable to the much higher
  value of $Q_{sat}$, the parameter of the third order term in the Taylor expansion
  of the energy per nucleon as a function of the deviation from the saturation density,
  whose value is higher for DDME2 than for DD2 \cite{Dutra2014}.
  We shall see in Sec. \ref{sec:res} that
  different density profiles of the various baryonic species inside NS
  will result 
  in significantly different thermal evolutions due to
  the density dependence of the symmetry energy and
  EoS stiffness. 
The different behaviors of
$P(\varepsilon)$ curves are also reflected
in the mass-radius curves for all NS masses, in particular, in  the strongly different values of the radii, third panel of
Fig. ~\ref{fig:eos}.
For the canonical  1.4$M_\odot$ NS,
the dispersion among the predictions of our models is large
$\Delta R_{1.4}=R_{1.4}^{\rm max}-R_{1.4}^{\rm min}=0.7$ km.
The stiffest EoS, NL3$\omega\rho$, provides the largest radii for all
NS masses, and the smallest radii are obtained by DD2, considering NS with $M \geq 1.4 M_{\odot}$.
NS with masses exceeding the threshold density for nucleation of $\Sigma$
also manifest a non-negligible dependence of the radii
on $U_{\Sigma}^{(N)}$: the larger the abundances of $\Sigma$, favored by attractive
$\Sigma$N potentials, the smaller the NS radii.
Also shown are the recent measurements of the mass and radius of PSR J0030+0451,
  as obtained by the NICER mission \cite{Miller_2019,Riley_2019}. The large uncertainties
  nevertheless do not allow to constraint the EoS. The only astrophysical constraints
  on the NS EoS available so far concern the low and intermediate
  density domains and have been provided by the tidal deformability measurements in the
  event GW170817 \cite{GW170817}. The extent in which our models comply with these data is
  further considered in the bottom panel, where the tidal deformability is plotted as a function
  of the NS gravitational mass. The vertical error bar corresponds to the $\Lambda=190_{-120}^{+390}$
constraint in \cite{GW170817}. This constraint is extracted from the effective tidal
deformability $\tilde \Lambda$, but  depends on
the analysis undertaken. In particular, it  was obtained
from a set of EoS that did not necessarily describe two solar mass
stars, and, therefore, should be taken with care.
FSU2H provides for $1.4M_{\odot}$ a tidal deformability at the upper limit of the
observational constraint, while the other EoS in this work overshoot the constraint from GW170817;
the largest overestimation corresponds to the stiffest EoS, NL3$\omega\rho$.

\begin{table*}
\label{tabii}
\caption{Properties of NS built upon relativistic density
  functional models considered in this work. For each model, we first indicate the properties of NS
  without hyperons (noY) and then for different values of $U_\Sigma^{(N)}$, as specified on column 2.
  $R_{1.4}$ represents the radius of the canonical $1.4M_{\odot}$ NS;
  $M_{\rm max}$ and $n_{c, \rm max}$ refer to the maximum NS mass and
  the corresponding central baryonic number density.
  Columns 6-14 list the hyperonic species that nucleate in stable stars,
  their onset densities and the associated NS mass.
  Columns 15-22 list the density and mass thresholds above which the
  nucleonic and some hyperonic dUrca processes operate.
  Particle number densities are expressed in fm$^{-3}$ and
  NS masses are expressed in $M_\odot$.
  $np$, $\Lambda p$, $\Sigma^- \Lambda$ and $\Sigma^- n$ are abbreviations for
  the dUrca processes that involve the specified baryons.}
{\scriptsize
\hspace*{-1cm}
\vspace*{0cm}
\begin{tabular*}
  %\begin{longtable}
    {\linewidth}{ @{\extracolsep{\fill}} l| c| c| c | c| lcc | lcc | lcc | cc | cc | cc | cc @{}}
\toprule
 \hline
 \hline
\multicolumn{5}{c|}{}&
\multicolumn{9}{c}{Y species} & \multicolumn{2}{c|}{np} & \multicolumn{2}{c|}{$\Lambda p$}
 & \multicolumn{2}{c|}{$\Sigma^- \Lambda$} & \multicolumn{2}{c}{$\Sigma^- n$}   \\
\hline
%\midrule \midrule \addlinespace \\

Model & $U_{\Sigma}^{(N)}$ & $R_{1.4}$ & $M_{\rm max}$ & $n_{c, \rm max}$&&$n_Y$&$M_Y$&&$n_Y$&$M_Y$&&$n_Y$&$M_Y$& $n_{DU}$ & $M_{DU}$ & $n_{DU}$ & $M_{DU}$& $n_{DU}$ & $M_{DU}$ & $n_{DU}$ & $M_{DU}$\\

& (MeV)& km & ($M_\odot$) & (fm$^{-3}$)& &  (fm$^{-3}$)& ($M_\odot$) & &
                                                                      (fm$^{-3}$)&
                                                                                   ($M_\odot$) & &  (fm$^{-3}$)& ($M_\odot$) & (fm$^{-3}$)& ($M_\odot$) &(fm$^{-3}$)& ($M_\odot$) &(fm$^{-3}$)& ($M_\odot$) &(fm$^{-3}$)& ($M_\odot$) \\
\hline
DD2 & no Y & 13.19 & 2.42  & 0.849  &&& &&& &&& && && && && \\ 
 & -10  &13.07 & 1.99  & 1.050  & $\Sigma^-$ &  0.295 &  1.00  & $\Lambda$ &  0.379 &  1.35  & $\Xi^-$ &  0.819 &  1.95       &        &            &  0.395 &  1.40   & 0.383 &  1.36      &     &    \\ 
 &  10 & 13.14&2.00  & 1.019  & $\Sigma^-$ &  0.325 &  1.21  & $\Lambda$ &  0.345 &  1.31  & $\Xi^-$ &  0.532 &  1.76       &        &            &  0.346 &  1.31   & 0.347 &  1.32      &     &    \\ 
 & 30   & 13.15&2.00  & 1.000  & $\Lambda$ &  0.337 &  1.29  & $\Sigma^-$ &  0.374 &  1.45  & $\Xi^-$ &  0.375 &  1.46       &        &            &  0.337 &  1.29   & 0.374 &  1.45      &     &    \\ 
DDME2 & no Y & 13.23 & 2.48  & 0.812  &&& &&& &&& && && && && \\ 
 & -10   & 13.16&2.06  & 0.981  & $\Sigma^-$ &  0.299 &  1.08  & $\Lambda$ &  0.378 &  1.45  & $\Xi^-$ &  0.771 &  2.03       &        &            &  0.393 &  1.50   & 0.382 &  1.46      &     &    \\ 
 & 10   & 13.20& 2.07  & 0.952  & $\Sigma^-$ &  0.328 &  1.30  & $\Lambda$ &  0.346 &  1.40  & $\Xi^-$ &  0.493 &  1.81       &        &            &  0.346 &  1.40   & 0.349 &  1.41      &     &    \\ 
 & 30   & 13.20& 2.08  & 0.935  & $\Lambda$ &  0.340 &  1.39  & $\Xi^-$ &  0.372 &  1.54  & $\Sigma^-$ &  0.386 &  1.59       &        &            &  0.340 &  1.39   & 0.386 &  1.59      &     &    \\ 
FSU2H & no Y & 13.20 & 2.38 & 0.801  &&& &&& &&& & 0.572 &2.27 && && && \\ 

 &  -10& 13.29&1.95  & 0.876  & $\Sigma^-$ &  0.299 &  1.16  & $\Lambda$ &  0.357 &  1.42      &         &        &    & 0.425 &  1.61    &  0.365 &  1.45   & 0.357 &  1.42  & 0.627 &  1.87   \\ 
 & 10  & 13.32&1.98  & 0.919  & $\Lambda$ &  0.333 &  1.42  & $\Sigma^-$ &  0.333 &  1.41      &         &        &    & 0.479 &  1.76    &  0.333 &  1.41   & 0.341 &  1.45      &     &    \\ 
 & 30   &13.32& 1.99  & 0.900  & $\Lambda$ &  0.332 &  1.41  & $\Sigma^-$ &  0.428 &  1.71  & $\Xi^-$ &  0.4939 &  1.81   & 0.534 &  1.86    &  0.332 &  1.41   & 0.446 &  1.74      &     &    \\ 
 NL3$\omega\rho$ & no Y & 13.75& 2.75& 0.688  &&& &&& &&& & 0.517 &2.55 && && && \\ 

 & -10   & 13.74&2.27  & 0.712  & $\Sigma^-$ &  0.283 &  1.35  & $\Lambda$ &  0.341 &  1.70      &         &        &    & 0.434 &  1.99    &  0.349 &  1.73   & 0.341 &  1.70  & 0.572 &  2.20   \\ 
 & 10   & 13.75&2.30  & 0.722  & $\Sigma^-$ &  0.309 &  1.62  & $\Lambda$ &  0.317 &  1.67  & $\Xi^-$ &  0.562 &  2.22   & 0.488 &  2.14    &  0.317 &  1.67   & 0.320 &  1.69      &     &    \\ 
 & 30   & 13.75&2.31  & 0.763  & $\Lambda$ &  0.317 &  1.68  & $\Xi^-$ &  0.363 &  1.89  & $\Sigma^-$ &  0.423 &  2.05   & 0.535 &  2.22    &  0.317 &  1.68       &      &               &     &    \\ 

 \hline
 \hline
\bottomrule
\end{tabular*}}
\end{table*}

Table I summarizes,  for each model, several NS properties 
both  for  purely
nucleonic models (noY) and for hyperonic models with different values
of $U_{\Sigma}^{(N)}$:
the radius of the canonical $1.4 M_{\odot}$ NS ($R_{1.4}$);
the maximum mass ($M_{\rm max}$) and respective central baryonic number density ($n_{c, \rm max}$);
the onset densities of the hyperonic species that nucleate in stable stars;
the threshold densities of nucleonic and various hyperonic dUrca processes.
and corresponding NS masses with these central baryonic densities
One of the EoS (NL3$\omega \rho$) provides a maximum NS mass
in excess to the most stringent astrophysical constraint
$2.14^{+0.10}_{-0.09}M_{\odot}$ \cite{Cromartie2019}.
DDME2 agrees, within error bars, with the commonly adopted lower limit
of the maximum mass, corresponding to the pulsars PSR J$0348+0432$,
with $2.01 \pm 0.04M_{\odot}$\cite{Antoniadis13}.
Finally, DD2 and FSU2H predict slightly lower maximum masses,
though marginally consistent with the
above constraint from PSR J$0348+0432$ \cite{Antoniadis13}.

Other information from Table I, relevant for the present discussion,
is as follows:

\begin{itemize}

\item threshold densities for the onset of hyperons depend on the
  nucleonic EoS; the dispersion among the predictions corresponding
  to different EoS increases with the magnitude of the onset density;
  this is a consequence of the increasing uncertainties that affect
  the nucleonic EoS away from $n_0$,

\item irrespective the nucleonic EoS, the only hyperonic
  species that nucleate in cold catalyzed matter are $\Lambda$, $\Xi^-$
  and $\Sigma^-$; the explanation relies on the attractive character
  of $\Lambda N$ and $\Xi N$ interactions and dominance of negatively
  charged particles; note that other models \cite{Gusakov_MNRAS2014,Fortin16,Li2018EPJA}
  of hyperonic compact stars also allow for $\Xi^0$,

\item threshold densities for the onset of $\Sigma$ strongly depends
  on $U_{\Sigma}^{(N)}$; strongly/poorly repulsive potentials
  favor, as expected, late/early onset of $\Sigma$,

\item attractive $U_{\Sigma}^{(N)}$ potentials are responsible for a reduction
  of $M_{\rm max}$; the effect is nevertheless small, as is the abundance of
  $\Sigma$ in star matter,
  
\item the onset  baryonic density  of the nucleonic dUrca depends on the nucleonic EoS;
  the two density dependent models (DD2 and DDME2) do not allow for
  nucleonic dUrca to operate in stable stars, while the two models with
  non-linear couplings do allow for this process;
  the lowest onset density of the nucleonic dUrca occurs for FSU2H,
  with $0.42 \leq n_{DU} \leq 0.53$ fm$^{-3}$, which corresponds to
  stars with masses
  $1.61 \leq M_{DU}/M_{\odot} \leq 1.86$,

\item attractive $U_{\Sigma}^{(N)}$ potentials shift the density threshold
  of nucleonic dUrca to lower values; the rationale is that, by partially
  replacing the electrons which compensate for the positive charge of protons,
  $\Sigma^-$ alter the beta-equilibrium condition which determines the relative
  abundances of neutrons and protons,
  
\item depending on the EoS and $U_{\Sigma}^{(N)}$ different hyperonic dUrca processes
  are active; for the most repulsive (attractive) $U_{\Sigma}^{(N)}$ potential,
  the first hyperonic dUrca which operates is $\Lambda \to p+l+\tilde \nu_l$
  ($\Sigma^- \to \Lambda +l+\tilde \nu_l$).

\end{itemize}

\section{Observational data}
\label{sec:obsdata}

Observation of thermal evolution of NS can potentially
provide information on NS interiors.
Typically two classes of objects are considered.
The first class corresponds to isolated middle-aged ($10^2 - 10^6$ yr) NS,
whose effective surface temperatures $T_s^{\infty} \approx 10^6~{\rm K}$ or,
equivalently, effective photon luminosities
$10^{32} \lesssim L^{\infty}_{\gamma} \lesssim 10^{34}$ ${\rm erg\,s^{-1}}$
are negatively correlated with the star age.
Preeminence of thermal radiation in the measured total radiation spectrum and
thermal relaxation throughout the volume are the main arguments which allow
one to bridge measured temperatures and/or photon luminosities to
the multi-source neutrino emission from the core.

We here consider the observational data of 19 INS compiled in Table 1 of \cite{BY15}.
They span the age range $0.33 \leq t \leq 1300 $ kyr
and the surface effective temperature range $ 0.42 \leq T_{s}^{\infty} \leq 2$ MK.
In all cases $T_{s}^{\infty}$ corresponds to thermal emission from the entire
surface. Depending on the particularities of the observed spectra
and/or the extracted values of the NS radii in the observational paper(s) related to each source,
the NS atmosphere was assumed to be made of Fe, C or H.

The second class of objects corresponds to transiently accreting
quasi-stationary NS in low-mass X-ray binaries.
Contrary to NS in the first class, NS in this class heat-up because of the
energy deposited in the bottom layers of the crust by the material
which is intermittently accreted from the low mass companion.
As such, the thermal evolution is followed as a function of the
average accreted mass rate, averaged over periods of accretion and quiescence, which covers the range $10^{-12} \leq \dot M/\left(M_{\odot}\,{\rm yr}^{-1}\right) \leq 10^{-8}$. 
We remind that the heating process is due to a series of nuclear reactions
(electron capture, neutron absorption and emission, pycnonuclear reactions)
and its rate is estimated at $\approx 1-2$ MeV per accreted nucleon
\cite{HZ90,HZ08}. In order to model such accreting NS we employ the model of accreted crust from Ref.\cite{HZ08}. It is inconsistent with the one employed in the core as no such model exists for the EoS we employ.

We here consider the observational data of 24 XRT compiled in \cite{BY15,Potekhin2019} .
They span the effective photon luminosity range $L_{\gamma}^{\infty}= 4.9 \times 10^{30} - 5.3 \times 10^{33}$ erg/s
and the average accretion rate range $\dot{M}= 2.5 \times 10^{-12} -5.2 \times 10^{-9} M_{\odot}\,{\rm yr}^{-1}$.
Despite a large dispersion of data, one can identify a positive correlation
between the surface photon luminosity and the accretion rate.
With the exception of two, particularly faint objects
(SAX J1808.4$-$3658 and 1H 1905+000),
observational data fill uniformly the domain in
$L_{\gamma}^{\infty} - \dot{M}$.

These two faintest objects are of great interest as, according to the present understanding,
  they are the only NS whose thermal states
  require the (unsuppressed) dUrca to operate over important fractions of volume.
For 1H 1905+00 only an upper limit on the accretion rate and luminosity exist.
  SAX J1808.4$-$3658 (here after SAX J1808) has an accurately estimated accretion rate and
  an upper-limit of the luminosity; we remind that it is the less luminous XRT observed.

\section{Cooling simulations}
\label{sec:cool}

The thermal evolution of NS is typically studied by confronting
calculated cooling/heating curves obtained for a certain EoS
and given superfluidity (SPF) assumptions with observational data.
NS with masses $1 \leq M/M_{\odot} \leq M_{\rm max}$ are considered.

In the most general case the cooling tracks can be separated
in three families depending on whether dUrca processes operate or not.
As long as no dUrca is active, NS cool slowly (i).
If SPF in the core is disregarded, the cooling curves are independent of
both EoS and NS mass. 
As soon as one dUrca channel opens up, the cooling is accelerated.
If at least one of the involved baryonic species
is superfluid throughout the whole core, the cooling is moderated (ii).
The efficiency of the cooling increases steeply with
the amount of matter which accommodates dUrca
and the quenching of the SPF gap at high densities.
Cooling curves in this regime manifest increased sensitivity to EoS,
NS mass and SPF properties.
Finally, if at least one dUrca is allowed and it is not regulated by
SPF over  at least a finite fraction
of the whole volume, the cooling is fast (iii).
Similarly to (ii), cooling tracks in (iii) manifest strong dependence on EoS,
NS mass and SPF properties.

Within the steady state approximation,
Yakovlev {\em et al.} \cite{Yakovlev_AA_2003}
demonstrated that heating of XRT is equivalent to INS cooling.  XRT are in the photon or neutrino emission
regimes
depending on whether the energy deposited in the deep layers is
transported to the surface from which it is radiated away by photon emission
or, alternatively, spread all over the volume from which it is carried
away by neutrino emission. In the photon emission regime the surface temperature depends on the
accretion rate and is independent of the internal structure of the NS.
In the neutrino emission regime the surface temperature depends on the
internal structure, i.e. 
neutrino emission mechanisms and their quenching by SPF.
High photon surface luminosities require high accretion rates
and slow neutrino emission;
low photon surface luminosities require small accretion rates
and/or fast neutrino emission.
Photon and neutrino emission regimes of XRT correspond to photon and neutrino
cooling regimes of INS.
According to \cite{Yakovlev_AA_2003},
the only difference between thermal evolution
of INS and XRT consists in the low dependence of the latter on the
heat capacity of matter and the thermal conductivity of the isothermal
interior.

To perform our cooling simulations we use
an upgraded version of the public domain code
{\tt NSCool}\footnote{http://www.astroscu.unam.mx/neutrones/NSCool} by D. Page.
Neutrino emission from the core, which is the dominant cooling mechanism
for middle-aged INS and XRT, occurs via the following reactions:
a) nucleonic and hyperonic dURCA, with emissivities as in \cite{DUY92};
if one or both baryonic species involved in dUrca is paired,
suppression of emissivity is implemented as in \cite{Levenfish_1994},
b) modified Urca involving nucleons, with emissivities and emissivity suppression
as in \cite{Yakovlev_AA_1995},
c) bremsstrahlung from nucleon-nucleon collisions \cite{Yakovlev_AA_1995},
and, in case of paired baryons,
d) Cooper PBF in $S$- and $P$- wave channels;
emissivities in these two channels are treated as in
\cite{Leison_PLB_2006,Leinson2016,Fortin2018} with the vector part of the
PBF process strongly suppressed and the axial part negligible for the
$^1S_0$ pairing and reduced for $^3P_2$ pairing.
Modified Urca involving hyperons and bremsstrahlung from hyperon-baryon collisions are
  disregarded as their emissivities are subdominant to dUrca \cite{Yakovlev_PhysRep_2001}.

Baryonic species with attractive interactions are known to
experience pairing.
Neutrons in the crust
and protons in the core manifest $^1S_0$ pairing, due to their relatively low densities;
besides, neutrons in the core manifest $^3P_2-^3F_2$ pairing.
Calculations performed using different interactions and many-body
techniques have
shown that the magnitude of the pairing gaps strongly depends on the
nucleon-nucleon interactions and many-body correlations;
for a recent review, see \cite{Sedrakian_EPJA_2019}.
Good constraints on the neutron-neutron interaction coming from scattering
data  determine  rather well the extension of the neutron pairing gap in the $^1S_0$ channel,
expressed in terms of the Fermi momentum.  
The size of this gap and the way in which it is quenched as the density increases
are affected by the treatment of many-body effects.
For instance, \cite{Ding_PRC_2016} gets a reduction of 1.2 MeV
(or a factor of 1.7) when
the effects of both short- and long-range correlations are employed,
with respect to the case in which the calculations are performed
within the Bardeen-Cooper-Schrieffer (BCS) theory. 
Proton $^1S_0$ and neutron $^3P_2-^3F_2$ pairing gaps are affected by larger errors.
For a review of proton $^1S_0$ pairing gaps in NS matter, see \cite{Page_2004}.
Neutron $^3P_2-^3F_2$ pairing in pure neutron matter has been recently addressed
in \cite{Ding_PRC_2016}, who showed that the size of the gap may be modified
by a factor of 3 (50) when different bare nucleon-nucleon
interactions (many-body effects) are employed.
When short- and long-range correlations are accounted for,
the largest neutron $^3P_2-^3F_2$ pairing gap obtained in \cite{Ding_PRC_2016}
corresponds to the Entem-Machleit N3LO \cite{N3LO} potential;
it extends over $1.2 \lesssim k_F \lesssim 2.7~{\rm fm}^{-1}$,
with a maximum of $\approx 0.2$ MeV,
obtained for $k_F\approx 1.9~{\rm fm}^{-1}$.

Neutron $^1S_0$ pairing in the crust regulates
the thermalization of the crust and can only have observable
effects before the crust and the core reach thermal equilibrium.
As this is not the case of NS of interest for us,
we do not explore its effects and use only one value,
calculated in \cite{SFB_2003} by accounting for
long-range correlations (polarization effects).

Proton $^1S_0$ pairing in the core suppresses
the neutrino emission from nucleonic and
hyperonic $\Lambda \to p+l+\tilde \nu$ dUrca reactions.
For internal temperatures lower than the critical temperature
for pairing, it also leads to neutrino emission from PBF.
Uncertainties related to it are accounted for in this paper
by considering
two {\it extreme} scenarios: BCLL from \cite{Baldo_NPA1992}
and CCDK from \cite{Chen_NPA1993}.
The BCLL \cite{Baldo_NPA1992} gap is calculated in the frame of the
standard BCS theory, using the first-order (bare) particle-particle
interaction as the pairing interaction;
for the nucleon-nucleon interaction the Argonne v14 potential
\cite{Argonne_v14} is used;
the gap extends over $0.1 \lesssim k_F \lesssim 1~ {\rm fm}^{-1}$,
with a maximum size of $\approx$0.8 MeV,
obtained for $k_F \approx 0.4~{\rm fm}^{-1}$.
The CCDK \cite{Chen_NPA1993} gap is calculated based
on the matrix elements extracted from Reid-soft-core potential
\cite{Chao_NPA_1972}; it extends over
$0 \lesssim k_F \lesssim 1.3~{\rm fm}^{-1}$,
with a maximum size of $\approx$1 MeV,
obtained for $k_F \approx 0.7~{\rm fm}^{-1}$.

Neutron $^3P_2-^3F_2$ pairing in the core has a large impact on the
thermal evolution, as neutrons represent the dominant species in NS cores.
For NS with baryonic particle densities smaller than the threshold
for nucleonic dUrca, neutron $^3P_2-^3F_2$ pairing will accelerate
the cooling by neutrino emission from PBF.
For NS which allow for nucleonic dUrca, it will contribute
to the suppression of the emissivity.
This effect is nevertheless expected to be negligible, considering
that nucleonic dUrca is already strongly quenched by proton $^1S_0$ pairing,
whose gap size is larger than the one of the neutrons.
Finally, stars in the photon cooling era will cool much faster,
due to the strong reduction of the heat capacity.
In order to restrain the parameter space of our study,
in this work we assume a vanishing pairing gap for neutrons in the
$^3P_2-^3F_2$ channel.
All the implications of this assumption are such that agreement
with thermal data of INS is facilitated:
larger surface temperatures will be obtained for 
youngest and warmest INS and
oldest and coolest INS.
We note that this assumption has  already  been employed in cooling studies
\cite{Grigorian2018,Taranto,Beznogov_PRC_2018}.

The measured surface temperatures of NS depend on the composition
of the atmosphere. Atmospheres made of light atoms
(e.g. hydrogen, helium and carbon)
are thought to correspond to young and accreting NS while
old NS are considered to have atmospheres made of 
heavy atoms (iron).
In the evolutionary stages in which the energy loss is due to neutrino emission,
and is thus dependent on the internal temperature of the star,
light atom atmospheres lead to higher effective surface temperatures
than heavy atom atmospheres.
When, on the contrary, the energy loss depends on the surface temperature
itself higher temperatures are obtained for atmospheres made of heavy elements.
In this work we shall consider, for all simulations, 
atmospheres  both entirely made of hydrogen and of iron.
The relations between the temperature of the outer boundary
of the isothermal internal region and the surface
are implemented as in \cite{PY03}.
As such, the effective surface temperature will correspond to the two limiting cases corresponding
to the absence of light elements (iron atmospheres)
and a maximum amount of them (fully accreted atmospheres),

We neglect effects due to rotation, magnetic field and late heating.

\section{Results}
\label{sec:res}

As specified in Sec.~\ref{sec:eos}, the four selected nucleonic EoS,
DDME2, DD2, FSU2H and NL3$\omega\rho$, comply with the two solar mass
constraint on the lower limit of the maximum mass when the hyperonic degrees of
freedom are allowed.
In this section we consider all these models and discuss to what extent
the cooling/heating curves predicted by each one are able to describe
the observational data of INS and XRT. 

For each EoS we solve the  Tolman–Oppenheimer–Volkoff equations and we build non-rotating spherically-symmetric
configurations
for masses ranging from just below the mass threshold for the first dUrca process to the maximum mass.
Cooling/heating curves of these stars are simulated
under three scenarios for proton $^1S_0$ SPF in the core:
no SPF;
narrow pairing gap (BCLL) \cite{Baldo_NPA1992}; wide pairing gap (CCDK) \cite{Chen_NPA1993}.
As explained in Sec.~\ref{sec:cool},  neutron $^1S_0$ SPF in the crust
is implemented according to \cite{SFB_2003}
and neutron $^3P_2-^3F_2$ SPF in the core is disregarded.
Hyperon SPF is disregarded as well.
For INS we use atmosphere models with the same composition as the one employed
when the temperature was determined from x-ray observations:
pure Fe, H and, C for two sources (Cas A and XMMU J1731$-$347).
For XRT we consider alternatively atmospheres entirely made of Fe and H.
For accreting stars, we consider average accretion rates in the domain
$10^{-13} \leq \dot M /\left(M_{\odot}\,{\rm yr}^{-1}\right)\leq 10^{-8}$.

\subsection{Nucleonic stars}

We briefly discuss the thermal evolution curves 
of purely nucleonic stars, as predicted by the four EoS models we employ:
DDME2, DD2, FSU2H and NL3$\omega\rho$.
As specified in Table I, nucleonic dUrca 
does not operate in stable stars built upon DDME2 and DD2 models;
stars built upon FSU2H and NL3$\omega\rho$ models allow for dUrca
but only in very
massive stars with masses in excess of 2.27$M_\odot$ and, respectively, 2.55$M_\odot$.
As a consequence, cooling/heating tracks of INS/XRT are expected to be unable to account for the
important dispersion manifested by the thermal data, even if proton superfluidity is accounted for.

\begin{figure}
\includegraphics[width=1\linewidth]{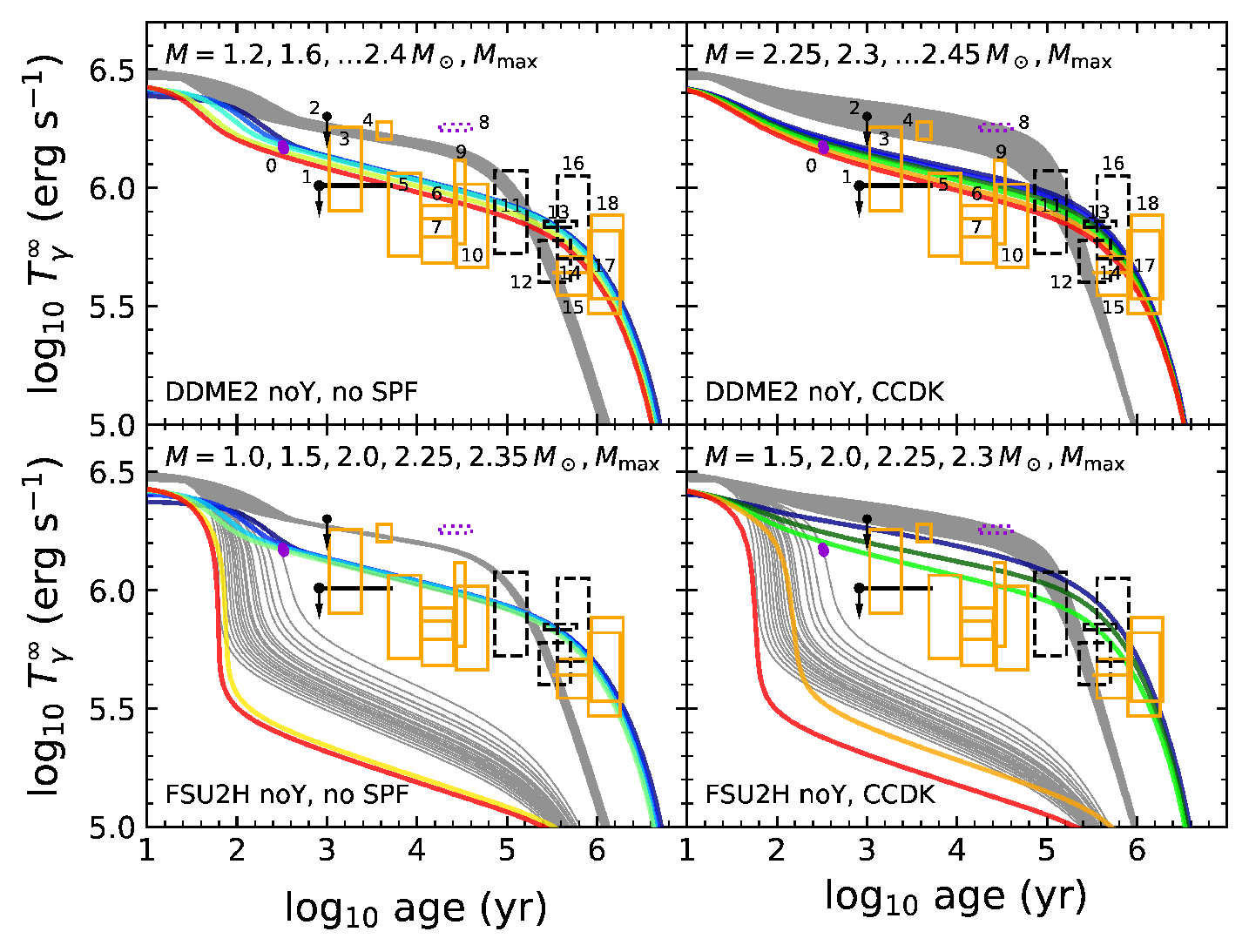}
\caption{Thermal states of INS built upon the DDME2 (top) and FSU2H (bottom)
  EoS and different scenarios for proton $S$-wave SPF: no SPF (left) and CCDK (right).
  The colored lines correspond to calculations employing an iron atmosphere for the
  masses indicated on each plot.
  The gray lines correspond to a hydrogen atmosphere and masses between 1$M_\odot$
  and the maximum mass, with a step of 0.005$M_\odot$.
  The observational data, taken from \cite{BY15}, are represented with  orange boxes
  for a hydrogen atmosphere, violet for a carbon one and black for an iron one.
  We employ $2-\sigma$ error bars, if these are available;
  otherwise a factor of 0.5 and 2 in both the temperature and the age (except for upper limits).
  The sources are 0 - CasA NS, 1 - PSR J0205+6449 (in 3C58), 2 - PSR B0531+21 (Crab), 3 - PSR J1119$-$6127, 4 - RX J0822-$-$300 (in PupA), 5 - PSR J1357$-$6429, 6 - PSR B1706$-$44, 7 - PSR B0833$-$45 (Vela), 8 - XMMU J1731$-$347, 9 - PSR J0538$+$2817, 10 - PSR B2334$+$61, 11 - PSR B0656$+$14, 12 - PSR B0633$+$1748 (Geminga), 13 - PSR J1741$-$2054, 14 - RX J1856.4$-$3754, 15 - PSR J0357+3205 (Morla), 16 - PSR B1055$-$52, 17 - PSR J2043+2740, 18 - RX J0720.4$-$3125.}
\label{fig:noY_INS}
\end{figure}

\begin{figure}
\includegraphics[width=1\linewidth]{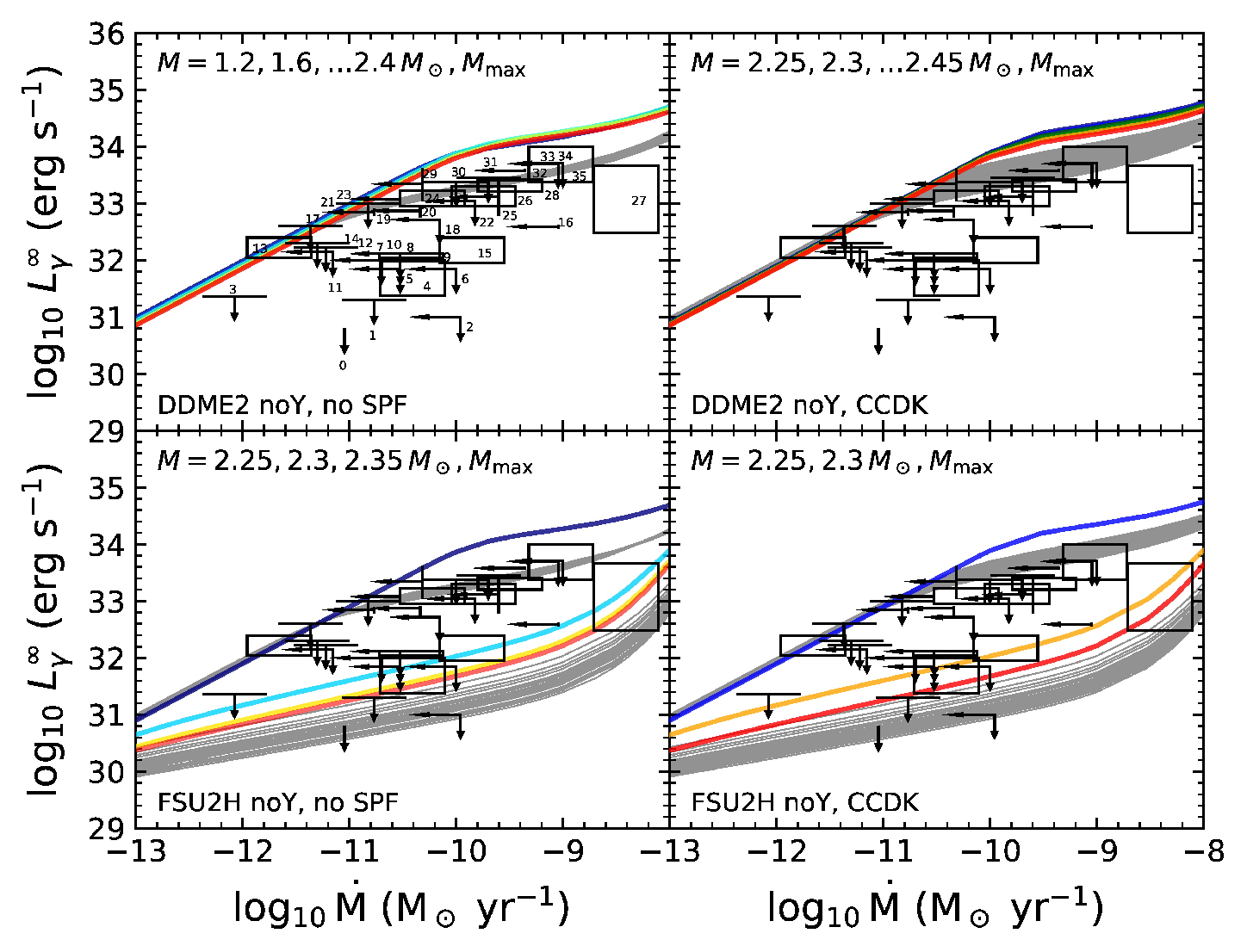}
\caption{Thermal states of XRT built upon the DDME2 (top) and FSU2H (bottom) EoS.
  Different scenarios are considered for proton $S$-wave SPF: no SPF (left) and
  CCDK gap (right).
  The colored lines correspond to calculations employing a hydrogen atmosphere
  and a mass sequences indicated on each plot.
  The gray lines correspond to an iron atmosphere and NS masses between 1$M_\odot$ and
  the maximum mass, with a step of 0.005$M_\odot$.
  The observational data, taken from \cite{Potekhin2019}, are represented with
  a factor of 0.5 and 2 in both the luminosity and the accretion rate
  (except for upper limits).
  For the source SAX J1808, we show data from \cite{BY15} labeled by
  0 and \cite{Potekhin2019} designated by 1.
  The other sources are: 2 - 1H 1905+00, 3 - NGC 6440 X$-$2, 4 - HETE J1900.1, 5 - EXO 1747, 6 - IGR J18245, 7 - XTE J0929, 8 - Ter 5 X$-$1, 9 - Cen X$-$4, 10 - XTE J1807, 11 - XTE 2123, 12 - XTE J1814, 13 - IGR J00291, 14 - SAX J1810.8, 15 - MXB 1659$-$29, 16 - KS 1731$-$260, 17 - XTE J1751, 18 - 2S 1803$-$245, 19 - Ter 5 X$-$2, 20 - 1RXS J180408, 21 - Swift J1756.9, 22 - XB 1732$-$304, 23 - Ter 5 XR3, 24 - NGC 6440 X$-$1, 25 - 1M 1716$-$315, 26 - XTE J1709, 27 - 4U 2129+47, 28 - Aql X$-$1, 29 - 4U 1730$-$22, 30 - GRS 1747, 31 - SAX J1750.8, 32 - EXO 0748, 33 - XTE J1701, 34 - MAXI J0556, 35 - 4U 1608$-$522.}
\label{fig:noY_XRT}
\end{figure}

This situation is illustrated
in Figs.~\ref{fig:noY_INS} and \ref{fig:noY_XRT}, where the predictions of
DDME2 and FSU2H EoS are confronted with
the thermal states of INS and, respectively, XRT.
Two {\it extreme} assumptions are made in what regards
the proton $S$-wave SPF gap:
vanishing gap (left panels) and wide and large gap (right panels);
for the latter situation we use the parameterization CCDK in \cite{Chen_NPA1993}.
Results for the two models not shown, DD2   and NL3$\omega\rho$, %\new{(not shown here)}
are similar to those corresponding, respectively, to DDME2  and FSU2H.

Independently of how the proton $S$-wave SPF gap is
varied between the above mentioned limiting cases,
the observational data can not be
described. Cooling/heating curves produced by EoS models that do not allow for nucleonic dUrca
form a unique narrow band. In the case of INS, this band would pass through most of the data
if all stars would have a Fe atmosphere, which seems not to be compatible with spectral analyses.
In the case of XRT, agreement is obtained only for stars with intermediate and high luminosity
(high luminosity) and low (medium to high) accretion rates if H (Fe) atmosphere models are assumed.

Thermal evolution tracks generated by models that allow for nucleonic dUrca split into two narrow
bands separated by a wide gap.
The high luminosity bands span the same $T_s-t$ and $T_s-\langle \dot
M \rangle$ domains as
those spanned by models that do not allow for dUrca;
for INS the low luminosity band explores a region where there is no experimental data;
for XRT, it allows one to describe the coolest stars, whose average accretion rates cover the whole
range of values. Agreement with these stars nevertheless requires,
for the presently considered models,
very unlikely masses: $\approx 2.25 M_\odot$ ($2.5 M_\odot$) for FSU2H (NL3$\omega\rho$).
Quantitative differences between results corresponding to various scenarios of proton $S$-wave SPF
regard the widening of the high luminosity band and a slight shift towards higher
luminosities of tracks corresponding to stars which allow for dUrca only
in a tiny fraction of the core.
The stronger the pairing gap the larger the amplitude of these modifications.
The first modification is due to the opening of Cooper PBF and the second
to the partial suppression of dUrca reactions by SPF.

The task of simultaneously describing the two lots of available data seem to indicate
that: i) dUrca is needed and the minimum NS mass that allows for it should be significantly
smaller than the present lower bound on NS maximum mass, $\approx 2M_{\odot}$,
ii) its efficiency should be regulated by SPF.
The case of purely nucleonic
stars has been successfully considered by \cite{BY15,Beloin_PRC_2019}.
As none of the EoS models described in Sec. \ref{sec:eos}
allows for nucleonic dUrca in stars with masses lower than the maximum measured NS mass,
we shall focus on hyperonic dUrca.  
  
\subsection{Hyperonic stars}

We next discuss the cooling/heating curves obtained for NS with hyperonic
degrees of freedom, whose properties are reviewed in Sec.~\ref{sec:NS}.
The same RMF models considered in the previous section are employed.
We will start the discussion taking DDME2 as reference, and will next complete the
discussion looking at FSU2H, NL3$\omega\rho$ and DD2.

\begin{figure}
\includegraphics[width=1\linewidth]{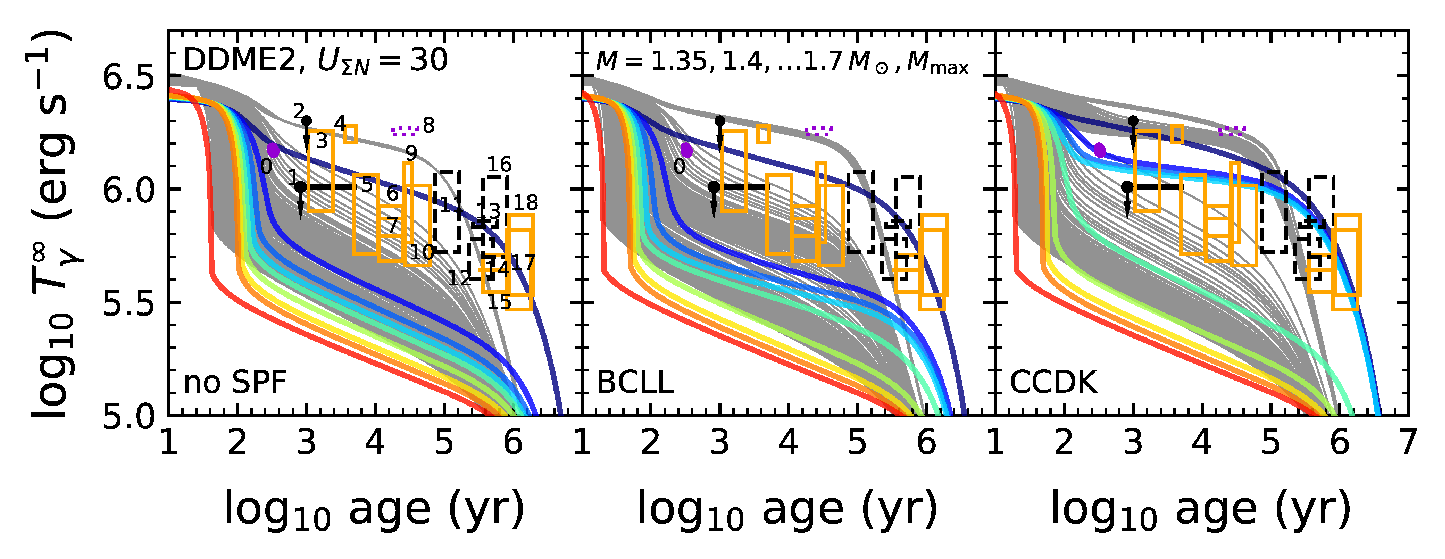}\\
\includegraphics[width=1\linewidth]{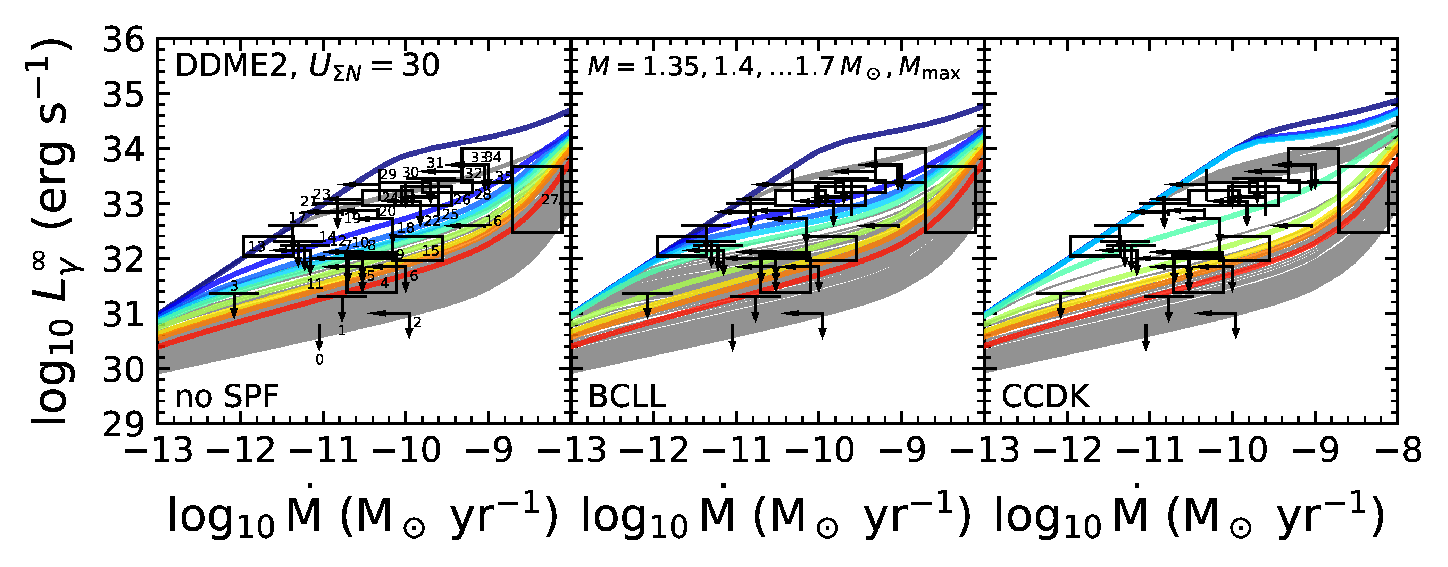}
\caption{Thermal states of INS (top) and XRT (bottom) for NS built upon
  the hyperonic DDME2 EoS with $U_\Sigma^{(N)}=+30$ MeV.
  Three superfluid scenarios are used for proton $^1S_0$ pairing:
  no gap (left), BCLL gap (middle) and CCDK gap (right).
  For INS (XRT) the mass sequences and color key are
  as explained in Fig. \ref{fig:noY_INS} (\ref{fig:noY_XRT}).
}
\label{fig:Usig30}
\end{figure}

The effect of including the proton $S$-wave pairing in the core is illustrated
in Fig.~\ref{fig:Usig30} for DDME2 with $U_\Sigma^{(N)}=30$ MeV, for
INS (top panels) and XRT (bottom panels).
Left, middle and right panels correspond, in this order, to vanishing, BCLL and
CCDK gaps.
When proton SPF is disregarded, the inclusion of hyperons makes a big difference
in the cooling/heating curves:
it modifies the evolution of stars with masses $\gtrsim 1.4\, M_\odot$,
which suffer fast cooling from an age of $\approx 10^2$yr.
The  main cooling agent is the $\Lambda \to p+l+\tilde \nu_l$ 
and, for $M \gtrsim 1.6M_{\odot}$ , also $\Sigma^- \to \Lambda+l+\tilde \nu_l$
(see Table I and comments in Sec.~\ref{sec:NS}).
Agreement with data requires that all NS have masses below 1.4-1.5 $M_\odot$.
  Even so thermal states of INS \#1, \#8, \#17 and \#18 can not be
  described.

If the proton $^1S_0$-pairing is implemented according to BCLL,
no improvement is obtained for the description of INS and XRT. 
  This means that proton Fermi momentum in the core of NS as small as $1.4 M_{\odot}$ 
  already exceeds the maximum value for which BCLL provides sizable pairing gaps,
  which translates into an insufficient suppression of neutrino emission
  from $\Lambda \to p+l+\tilde \nu_l$.
  If the $^1S_0$-pairing of protons is implemented according to CCDK,
  the cooling of stars with $M/M_{\odot}=1.4, 1.45$ is slowed down significantly.
  Thermal tracks of NS models with $1.5 \lesssim M/M_{\odot} \lesssim 1.6$ 
  are marginally better or identical to those obtained for BCLL, which means
  that the innermost shells feature unpaired protons.
  Stars with $M/M_{\odot} \gtrsim 1.6$ also accommodate for
  $\Sigma^- \to \Lambda+l+\tilde \nu_l$ dUrca which further enhances their cooling.
  Not accounting for hyperonic pairing, this second reaction operates at full power.

The best overall agreement with INS data is obtained for CCDK.
We notice that if an hydrogen atmosphere is employed for INS \#14, \#17 and \#18,
as suggested by spectral analyses, our simulations undershoot the observational data.
Similarly, if a carbon atmosphere is employed for INS \#8, its thermal state can
only be reproduced by models with $M/M_\odot \lesssim 1.35$.

For XRT the situation is more delicate as little is known about their
atmospheres.
If one assumes a H atmosphere our NS models require narrow proton SPF gaps (BCLL)
in order to reproduce the data.
If, alternatively, wide proton SPF gaps (CCDK) are used, the agreement with data requires
quite high masses, above 1.5$M_\odot$, for a large number of stars.
The particular case of SAX J1808 requires a more careful discussion and is,
thus, left for later.

\begin{figure*}
\includegraphics[width=1\linewidth]{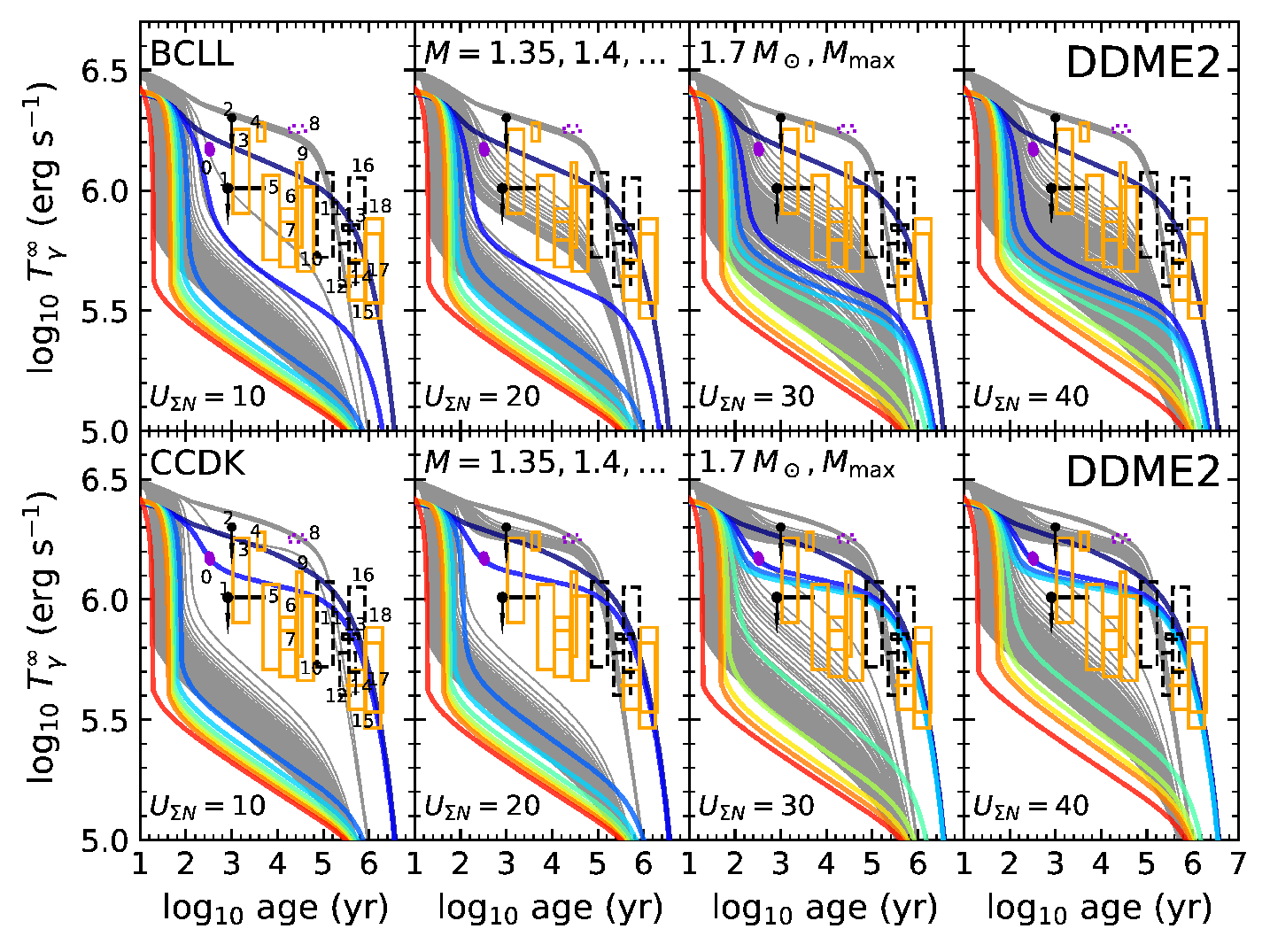}
\caption{Thermal states of INS built upon the hyperonic DDME2 EoS with
  $U_\Sigma^{(N)}=10,\, 20,\, 30,\,40$ MeV and the BCLL (top) and  CCDK (bottom) pairing models.
Mass sequences and color key are as explained in Fig. \ref{fig:noY_INS}.}
\label{fig:UsigINS}
\end{figure*}

\begin{figure*}
\includegraphics[width=1\linewidth]{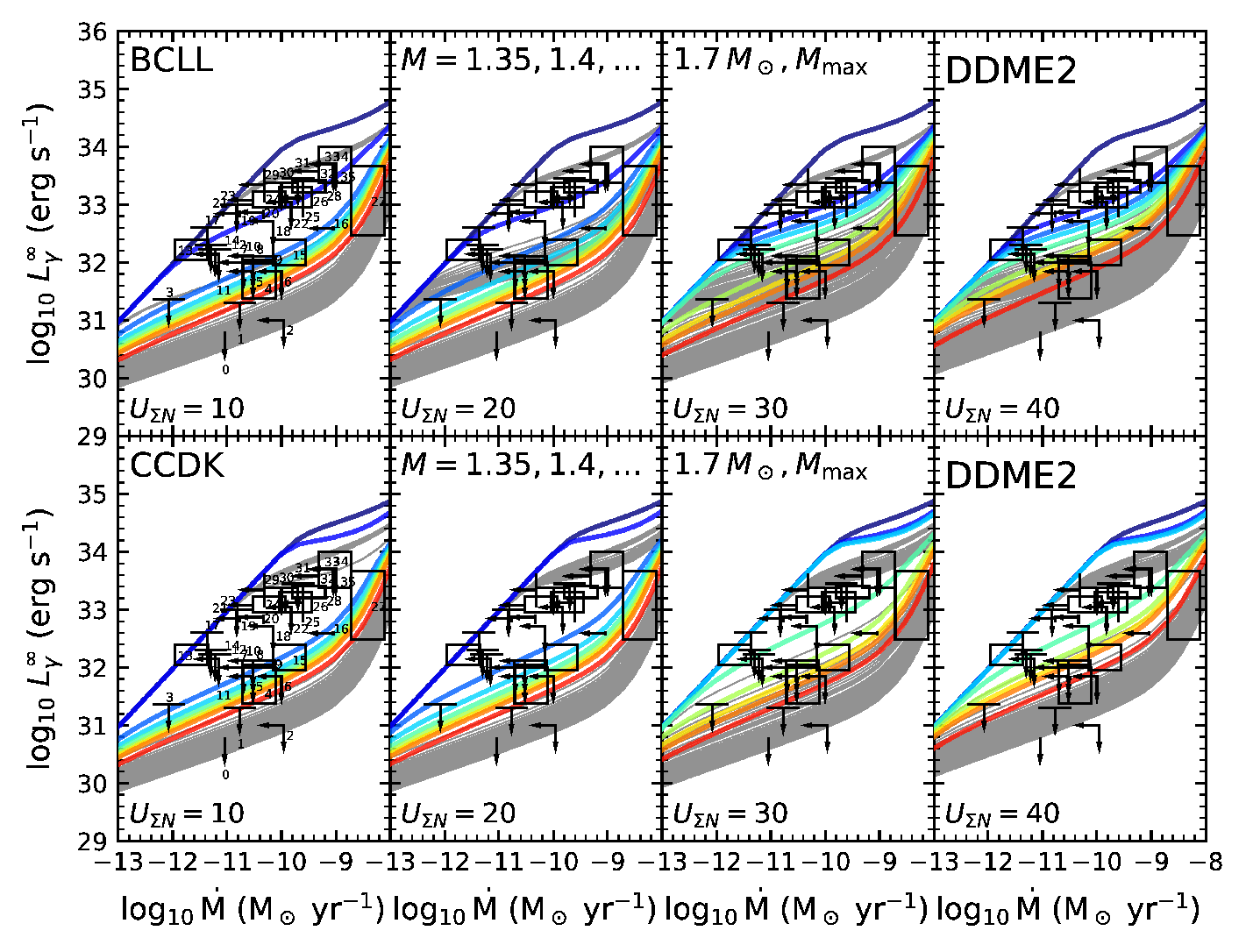}
\caption{Thermal states of XRT built upon the hyperonic DDME2 EoS with
  $U_\Sigma^{(N)}=10,\, 20,\, 30,\,40$ MeV and the BCLL (top) and CCDK (bottom) 
  models for proton $S$-wave pairing.
  Mass sequences and color key are as explained in Fig. \ref{fig:noY_XRT}.}
\label{fig:UsigSXT}
\end{figure*}

In order to discuss the effect of the magnitude of the $\Sigma N$ potential,
Figs.~\ref{fig:UsigINS} and \ref{fig:UsigSXT} illustrate
  thermal evolution tracks of, respectively, INS and XRT built upon the DDME2
  model when different values of the $U_\Sigma^{(N)}$ potential between $10$ and $40$ MeV
  are employed.
Less repulsive potentials are not shown because they give results even less compatible
with observations.
For INS the effect of $U_\Sigma^{(N)}$ is distinctly seen:
the more repulsive the $U_\Sigma^{(N)}$ the better is the temperature-age plane covered.
Irrespective the value of $U_\Sigma^{(N)}$ and the proton SPF model,
INS \#14, \#17 and \#18 fail to be described.
Under the assumption of a narrow proton SPF gap, also INS \#8 fails to be described.
The most consistent reproduction of INS data is obtained for
$U_\Sigma^{(N)}=30$ MeV and the CCDK model for proton SPF.
A higher value, {\it e.g.} $U_\Sigma^{(N)}=40$ MeV, can not be used
as it would result in 
strong constraints on the mass of half of the INS: $ M\gtrsim 1.5 M_\odot$.

In what regards XRT one notes that
the narrow proton SPF gap makes
evolutionary tracks more consistent with data in the sense that
the heating curves cover more
uniformly the luminosity-accretion rate plane.
Large values of $U_\Sigma^{(N)}$ improve the description but, as in the case of INS,
they should not exceed 40 MeV.
The largest considered value of $U_\Sigma^{(N)}$ 
requires relatively small masses for a
large number of stars and does not allow the observational point \#1 of SAX J1808 to be described.
We will come back to SAX J1808 later in the discussion.

\begin{figure}
\includegraphics[width=1\linewidth]{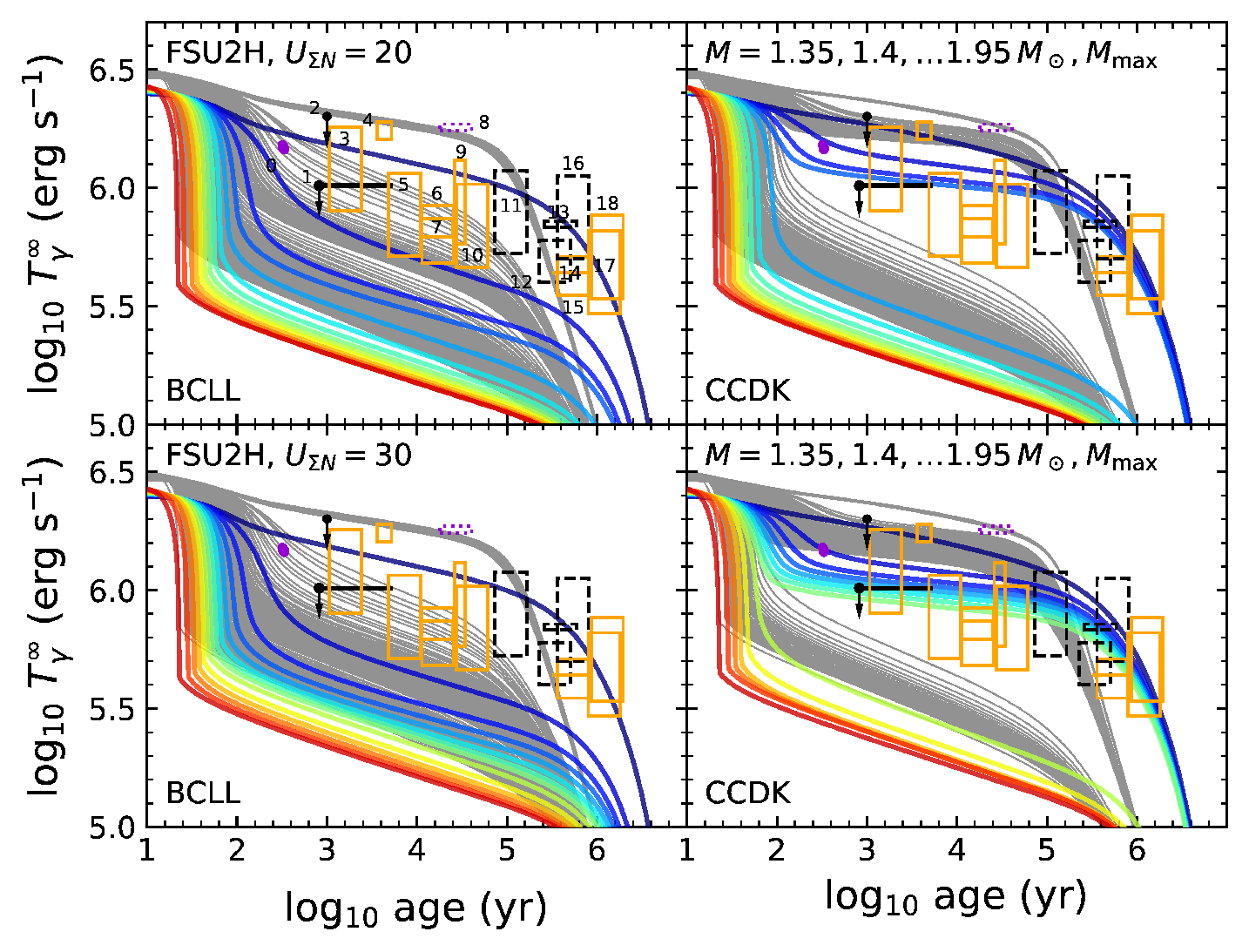}
\caption{Thermal states of INS built upon the hyperonic FSU2H EoS with
  $U_\Sigma^{(N)}=20$ (top) and $30$ MeV (bottom) for the BCLL (left) and  CCDK (right)
  models for proton $S$-wave pairing.
  Mass sequences and color key are as explained in Fig. \ref{fig:noY_INS}.}
\label{fig:fsu2hINS}
\end{figure}

\begin{figure}
\includegraphics[width=1\linewidth]{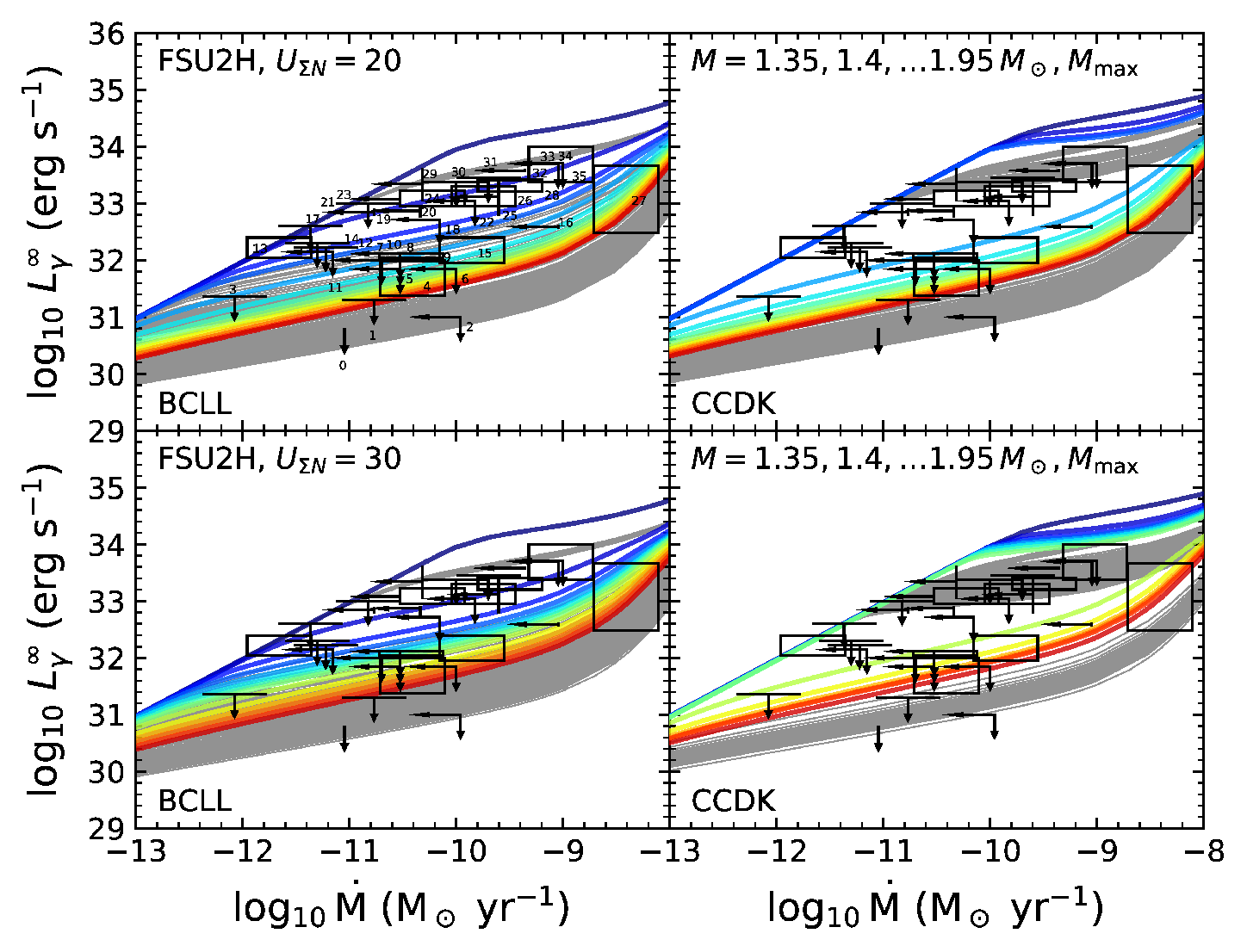}
\caption{Thermal states of XRT built upon the hyperonic FSU2H EoS with
  $U_\Sigma^{(N)}=20$ (top) and $30$ MeV (bottom) for the BCLL (left) and CCDK (right)
  models for proton $S$-wave pairing.
  Mass sequences and color key are as explained in Fig. \ref{fig:noY_XRT}.
}
\label{fig:fsu2hSXT}
\end{figure}

We next consider FSU2H.
Data in Table I show that for  $U_\Sigma^{(N)}=-10$ and 10 MeV
  $\Lambda \to p+e+\tilde \nu_e$ and $\Sigma^- \to \Lambda + e+\tilde \nu_e$ dUrca processes
  start operating in NS with masses similar to those obtained when DDME2 is employed;
  for $U_\Sigma^{(N)}=30$ MeV  $\Sigma^- \to \Lambda + e+\tilde \nu_e$ dUrca is allowed only in
  relatively massive stars; for $U_\Sigma^{(N)}=-10$ MeV also $\Sigma^- \to n+e+\tilde \nu_e$
  is allowed, but only in NS with masses near the largest NS masses measured so far;
  the most notable difference with respect to DDME2 is that the nucleonic dUrca is now allowed.
  For the most repulsive $U_\Sigma^{(N)}$ potentials it nevertheless operates in NS massive enough
  to not significantly change the picture obtained when using DDME2.
  Thermal evolution tracks provided by FSU2H are compared with INS and XRT data in Figs.
  \ref{fig:fsu2hINS} and \ref{fig:fsu2hSXT}. Top (bottom) panels correspond to
  $U_\Sigma^{(N)}=20$ MeV ($U_\Sigma^{(N)}=30$ MeV); left (right) panels correspond to BCLL (CCDK)
  proton SPF gaps. As it was the case with DDME2, INS data favor $U_\Sigma^{(N)}=30$ MeV and CCDK.
  We note that in this case also \#14 is well reproduced and that
  agreement with data requires models with $M \lesssim 1.5M_\odot$;
  narrow proton SPF gaps (BCLL) do now allow thermal states of stars \#8, \#14, \#17 and \#18 to be
  reproduced.
  In what regards XRT, the situation is similar to the one obtained for DDME2:
  agreement with data is favored for narrow proton SPF gaps irrespective the value of $U_\Sigma^{(N)}$.
  Wide proton SPF gaps and repulsive $U_\Sigma^{(N)}$ potentials would require $M/M_\odot \gtrsim 1.75 $
  in order to describe the data.

\begin{figure}
\includegraphics[width=1\linewidth]{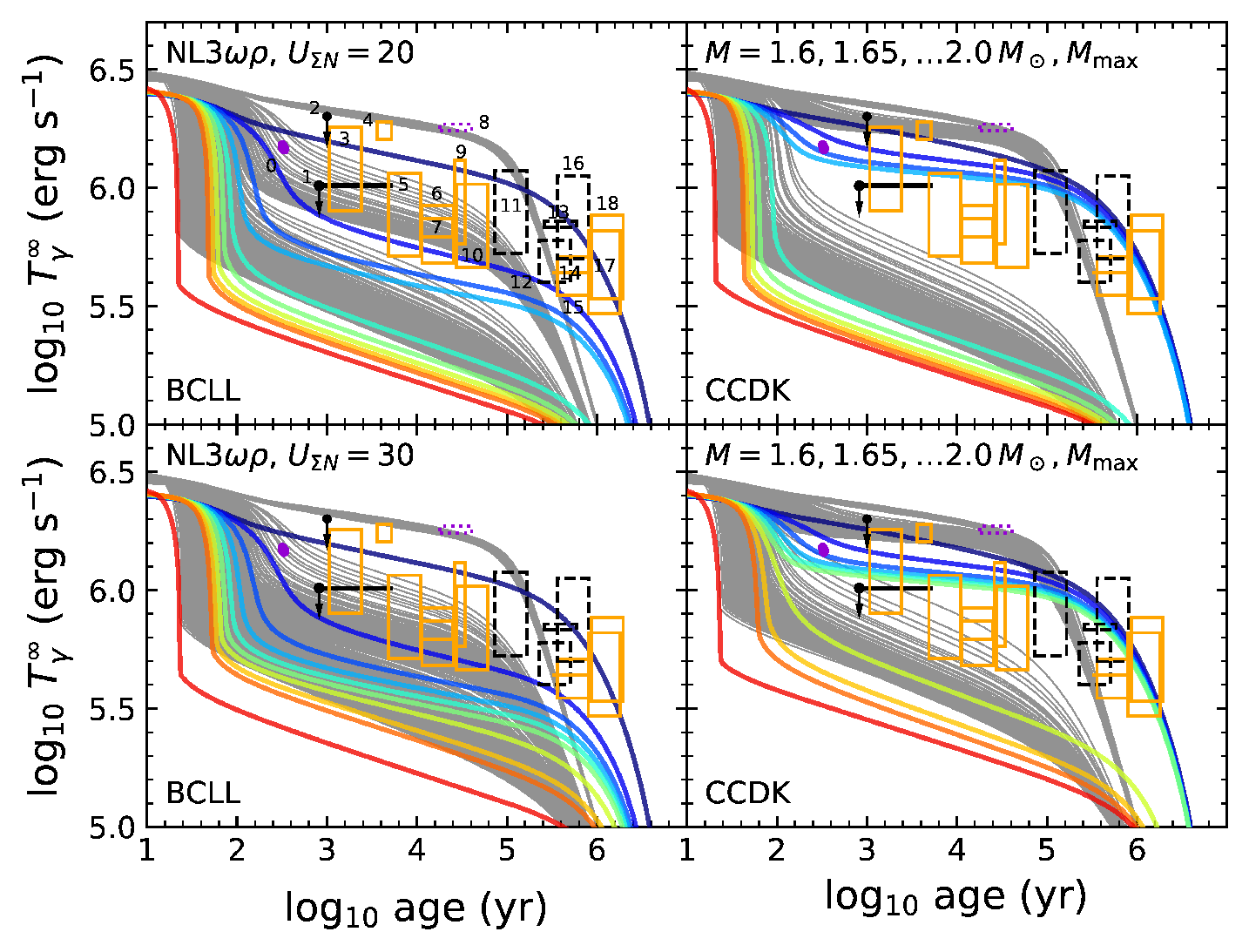}
\caption{Thermal states of INS built upon the hyperonic NL3$\omega\rho$ EoS with
  $U_\Sigma^{(N)}=20$ (top) and $30$ MeV (bottom) for the BCLL (left) and CCDK (right)
  models for proton $S$-wave pairing.
  Mass sequences and color key are as explained in Fig. \ref{fig:noY_INS}.}
\label{fig:nl3wrINS}
\end{figure}

\begin{figure}
\includegraphics[width=1\linewidth]{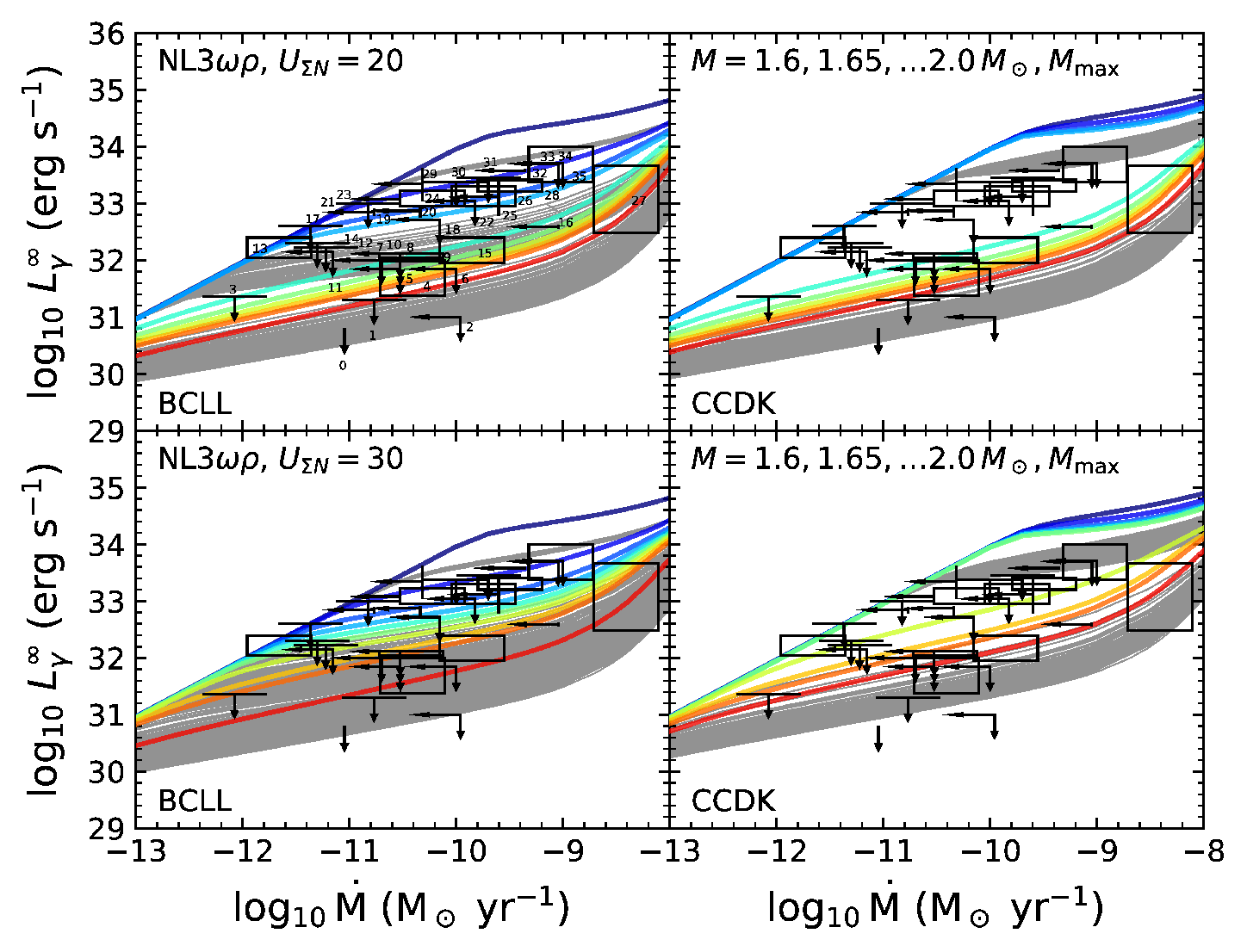}
\caption{Thermal states of XRT built upon the hyperonic NL3$\omega\rho$ EoS with
  $U_\Sigma^{(N)}=20$ (top) and $30$ MeV (bottom) for the BCLL (left) and
  CCDK (right) models for proton $S$-wave pairing.
  Mass sequences and color key are as explained in Fig. \ref{fig:noY_XRT}.}
\label{fig:nl3wrSXT}
\end{figure}

We now switch to NL3$\omega\rho$. Data in Table I show that the same dUrca processes
  allowed by FSU2H may operate but, irrespective the value of $U_\Sigma^{(N)}$,
  they are active in stars more massive than those obtained for FSU2H.
  Comparison with data corresponding to INS and XRT is provided in Figs. \ref{fig:nl3wrINS} and,
  respectively, \ref{fig:nl3wrSXT}. As it was the case with FSU2H, top (bottom) panels
  illustrate results corresponding to $U_\Sigma^{(N)}=20$ MeV ($U_\Sigma^{(N)}=30$ MeV),
  while left (right) panels consider BCLL (CCDK) proton SPF gaps.
  As before, agreement with INS data is favored for CCDK;
    more precisely, for CCDK the only INS whose states are not reproduced by
    NL3$\omega\rho$ are \#17 and \#18;
    when BCLL is employed also INS \#8 fails to be described and reproduction of thermal states
    of many other stars is possible only assuming large masses, $M/M_\odot \gtrsim 1.65$.
    We note that the mass distribution is similar for $U_\Sigma^{(N)}=20$ and 30 MeV,
    which seems to be more affected by the superfluid model.
    In what regards XRT, the situation is qualitatively similar to those
    obtained above: better agreement with data is obtained for BCLL and, with the exception
    of SAX J1808,
    $U_\Sigma^{(N)}=20$ and 30 MeV provide similar results.  

\begin{figure}
\includegraphics[width=1\linewidth]{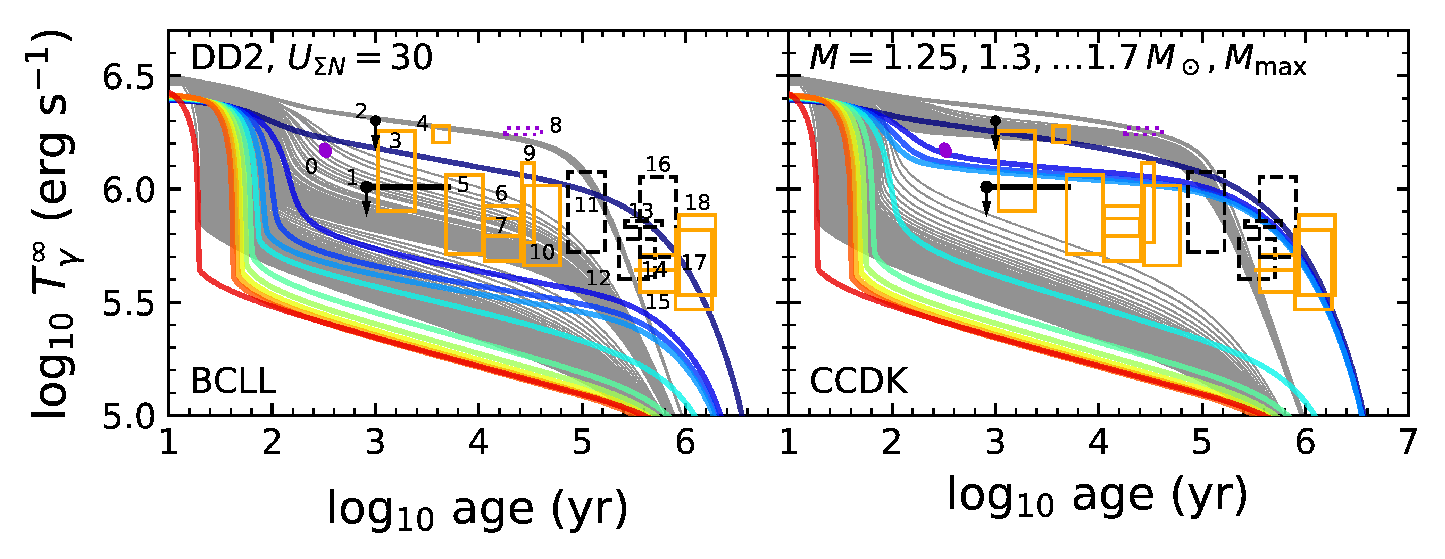}
\includegraphics[width=1\linewidth]{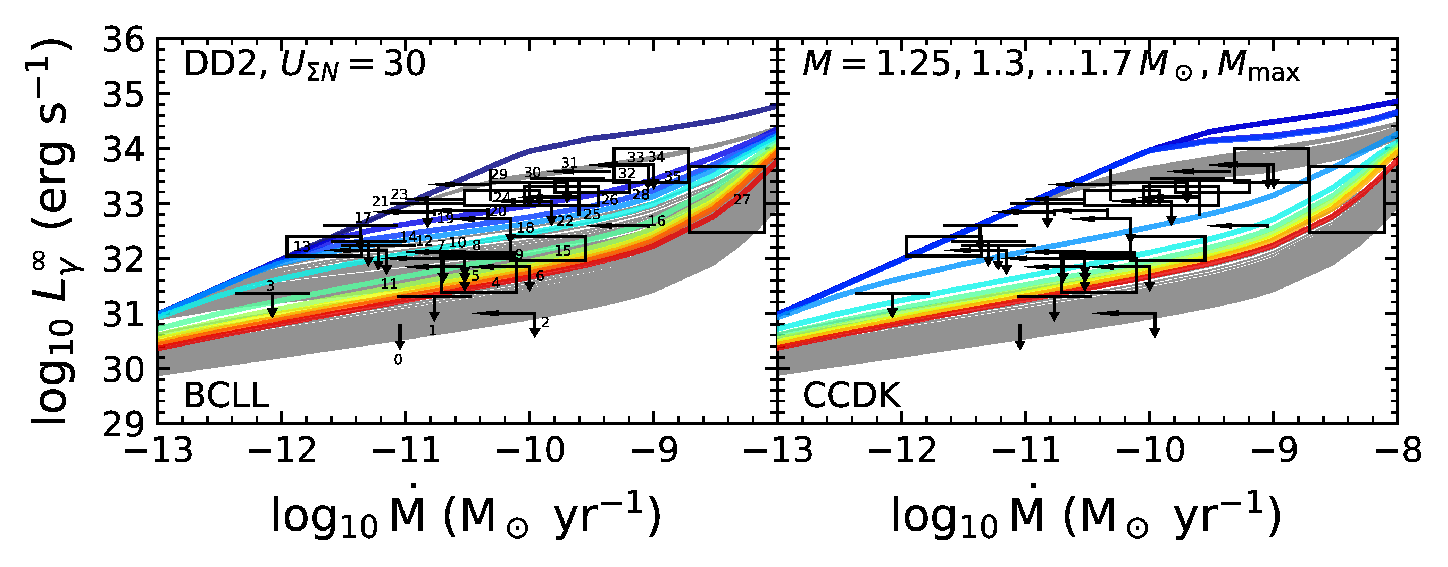}
\caption{Thermal states of INS (top) and XRT (bottom) built upon
  the hyperonic DD2 EoS with $U_\Sigma^{(N)}=+30$ MeV.
  The BCLL (left) and CCDK (right) proton pairing models are employed.}
\label{fig:dd2}
\end{figure}

Thermal evolution tracks predicted by DD2 for $U_\Sigma^{(N)}=30$ MeV are
  illustrated in Fig.~\ref{fig:dd2}. Top (bottom) panels correspond to INS (XRT);
  left (right) panels show results corresponding to the two extreme
  scenarios of proton SPF. One notes that this EoS provides the poorest agreement
  with data.
  This situation is surprising as, based on the values of the nuclear matter parameters
  customarily used to diagnose the density dependence of the EoS,
  one would expect DD2 to behave like DDME2. Its different performances can nevertheless
  be understood
  considering the third order terms in the Taylor expansion of the energy per nucleon, whose
  values are very different from those of DDME2, especially in the isoscalar channel
  (for details, see \cite{Dutra2014}).
  DD2 also shows limited sensitivity to $U_\Sigma^{(N)}$, see Table I.
  Still, as for the previously discussed EoS, the best agreement with INS (XRT)
  data is obtained for CCDK (BCLL).
  The only INS impossible to describe when CCDK is used are \#14,
  \#17 and \#18. This was also the case for DDME2.

\begin{figure}
\includegraphics[width=1\linewidth]{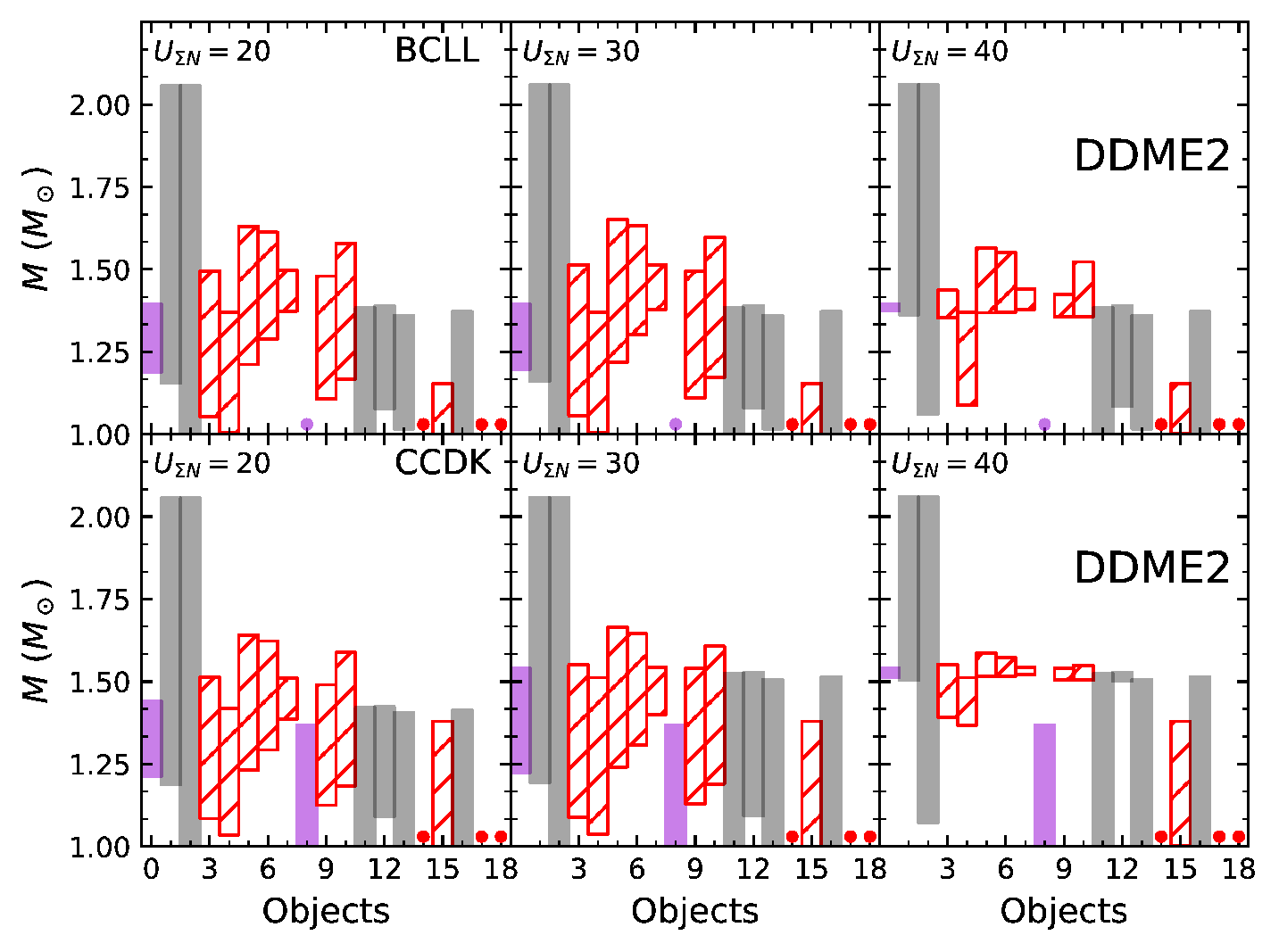}
\caption{Predicted mass range for the observed INS built upon the hyperonic DDME2 EoS
  with $U_{\Sigma}^{(N)}=20, 30, 40$ MeV (left, middle and right panels, respectively).
  BCLL and CCDK pairing models are considered in top and bottom panels, respectively.
  The color coding is the following: red for a hydrogen atmosphere,
  violet for a carbon one and gray for an iron one.
  For a given star, either a vertical bar or a dot on the X-axis is plotted.
  The dot indicates that the model is not consistent with
  the observational data;
  the vertical bar indicates the predicted mass range for the star under consideration.}
\label{fig:massDDME2}
\end{figure}

\begin{figure}
\includegraphics[width=1\linewidth]{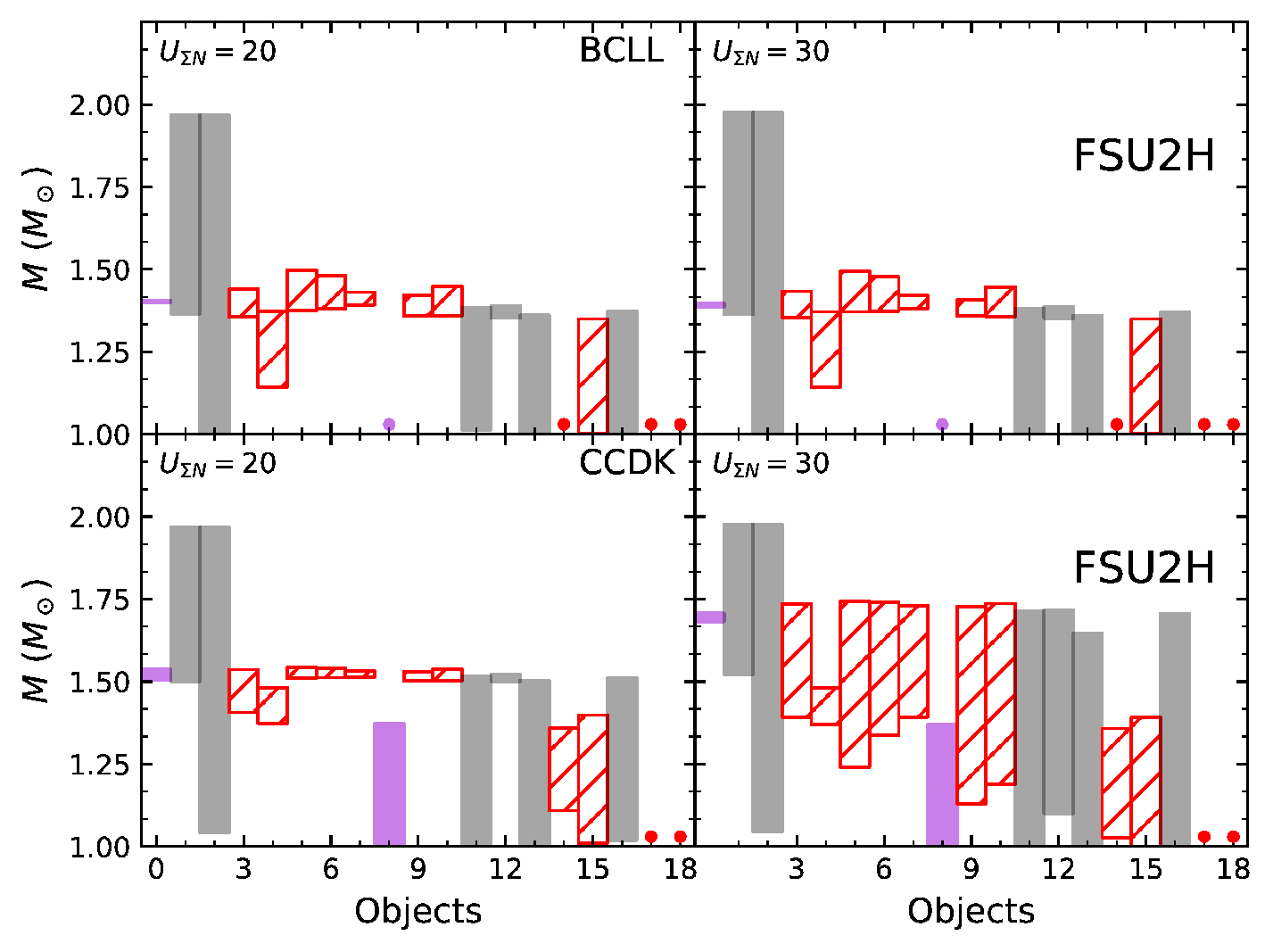}
\caption{Same as Fig.~\ref{fig:massDDME2} for the FSU2H EoS with
  $U_{\Sigma}^{(N)}=20$ and $30$ MeV (left and right panels, respectively)
  and the BCLL (top) and CCDK (bottom) pairing models.}
\label{fig:massFSU2H}
\end{figure}

\begin{figure}
\includegraphics[width=1\linewidth]{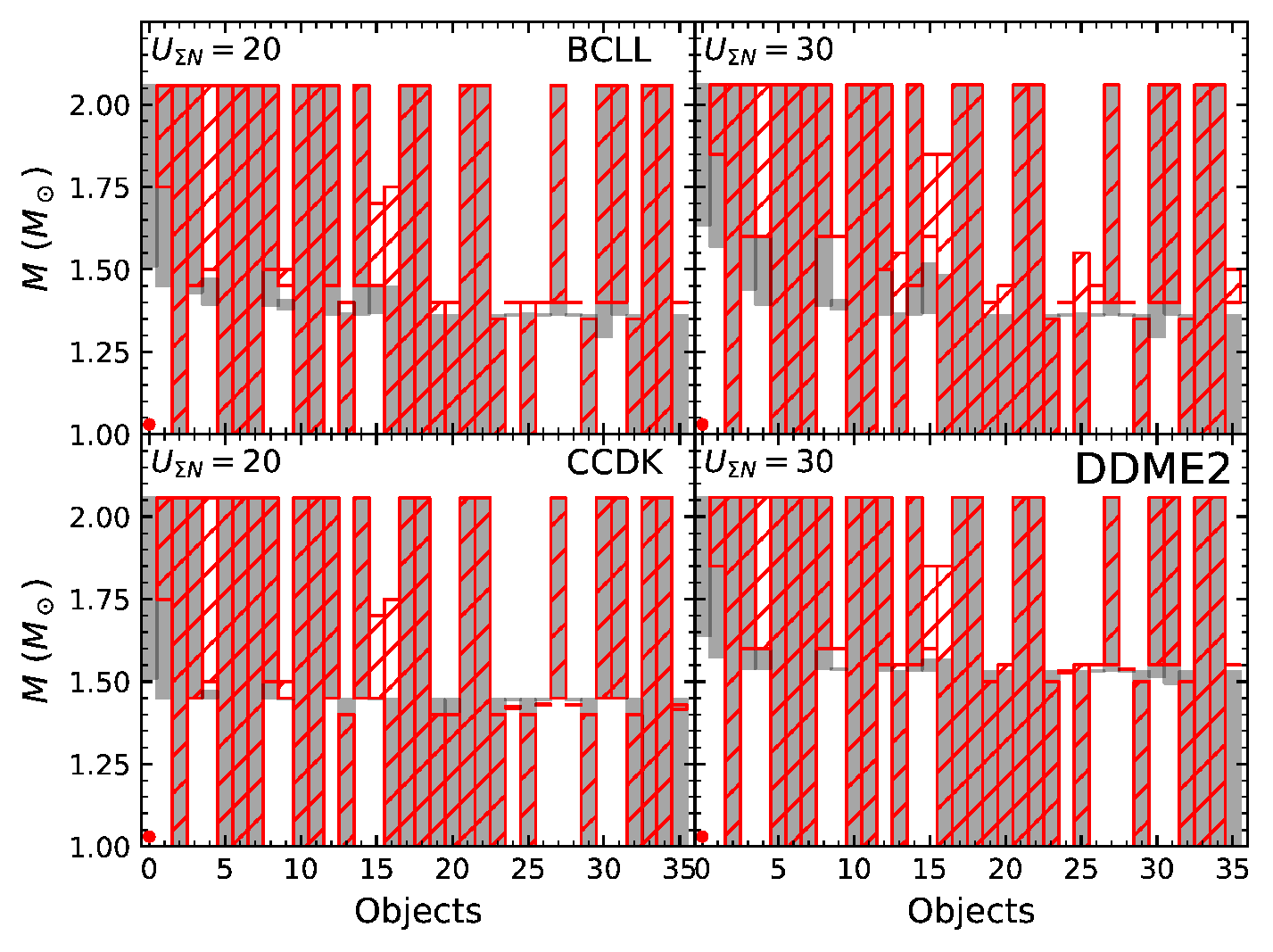}
\includegraphics[width=1\linewidth]{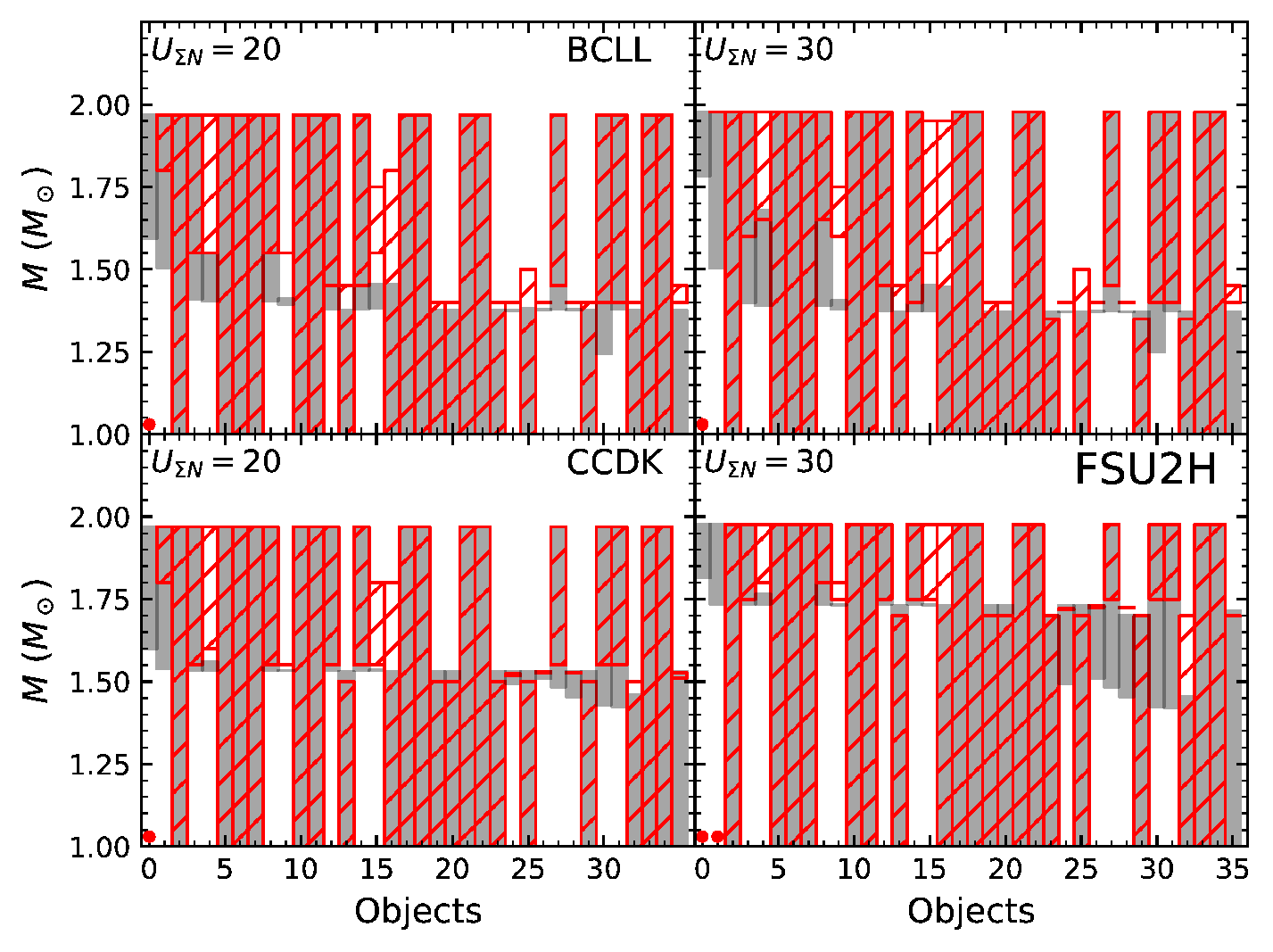}
\caption{Analog of Fig.~\ref{fig:massDDME2} but for XRT.
  The red (gray) dots or bars correspond to a H (Fe) atmosphere.
  The upper (lower) plots correspond to the DDME2 (FSU2H) EoS.
  In each plot the results for $U_\Sigma^{(N)}=20$ MeV (30 MeV) are in the
  left (right) panels.
  The CCDK (BCLL) pairing gaps are used in the upper (lower) panels of each plot.}
\label{fig:massSXT}
\end{figure}

In order to better analyze the ability of the different models to describe observational data
we have developed a set of plots that show the mass constraints that are imposed
by each INS or XRT star.
Constraints imposed by INS are depicted in
Figs.~\ref{fig:massDDME2}  and~\ref{fig:massFSU2H};
constraints imposed by XRT are shown in Fig. \ref{fig:massSXT};
  the results correspond to DDME2 and FSU2H; different values of $U_\Sigma^{(N)}$
  and proton SPF gaps (mentioned in each panel) are accounted for.
  Inspection of Figs.  \ref{fig:massDDME2} and~\ref{fig:massFSU2H} reinforces the
  conclusions drawn before, {\it i.e.} that the best description of INS is provided by
  $U_\Sigma^{(N)}=30$ MeV and CCDK.  
  Inspection of Fig. \ref{fig:massSXT} reveals that if the proton SPF gap is large,
  the number of NS which need to have $M/M_\odot \gtrsim 1.7$ is larger than the 
  one expected based on the mass distribution of observed NS.
  Overall, for XRT the best agreement with data is obtained for BCLL.
  Identical conclusions are obtained when DD2 and NL3$\omega\rho$ (not shown) are employed.

\subsection{The faint SAX J1808}
\label{ssec:SAX}

Let us finish this section with a comment on SAX J1808.
We have included in the XRT plots results from two different estimates for the
accretion rate and luminosity of this star, all taken from Ref.~\cite{Heinke09}.
The luminosity corresponding to point \#0 was obtained using the standard choice
of spectral model when fitting the observations (power-law plus black body, to be specific).
For point \#1, in addition to the black body component, also a plasma component
was considered; this
provided a higher value of the luminosity which, as mentioned in
Ref.~\cite{Potekhin_AA_2019}, is more conservative.
In addition, in the latter reference a revised value of the accretion rate is used,
obtained using a model for the instability in the accretion disk around the
NS \cite{Coriat_MNRAS_2012} which is larger than the one derived
in Ref.~\cite{Heinke09}.
SAX J1808 is a particularly promising object as -
whichever set of observational data is used -
a source of efficient cooling is necessary in order
to explain its low luminosity. This condition can only be met if
at least one dUrca process operates at full efficiency
({\em i.e.} with neutrino emissivity not suppressed by SPF) over a significant
fraction of NS volume.

In Table \ref{tab:for_us} we provide constraints on the minimum mass of SAX J1808,
  as obtained by requiring that its thermal state is reproduced by our simulations.
  Results corresponding to two
  extreme hypothesis on the composition of the atmosphere (H are Fe), different EoS models
  and values of $U_\Sigma^{(N)}$ potentials as well as proton SPF gaps are reported
  for each of the two above cited measurements (\#0 and \#1).
  {\em When data \#0 are considered}, agreement is possible only if a Fe atmosphere
  is used.
  The only combination of hyperonic-EoS/proton pairing gap unable to account for \#0 corresponds to:
  NL3$\omega\rho$, $U_\Sigma^{(N)}=30$ MeV and CCDK.
  The result for NL3$\omega\rho$, $U_\Sigma^{(N)}=30$ MeV and BCLL may also be ruled out
  because of the unrealistically large mass it predicts.
  The remaining NS models predict masses larger than $1.5 - 1.9 M_{\odot}$.\\
  {\em When data \#1 are considered,} agreement is possible for both
  hypothesis on the composition of the atmosphere;
  in the case of Fe-atmosphere, the minimum mass of SAX J1808 is $1.4-1.9 M_{\odot}$;
  when a H-atmosphere is employed, some of the models fail to reproduce the data;
  they are DDME2 with $U_{\Sigma}^{(N)}=40$ MeV, irrespective the proton gap;
  FSU2H with $U_{\Sigma}^{(N)}=30$ MeV and CCDK;
  NL3-$\omega \rho$ with $U_{\Sigma}^{(N)}=30$ MeV and CCDK; the remaining models
  predict minimum values significantly larger than those obtained under the assumption
  of Fe-atmosphere, {\em i.e.} $1.7-2.1  M_{\odot}$.
  For both \#0 and \#1 purely nucleonic models are either unable to reproduce
  the state of SAX J1808 or provide for it masses larger than the largest mass measured so far.
  This is obviously due to the density dependence of symmetry energy in the models
  we have selected.

\begin{table}
  \begin{tabular}{l ccc | c | c | c | c}
    \hline\hline
  EoS & core   & $U_{\Sigma}^{(N)}$ & proton    & set \#0 & set \#0 & set \#1 & set \#1 \\
      & compo. &  (MeV)          & $^1S_0$ gap&  Fe   &  H   &  Fe  & H   \\
      &        &                 &           & M($M_{\odot}$)& M($M_{\odot}$)& M($M_{\odot}$)& M($M_{\odot}$)\\
  \hline 
  \hline
DDME2 & N & - &  -   &  / &   / &  / &  /\\
DDME2 & N & - & CCDK &   / &    / &  / &  /\\
FSU2H & N & - &  -   &  2.3 &    / &  2.3 &  2.4\\
FSU2H & N & - & CCDK &   2.3 &    / &  2.3 &  2.4\\
\hline
DDME2 & NY & 30 & - &   1.6 &    / &  1.4 &  1.9\\
DDME2 & NY & 30 & CCDK &    1.6 &    / &  1.6 &  1.9\\
DDME2 & NY & 30 & BCLL &    1.6 &    / &  1.6 &  1.9\\
\hline
DDME2 & NY & 10 & CCDK &    1.5 &    / &  1.4 &  1.8\\
DDME2 & NY & 10 & BCLL &    1.5 &    / &  1.4 &  1.8\\
DDME2 & NY & 20 & CCDK &    1.5 &    / &  1.5 &  1.8\\
DDME2 & NY & 20 & BCLL &    1.5 &    / &  1.5 &  1.8\\
DDME2 & NY & 40 & CCDK &    1.9 &    / &  1.6 &  /\\
DDME2 & NY & 40 & BCLL &    1.8 &    / &  1.6 &  /\\
\hline
FSU2H & NY & 20 & CCDK &    1.6 &    / &  1.5 &  1.8\\
FSU2H & NY & 20 & BCLL &    1.6 &    / &  1.5 &  1.8\\
FSU2H & NY & 30 & CCDK &    1.8 &    / &  1.7 &  /\\
FSU2H & NY & 30 & BCLL &    1.8 &    / &  1.5 &  2.0\\
\hline
NL3$\omega\rho$ & NY & 20 & CCDK &    1.8 &    / &  1.8 &  2.1\\
NL3$\omega\rho$ & NY & 20 & BCLL &    1.8 &    / &  1.8 &  2.1\\
NL3$\omega\rho$ & NY & 30 & CCDK &    / &    / &  1.9 &  /\\
NL3$\omega\rho$ & NY & 30 & BCLL &    2.2 &    / &  1.9 &  2.0\\
\hline
DD2 & NY & 30 & CCDK &      1.5 &    / &  1.5 &  1.7\\
DD2 & NY & 30 & BCLL &      1.5 &    / &  1.5 &  1.7\\
\hline
\hline
\end{tabular} 
\caption{Predictions on the minimum mass of SAX J1808 (in $M_{\odot}$)
  based on observational data \#0 and \#1, for different EoS models, particle compositions 
  of the core [purely nucleonic (N) and with admixture of hyperons (NY)],
  values of $U_{\Sigma}^{(N)}$, proton $^1S_0$ pairing gaps
  and models of atmosphere (pure iron and pure hydrogen).
  "/" means that data reproduction is not possible.}
\label{tab:for_us}
\end{table}

\section{Conclusions}
\label{sec:concl}

The main objective of the present study was to verify whether simultaneous comparison
  with thermal states of INS and XRT may single out {\it the most probable EoS}
  of hyperonic compact stars. The
  candidate EoS in our set have been obtained in \cite{Fortin_PRD_2020}
  and comply with well accepted nuclear and hypernuclear data as well as constraints from
  astrophysical measurements, in particular the lower limit $\approx 2M_{\odot}$
  of the maximum NS mass. Uncertainties in the proton $^1S_0$ pairing, known to play
  a major role in regulating the neutrino emission from the core, were accounted for by
  considering two {\it extreme} scenarios. Uncertainties in the composition of the atmosphere have
  been accounted for by considering two extreme hypothesis (pure H and pure Fe).

  As previously shown by many studies, agreement with most data corresponding to one category of NS,
  INS or XRT, can be obtained for several combinations of EoS, phenomenological
  pairing gaps of baryons in the core, composition of the atmosphere and, in our case,
  also the well depth of $\Sigma$-hyperon at rest in symmetric saturated nuclear matter.
  Nevertheless none of the particular combinations we built allows to describe {\it all} data.
  Even more, a tension exists between "the most suitable" combination EoS plus proton-SPF gap,
  extracted from the fit of INS or, alternatively, XRT data.
  This result means that either at least one of the standard scenarios that we have employed here
  misses one or more essential ingredients or none of the EoS is realistic.
  Use of neutron $^3P_2-^3F_2$ pairing, here disregarded, would not improve the overall
  agreement with data. The reason is that, given the huge number of neutrons in the core, even a
  tiny gap diminishes the NS heat capacity and, thus, leads
  to accelerated cooling in the photon cooling era \cite{Raduta19}, which is not supported by INS data.
  Pairing of hyperons, here disregarded, would, in principle, improve the agreement with all data
  except those of coldest XRT.

  Finally we have tried to constrain the $U_{\Sigma}^{(N)}$ potential on the thermal state(s) of the faintest XRT,
  SAX J1808.
  For the particular set of EoS considered here, that do not allow for nucleonic dUrca or
  allow for it only in very massive NS, the conditions required by the low luminosity of SAX J1808
  can only be met under certain hypothesis on the $U_{\Sigma}^{(N)}$ potential, proton pairing gap,
  composition of the atmosphere {\it and} for masses exceeding certain threshold values.
  In most cases a repulsive potential, $U_{\Sigma}^{(N)} \approx 10-30$ MeV, is needed,
  in agreement with hypernuclear spectroscopic data.
  This range of values is supported also by the criterion of best agreement with {\it all}
  available data from INS and XRT.

{\bf Acknowledgments}:
This  work  was  supported  by  Fundação  para  a Ciência e Tecnologia,  Portugal,
under the projects UID/FIS/04564/2016 and POCI-01-0145-FEDER-029912
with financial support from POCI, in its FEDER component,
and by the FCT/MCTES budget through national funds  (OE),
and by the Polish National Science Centre (NCN) under 759 grant No. UMO-2014/13/B/ST9/02621.
A. R. R. acknowledges the support provided by the European
COST Action “PHAROS” (CA16214), through a STSM grant as well as the kind hospitality
of the Department of Physics, University of Coimbra. CP acknowledges
the support of THEIA networking activity of the Strong 2020 Project.

\bibliography{biblio.bib}

\begin{thebibliography}{98}
\expandafter\ifx\csname natexlab\endcsname\relax\def\natexlab#1{#1}\fi
\expandafter\ifx\csname bibnamefont\endcsname\relax
  \def\bibnamefont#1{#1}\fi
\expandafter\ifx\csname bibfnamefont\endcsname\relax
  \def\bibfnamefont#1{#1}\fi
\expandafter\ifx\csname citenamefont\endcsname\relax
  \def\citenamefont#1{#1}\fi
\expandafter\ifx\csname url\endcsname\relax
  \def\url#1{\texttt{#1}}\fi
\expandafter\ifx\csname urlprefix\endcsname\relax\def\urlprefix{URL }\fi
\providecommand{\bibinfo}[2]{#2}
\providecommand{\eprint}[2][]{\url{#2}}

\bibitem[{\citenamefont{Demorest et~al.}(2010)\citenamefont{Demorest, Pennucci,
  Ransom, Roberts, and Hessels}}]{Demorest10}
\bibinfo{author}{\bibfnamefont{P.}~\bibnamefont{Demorest}},
  \bibinfo{author}{\bibfnamefont{T.}~\bibnamefont{Pennucci}},
  \bibinfo{author}{\bibfnamefont{S.}~\bibnamefont{Ransom}},
  \bibinfo{author}{\bibfnamefont{M.}~\bibnamefont{Roberts}}, \bibnamefont{and}
  \bibinfo{author}{\bibfnamefont{J.}~\bibnamefont{Hessels}},
  \bibinfo{journal}{Nature} \textbf{\bibinfo{volume}{467}},
  \bibinfo{pages}{1081} (\bibinfo{year}{2010}), \eprint{1010.5788}.

\bibitem[{\citenamefont{Antoniadis et~al.}(2013)}]{Antoniadis13}
\bibinfo{author}{\bibfnamefont{J.}~\bibnamefont{Antoniadis}}
  \bibnamefont{et~al.}, \bibinfo{journal}{Science}
  \textbf{\bibinfo{volume}{340}}, \bibinfo{pages}{6131} (\bibinfo{year}{2013}),
  \eprint{1304.6875}.

\bibitem[{\citenamefont{Cromartie et~al.}(2019)}]{Cromartie2019}
\bibinfo{author}{\bibfnamefont{H.~T.} \bibnamefont{Cromartie}}
  \bibnamefont{et~al.} (\bibinfo{year}{2019}), \eprint{1904.06759}.

\bibitem[{\citenamefont{Weissenborn et~al.}(2012)\citenamefont{Weissenborn,
  Chatterjee, and Schaffner-Bielich}}]{Weissenborn_PRC_2012}
\bibinfo{author}{\bibfnamefont{S.}~\bibnamefont{Weissenborn}},
  \bibinfo{author}{\bibfnamefont{D.}~\bibnamefont{Chatterjee}},
  \bibnamefont{and}
  \bibinfo{author}{\bibfnamefont{J.}~\bibnamefont{Schaffner-Bielich}},
  \bibinfo{journal}{Phys. Rev.} \textbf{\bibinfo{volume}{C85}},
  \bibinfo{pages}{065802} (\bibinfo{year}{2012}), \bibinfo{note}{[Erratum:
  Phys. Rev.C90,no.1,019904(2014)]}, \eprint{1112.0234}.

\bibitem[{\citenamefont{Weissenborn et~al.}(2013)\citenamefont{Weissenborn,
  Chatterjee, and Schaffner-Bielich}}]{Weissenborn_NPA_2013}
\bibinfo{author}{\bibfnamefont{S.}~\bibnamefont{Weissenborn}},
  \bibinfo{author}{\bibfnamefont{D.}~\bibnamefont{Chatterjee}},
  \bibnamefont{and}
  \bibinfo{author}{\bibfnamefont{J.}~\bibnamefont{Schaffner-Bielich}},
  \bibinfo{journal}{Nucl. Phys.} \textbf{\bibinfo{volume}{A914}},
  \bibinfo{pages}{421} (\bibinfo{year}{2013}).

\bibitem[{\citenamefont{Bonanno and Sedrakian}(2012)}]{Bonanno2012}
\bibinfo{author}{\bibfnamefont{L.}~\bibnamefont{Bonanno}} \bibnamefont{and}
  \bibinfo{author}{\bibfnamefont{A.}~\bibnamefont{Sedrakian}},
  \bibinfo{journal}{Astron. Astrophys.} \textbf{\bibinfo{volume}{539}},
  \bibinfo{pages}{A16} (\bibinfo{year}{2012}), \eprint{1108.0559}.

\bibitem[{\citenamefont{{Bednarek, I.} et~al.}(2012)\citenamefont{{Bednarek,
  I.}, {Haensel, P.}, {Zdunik, J. L.}, {Bejger, M.}, and {Ma\'{}nka,
  R.}}}]{Bednarek2012}
\bibinfo{author}{\bibnamefont{{Bednarek, I.}}},
  \bibinfo{author}{\bibnamefont{{Haensel, P.}}},
  \bibinfo{author}{\bibnamefont{{Zdunik, J. L.}}},
  \bibinfo{author}{\bibnamefont{{Bejger, M.}}}, \bibnamefont{and}
  \bibinfo{author}{\bibnamefont{{Ma\'{}nka, R.}}}, \bibinfo{journal}{A\&A}
  \textbf{\bibinfo{volume}{543}}, \bibinfo{pages}{A157} (\bibinfo{year}{2012}),
  \urlprefix\url{https://doi.org/10.1051/0004-6361/201118560}.

\bibitem[{\citenamefont{Long et~al.}(2012)\citenamefont{Long, Sun, Hagino, and
  Sagawa}}]{Long2012}
\bibinfo{author}{\bibfnamefont{W.~H.} \bibnamefont{Long}},
  \bibinfo{author}{\bibfnamefont{B.~Y.} \bibnamefont{Sun}},
  \bibinfo{author}{\bibfnamefont{K.}~\bibnamefont{Hagino}}, \bibnamefont{and}
  \bibinfo{author}{\bibfnamefont{H.}~\bibnamefont{Sagawa}},
  \bibinfo{journal}{Phys. Rev. C} \textbf{\bibinfo{volume}{85}},
  \bibinfo{pages}{025806} (\bibinfo{year}{2012}).

\bibitem[{\citenamefont{Providencia and Rabhi}(2013)}]{Providencia13}
\bibinfo{author}{\bibfnamefont{C.}~\bibnamefont{Providencia}} \bibnamefont{and}
  \bibinfo{author}{\bibfnamefont{A.}~\bibnamefont{Rabhi}},
  \bibinfo{journal}{Phys. Rev.} \textbf{\bibinfo{volume}{C87}},
  \bibinfo{pages}{055801} (\bibinfo{year}{2013}), \eprint{1212.5911}.

\bibitem[{\citenamefont{{Colucci} and {Sedrakian}}(2013)}]{Colucci_13}
\bibinfo{author}{\bibfnamefont{G.}~\bibnamefont{{Colucci}}} \bibnamefont{and}
  \bibinfo{author}{\bibfnamefont{A.}~\bibnamefont{{Sedrakian}}},
  \bibinfo{journal}{Phys. Rev. C} \textbf{\bibinfo{volume}{87}},
  \bibinfo{eid}{055806} (\bibinfo{year}{2013}).

\bibitem[{\citenamefont{Miyatsu et~al.}(2013)\citenamefont{Miyatsu, Cheoun, and
  Saito}}]{Miyatsu_PRC2013}
\bibinfo{author}{\bibfnamefont{T.}~\bibnamefont{Miyatsu}},
  \bibinfo{author}{\bibfnamefont{M.-K.} \bibnamefont{Cheoun}},
  \bibnamefont{and} \bibinfo{author}{\bibfnamefont{K.}~\bibnamefont{Saito}},
  \bibinfo{journal}{Phys. Rev. C} \textbf{\bibinfo{volume}{88}},
  \bibinfo{pages}{015802} (\bibinfo{year}{2013}),
  \urlprefix\url{https://link.aps.org/doi/10.1103/PhysRevC.88.015802}.

\bibitem[{\citenamefont{{van Dalen} et~al.}(2014)\citenamefont{{van Dalen},
  {Colucci}, and {Sedrakian}}}]{vandalen_14}
\bibinfo{author}{\bibfnamefont{E.~N.~E.} \bibnamefont{{van Dalen}}},
  \bibinfo{author}{\bibfnamefont{G.}~\bibnamefont{{Colucci}}},
  \bibnamefont{and}
  \bibinfo{author}{\bibfnamefont{A.}~\bibnamefont{{Sedrakian}}},
  \bibinfo{journal}{Phys. Lett. B} \textbf{\bibinfo{volume}{734}},
  \bibinfo{pages}{383} (\bibinfo{year}{2014}).

\bibitem[{\citenamefont{Gusakov et~al.}(2014)\citenamefont{Gusakov, Haensel,
  and Kantor}}]{Gusakov_MNRAS2014}
\bibinfo{author}{\bibfnamefont{M.~E.} \bibnamefont{Gusakov}},
  \bibinfo{author}{\bibfnamefont{P.}~\bibnamefont{Haensel}}, \bibnamefont{and}
  \bibinfo{author}{\bibfnamefont{E.~M.} \bibnamefont{Kantor}},
  \bibinfo{journal}{MNRAS} \textbf{\bibinfo{volume}{439}}, \bibinfo{pages}{318}
  (\bibinfo{year}{2014}).

\bibitem[{\citenamefont{Oertel et~al.}(2015)\citenamefont{Oertel, Providência,
  Gulminelli, and Raduta}}]{Oertel:2014qza}
\bibinfo{author}{\bibfnamefont{M.}~\bibnamefont{Oertel}},
  \bibinfo{author}{\bibfnamefont{C.}~\bibnamefont{Providência}},
  \bibinfo{author}{\bibfnamefont{F.}~\bibnamefont{Gulminelli}},
  \bibnamefont{and} \bibinfo{author}{\bibfnamefont{A.~R.}
  \bibnamefont{Raduta}}, \bibinfo{journal}{J. Phys.}
  \textbf{\bibinfo{volume}{G42}}, \bibinfo{pages}{075202}
  (\bibinfo{year}{2015}).

\bibitem[{\citenamefont{Maslov et~al.}(2015)\citenamefont{Maslov, Kolomeitsev,
  and Voskresensky}}]{Maslov2015}
\bibinfo{author}{\bibfnamefont{K.}~\bibnamefont{Maslov}},
  \bibinfo{author}{\bibfnamefont{E.}~\bibnamefont{Kolomeitsev}},
  \bibnamefont{and}
  \bibinfo{author}{\bibfnamefont{D.}~\bibnamefont{Voskresensky}},
  \bibinfo{journal}{Phys. Lett. B} \textbf{\bibinfo{volume}{748}},
  \bibinfo{pages}{369} (\bibinfo{year}{2015}).

\bibitem[{\citenamefont{Fortin et~al.}(2016{\natexlab{a}})\citenamefont{Fortin,
  Providencia, Raduta, Gulminelli, Zdunik, Haensel, and Bejger}}]{Fortin2016}
\bibinfo{author}{\bibfnamefont{M.}~\bibnamefont{Fortin}},
  \bibinfo{author}{\bibfnamefont{C.}~\bibnamefont{Providencia}},
  \bibinfo{author}{\bibfnamefont{A.~R.} \bibnamefont{Raduta}},
  \bibinfo{author}{\bibfnamefont{F.}~\bibnamefont{Gulminelli}},
  \bibinfo{author}{\bibfnamefont{J.~L.} \bibnamefont{Zdunik}},
  \bibinfo{author}{\bibfnamefont{P.}~\bibnamefont{Haensel}}, \bibnamefont{and}
  \bibinfo{author}{\bibfnamefont{M.}~\bibnamefont{Bejger}},
  \bibinfo{journal}{Phys. Rev.} \textbf{\bibinfo{volume}{C94}},
  \bibinfo{pages}{035804} (\bibinfo{year}{2016}{\natexlab{a}}),
  \eprint{1604.01944}.

\bibitem[{\citenamefont{Tolos et~al.}(2016)\citenamefont{Tolos, Centelles, and
  Ramos}}]{Tolos2016}
\bibinfo{author}{\bibfnamefont{L.}~\bibnamefont{Tolos}},
  \bibinfo{author}{\bibfnamefont{M.}~\bibnamefont{Centelles}},
  \bibnamefont{and} \bibinfo{author}{\bibfnamefont{A.}~\bibnamefont{Ramos}},
  \bibinfo{journal}{ApJ} \textbf{\bibinfo{volume}{834}}, \bibinfo{pages}{3}
  (\bibinfo{year}{2016}).

\bibitem[{\citenamefont{Fortin et~al.}(2017)\citenamefont{Fortin, Avancini,
  Providência, and Vidaña}}]{Fortin17}
\bibinfo{author}{\bibfnamefont{M.}~\bibnamefont{Fortin}},
  \bibinfo{author}{\bibfnamefont{S.~S.} \bibnamefont{Avancini}},
  \bibinfo{author}{\bibfnamefont{C.}~\bibnamefont{Providência}},
  \bibnamefont{and} \bibinfo{author}{\bibfnamefont{I.}~\bibnamefont{Vidaña}},
  \bibinfo{journal}{Phys. Rev. C} \textbf{\bibinfo{volume}{95}},
  \bibinfo{eid}{065803} (\bibinfo{year}{2017}).

\bibitem[{\citenamefont{Li et~al.}(2018)\citenamefont{Li, Long, and
  Sedrakian}}]{Li2018EPJA}
\bibinfo{author}{\bibfnamefont{J.~J.} \bibnamefont{Li}},
  \bibinfo{author}{\bibfnamefont{W.~H.} \bibnamefont{Long}}, \bibnamefont{and}
  \bibinfo{author}{\bibfnamefont{A.}~\bibnamefont{Sedrakian}},
  \bibinfo{journal}{Eur. Phys. J. A} \textbf{\bibinfo{volume}{54}},
  \bibinfo{pages}{133} (\bibinfo{year}{2018}), ISSN \bibinfo{issn}{1434-601X},
  \urlprefix\url{https://doi.org/10.1140/epja/i2018-12566-6}.

\bibitem[{\citenamefont{{Fortin} et~al.}(2020)\citenamefont{{Fortin}, {Raduta},
  {Avancini}, and {Provid{\^e}ncia}}}]{Fortin_PRD_2020}
\bibinfo{author}{\bibfnamefont{M.}~\bibnamefont{{Fortin}}},
  \bibinfo{author}{\bibfnamefont{A.~R.} \bibnamefont{{Raduta}}},
  \bibinfo{author}{\bibfnamefont{S.}~\bibnamefont{{Avancini}}},
  \bibnamefont{and}
  \bibinfo{author}{\bibfnamefont{C.}~\bibnamefont{{Provid{\^e}ncia}}},
  \bibinfo{journal}{Phys. Rev. D} \textbf{\bibinfo{volume}{101}},
  \bibinfo{eid}{034017} (\bibinfo{year}{2020}), \eprint{2001.08036}.

\bibitem[{\citenamefont{{Haensel} and {Gnedin}}(1994)}]{Haensel_AA_1994}
\bibinfo{author}{\bibfnamefont{P.}~\bibnamefont{{Haensel}}} \bibnamefont{and}
  \bibinfo{author}{\bibfnamefont{O.~Y.} \bibnamefont{{Gnedin}}},
  \bibinfo{journal}{A\&A} \textbf{\bibinfo{volume}{290}}, \bibinfo{pages}{458}
  (\bibinfo{year}{1994}).

\bibitem[{\citenamefont{Schaab et~al.}(1996)\citenamefont{Schaab, Weber,
  Weigel, and Glendenning}}]{Schaab_NPA_1996}
\bibinfo{author}{\bibfnamefont{C.}~\bibnamefont{Schaab}},
  \bibinfo{author}{\bibfnamefont{F.}~\bibnamefont{Weber}},
  \bibinfo{author}{\bibfnamefont{M.~K.} \bibnamefont{Weigel}},
  \bibnamefont{and} \bibinfo{author}{\bibfnamefont{N.~K.}
  \bibnamefont{Glendenning}}, \bibinfo{journal}{Nuclear Physics A}
  \textbf{\bibinfo{volume}{605}}, \bibinfo{pages}{531 } (\bibinfo{year}{1996}),
  ISSN \bibinfo{issn}{0375-9474},
  \urlprefix\url{http://www.sciencedirect.com/science/article/pii/0375947496001649}.

\bibitem[{\citenamefont{Schaab et~al.}(1998)\citenamefont{Schaab, Balberg, and
  Schaffner-Bielich}}]{Schaab_1998}
\bibinfo{author}{\bibfnamefont{C.}~\bibnamefont{Schaab}},
  \bibinfo{author}{\bibfnamefont{S.}~\bibnamefont{Balberg}}, \bibnamefont{and}
  \bibinfo{author}{\bibfnamefont{J.}~\bibnamefont{Schaffner-Bielich}},
  \bibinfo{journal}{The Astrophysical Journal Letters}
  \textbf{\bibinfo{volume}{504}}, \bibinfo{pages}{L99} (\bibinfo{year}{1998}),
  \urlprefix\url{http://stacks.iop.org/1538-4357/504/i=2/a=L99}.

\bibitem[{\citenamefont{Yakovlev and Pethick}(2004)}]{Yakovlev04}
\bibinfo{author}{\bibfnamefont{D.~G.} \bibnamefont{Yakovlev}} \bibnamefont{and}
  \bibinfo{author}{\bibfnamefont{C.~J.} \bibnamefont{Pethick}},
  \bibinfo{journal}{Ann. Rev. Astron. Astrophys.}
  \textbf{\bibinfo{volume}{42}}, \bibinfo{pages}{169} (\bibinfo{year}{2004}),
  \eprint{astro-ph/0402143}.

\bibitem[{\citenamefont{Page and Reddy}(2006)}]{Page_ARNPS_2006}
\bibinfo{author}{\bibfnamefont{D.}~\bibnamefont{Page}} \bibnamefont{and}
  \bibinfo{author}{\bibfnamefont{S.}~\bibnamefont{Reddy}},
  \bibinfo{journal}{Annual Review of Nuclear and Particle Science}
  \textbf{\bibinfo{volume}{56}}, \bibinfo{pages}{327} (\bibinfo{year}{2006}),
  \eprint{https://doi.org/10.1146/annurev.nucl.56.080805.140600},
  \urlprefix\url{https://doi.org/10.1146/annurev.nucl.56.080805.140600}.

\bibitem[{\citenamefont{Page et~al.}(2006)\citenamefont{Page, Geppert, and
  Weber}}]{Page_NPA_2006}
\bibinfo{author}{\bibfnamefont{D.}~\bibnamefont{Page}},
  \bibinfo{author}{\bibfnamefont{U.}~\bibnamefont{Geppert}}, \bibnamefont{and}
  \bibinfo{author}{\bibfnamefont{F.}~\bibnamefont{Weber}},
  \bibinfo{journal}{Nuclear Physics A} \textbf{\bibinfo{volume}{777}},
  \bibinfo{pages}{497 } (\bibinfo{year}{2006}), ISSN \bibinfo{issn}{0375-9474},
  \bibinfo{note}{special Isseu on Nuclear Astrophysics},
  \urlprefix\url{http://www.sciencedirect.com/science/article/pii/S0375947405011164}.

\bibitem[{\citenamefont{Tsuruta et~al.}(2009)\citenamefont{Tsuruta, Sadino,
  Kobelski, Teter, Liebmann, Takatsuka, K.nomoto, and Umeda}}]{Tsuruta_2009}
\bibinfo{author}{\bibfnamefont{S.}~\bibnamefont{Tsuruta}},
  \bibinfo{author}{\bibfnamefont{J.}~\bibnamefont{Sadino}},
  \bibinfo{author}{\bibfnamefont{A.}~\bibnamefont{Kobelski}},
  \bibinfo{author}{\bibfnamefont{M.~A.} \bibnamefont{Teter}},
  \bibinfo{author}{\bibfnamefont{A.~C.} \bibnamefont{Liebmann}},
  \bibinfo{author}{\bibfnamefont{T.}~\bibnamefont{Takatsuka}},
  \bibinfo{author}{\bibnamefont{K.nomoto}}, \bibnamefont{and}
  \bibinfo{author}{\bibfnamefont{H.}~\bibnamefont{Umeda}},
  \bibinfo{journal}{The Astrophysical Journal} \textbf{\bibinfo{volume}{691}},
  \bibinfo{pages}{621} (\bibinfo{year}{2009}),
  \urlprefix\url{http://stacks.iop.org/0004-637X/691/i=1/a=621}.

\bibitem[{\citenamefont{{Yakovlev, D. G.} et~al.}(2003)\citenamefont{{Yakovlev,
  D. G.}, {Levenfish, K. P.}, and {Haensel, P.}}}]{Yakovlev_AA_2003}
\bibinfo{author}{\bibnamefont{{Yakovlev, D. G.}}},
  \bibinfo{author}{\bibnamefont{{Levenfish, K. P.}}}, \bibnamefont{and}
  \bibinfo{author}{\bibnamefont{{Haensel, P.}}}, \bibinfo{journal}{A\&A}
  \textbf{\bibinfo{volume}{407}}, \bibinfo{pages}{265} (\bibinfo{year}{2003}),
  \urlprefix\url{https://doi.org/10.1051/0004-6361:20030830}.

\bibitem[{\citenamefont{{Potekhin, A. Y.} et~al.}(2019)\citenamefont{{Potekhin,
  A. Y.}, {Chugunov, A. I.}, and {Chabrier, G.}}}]{Potekhin_AA_2019}
\bibinfo{author}{\bibnamefont{{Potekhin, A. Y.}}},
  \bibinfo{author}{\bibnamefont{{Chugunov, A. I.}}}, \bibnamefont{and}
  \bibinfo{author}{\bibnamefont{{Chabrier, G.}}}, \bibinfo{journal}{A\&A}
  \textbf{\bibinfo{volume}{629}}, \bibinfo{pages}{A88} (\bibinfo{year}{2019}),
  \urlprefix\url{https://doi.org/10.1051/0004-6361/201936003}.

\bibitem[{\citenamefont{{Prakash} et~al.}(1992)\citenamefont{{Prakash},
  {Prakash}, {Lattimer}, and {Pethick}}}]{DUY92}
\bibinfo{author}{\bibfnamefont{M.}~\bibnamefont{{Prakash}}},
  \bibinfo{author}{\bibfnamefont{M.}~\bibnamefont{{Prakash}}},
  \bibinfo{author}{\bibfnamefont{J.~M.} \bibnamefont{{Lattimer}}},
  \bibnamefont{and} \bibinfo{author}{\bibfnamefont{C.~J.}
  \bibnamefont{{Pethick}}}, \bibinfo{journal}{The Astrophysical Journal
  Letters} \textbf{\bibinfo{volume}{390}}, \bibinfo{pages}{L77}
  (\bibinfo{year}{1992}).

\bibitem[{\citenamefont{Fortin et~al.}(2016{\natexlab{b}})\citenamefont{Fortin,
  Providencia, Raduta, Gulminelli, Zdunik, Haensel, and Bejger}}]{Fortin16}
\bibinfo{author}{\bibfnamefont{M.}~\bibnamefont{Fortin}},
  \bibinfo{author}{\bibfnamefont{C.}~\bibnamefont{Providencia}},
  \bibinfo{author}{\bibfnamefont{A.~R.} \bibnamefont{Raduta}},
  \bibinfo{author}{\bibfnamefont{F.}~\bibnamefont{Gulminelli}},
  \bibinfo{author}{\bibfnamefont{J.~L.} \bibnamefont{Zdunik}},
  \bibinfo{author}{\bibfnamefont{P.}~\bibnamefont{Haensel}}, \bibnamefont{and}
  \bibinfo{author}{\bibfnamefont{M.}~\bibnamefont{Bejger}},
  \bibinfo{journal}{Phys. Rev.} \textbf{\bibinfo{volume}{C94}},
  \bibinfo{pages}{035804} (\bibinfo{year}{2016}{\natexlab{b}}),
  \eprint{1604.01944}.

\bibitem[{\citenamefont{Providência et~al.}(2019)\citenamefont{Providência,
  Fortin, Pais, and Rabhi}}]{Providencia19}
\bibinfo{author}{\bibfnamefont{C.}~\bibnamefont{Providência}},
  \bibinfo{author}{\bibfnamefont{M.}~\bibnamefont{Fortin}},
  \bibinfo{author}{\bibfnamefont{H.}~\bibnamefont{Pais}}, \bibnamefont{and}
  \bibinfo{author}{\bibfnamefont{A.}~\bibnamefont{Rabhi}},
  \bibinfo{journal}{Frontiers in Astronomy and Space Sciences}
  \textbf{\bibinfo{volume}{6}}, \bibinfo{pages}{13} (\bibinfo{year}{2019}),
  ISSN \bibinfo{issn}{2296-987X},
  \urlprefix\url{https://www.frontiersin.org/article/10.3389/fspas.2019.00013}.

\bibitem[{\citenamefont{{Beznogov} and {Yakovlev}}(2015)}]{BY15}
\bibinfo{author}{\bibfnamefont{M.~V.} \bibnamefont{{Beznogov}}}
  \bibnamefont{and} \bibinfo{author}{\bibfnamefont{D.~G.}
  \bibnamefont{{Yakovlev}}}, \bibinfo{journal}{Monthly Notices of the Royal
  Astronomical Society} \textbf{\bibinfo{volume}{447}}, \bibinfo{pages}{1598}
  (\bibinfo{year}{2015}), \eprint{1411.6803}.

\bibitem[{\citenamefont{Beloin et~al.}(2018)\citenamefont{Beloin, Han, Steiner,
  and Page}}]{Beloin_PRC_2018}
\bibinfo{author}{\bibfnamefont{S.}~\bibnamefont{Beloin}},
  \bibinfo{author}{\bibfnamefont{S.}~\bibnamefont{Han}},
  \bibinfo{author}{\bibfnamefont{A.~W.} \bibnamefont{Steiner}},
  \bibnamefont{and} \bibinfo{author}{\bibfnamefont{D.}~\bibnamefont{Page}},
  \bibinfo{journal}{Phys. Rev. C} \textbf{\bibinfo{volume}{97}},
  \bibinfo{pages}{015804} (\bibinfo{year}{2018}),
  \urlprefix\url{https://link.aps.org/doi/10.1103/PhysRevC.97.015804}.

\bibitem[{\citenamefont{Beloin et~al.}(2019)\citenamefont{Beloin, Han, Steiner,
  and Odbadrakh}}]{Beloin_PRC_2019}
\bibinfo{author}{\bibfnamefont{S.}~\bibnamefont{Beloin}},
  \bibinfo{author}{\bibfnamefont{S.}~\bibnamefont{Han}},
  \bibinfo{author}{\bibfnamefont{A.~W.} \bibnamefont{Steiner}},
  \bibnamefont{and}
  \bibinfo{author}{\bibfnamefont{K.}~\bibnamefont{Odbadrakh}},
  \bibinfo{journal}{Phys. Rev. C} \textbf{\bibinfo{volume}{100}},
  \bibinfo{pages}{055801} (\bibinfo{year}{2019}),
  \urlprefix\url{https://link.aps.org/doi/10.1103/PhysRevC.100.055801}.

\bibitem[{\citenamefont{Yakovlev et~al.}(2001)\citenamefont{Yakovlev, Kaminker,
  Gnedin, and Haensel}}]{Yakovlev_PhysRep_2001}
\bibinfo{author}{\bibfnamefont{D.~G.} \bibnamefont{Yakovlev}},
  \bibinfo{author}{\bibfnamefont{A.~D.} \bibnamefont{Kaminker}},
  \bibinfo{author}{\bibfnamefont{O.~Y.} \bibnamefont{Gnedin}},
  \bibnamefont{and} \bibinfo{author}{\bibfnamefont{P.}~\bibnamefont{Haensel}},
  \bibinfo{journal}{Phys. Rept.} \textbf{\bibinfo{volume}{354}},
  \bibinfo{pages}{1} (\bibinfo{year}{2001}), \eprint{astro-ph/0012122}.

\bibitem[{\citenamefont{Raduta et~al.}(2018)\citenamefont{Raduta, Sedrakian,
  and Weber}}]{Raduta17}
\bibinfo{author}{\bibfnamefont{A.~R.} \bibnamefont{Raduta}},
  \bibinfo{author}{\bibfnamefont{A.}~\bibnamefont{Sedrakian}},
  \bibnamefont{and} \bibinfo{author}{\bibfnamefont{F.}~\bibnamefont{Weber}},
  \bibinfo{journal}{Mon. Not. Roy. Astron. Soc.}
  \textbf{\bibinfo{volume}{475}}, \bibinfo{pages}{4347} (\bibinfo{year}{2018}),
  \eprint{1712.00584}.

\bibitem[{\citenamefont{Grigorian et~al.}(2018)\citenamefont{Grigorian,
  Voskresensky, and Maslov}}]{Grigorian2018}
\bibinfo{author}{\bibfnamefont{H.}~\bibnamefont{Grigorian}},
  \bibinfo{author}{\bibfnamefont{D.~N.} \bibnamefont{Voskresensky}},
  \bibnamefont{and} \bibinfo{author}{\bibfnamefont{K.~A.}
  \bibnamefont{Maslov}}, \bibinfo{journal}{Nucl. Phys.}
  \textbf{\bibinfo{volume}{A980}}, \bibinfo{pages}{105} (\bibinfo{year}{2018}),
  \eprint{1808.01819}.

\bibitem[{\citenamefont{Negreiros et~al.}(2018)\citenamefont{Negreiros, Tolos,
  Centelles, Ramos, and Dexheimer}}]{Negreiros18}
\bibinfo{author}{\bibfnamefont{R.}~\bibnamefont{Negreiros}},
  \bibinfo{author}{\bibfnamefont{L.}~\bibnamefont{Tolos}},
  \bibinfo{author}{\bibfnamefont{M.}~\bibnamefont{Centelles}},
  \bibinfo{author}{\bibfnamefont{A.}~\bibnamefont{Ramos}}, \bibnamefont{and}
  \bibinfo{author}{\bibfnamefont{V.}~\bibnamefont{Dexheimer}},
  \bibinfo{journal}{Astrophys. J.} \textbf{\bibinfo{volume}{863}},
  \bibinfo{pages}{104} (\bibinfo{year}{2018}), \eprint{1804.00334}.

\bibitem[{\citenamefont{Raduta et~al.}(2019)\citenamefont{Raduta, Li,
  Sedrakian, and Weber}}]{Raduta19}
\bibinfo{author}{\bibfnamefont{A.~R.} \bibnamefont{Raduta}},
  \bibinfo{author}{\bibfnamefont{J.~J.} \bibnamefont{Li}},
  \bibinfo{author}{\bibfnamefont{A.}~\bibnamefont{Sedrakian}},
  \bibnamefont{and} \bibinfo{author}{\bibfnamefont{F.}~\bibnamefont{Weber}},
  \bibinfo{journal}{Mon. Not. Roy. Astron. Soc.}
  \textbf{\bibinfo{volume}{487}}, \bibinfo{pages}{2639} (\bibinfo{year}{2019}),
  \eprint{1903.01295}.

\bibitem[{\citenamefont{Maslov et~al.}(2016)\citenamefont{Maslov, Kolomeitsev,
  and Voskresensky}}]{Maslov2016}
\bibinfo{author}{\bibfnamefont{K.~A.} \bibnamefont{Maslov}},
  \bibinfo{author}{\bibfnamefont{E.~E.} \bibnamefont{Kolomeitsev}},
  \bibnamefont{and} \bibinfo{author}{\bibfnamefont{D.~N.}
  \bibnamefont{Voskresensky}}, \bibinfo{journal}{Nucl. Phys.}
  \textbf{\bibinfo{volume}{A950}}, \bibinfo{pages}{64} (\bibinfo{year}{2016}),
  \eprint{1509.02538}.

\bibitem[{\citenamefont{Tolos et~al.}(2017)\citenamefont{Tolos, Centelles, and
  Ramos}}]{Tolos17}
\bibinfo{author}{\bibfnamefont{L.}~\bibnamefont{Tolos}},
  \bibinfo{author}{\bibfnamefont{M.}~\bibnamefont{Centelles}},
  \bibnamefont{and} \bibinfo{author}{\bibfnamefont{A.}~\bibnamefont{Ramos}},
  \bibinfo{journal}{Publ. Astron. Soc. Austral.} \textbf{\bibinfo{volume}{34}},
  \bibinfo{pages}{e065} (\bibinfo{year}{2017}), \eprint{1708.08681}.

\bibitem[{\citenamefont{Chen and Piekarewicz}(2014)}]{Chen2014}
\bibinfo{author}{\bibfnamefont{W.-C.} \bibnamefont{Chen}} \bibnamefont{and}
  \bibinfo{author}{\bibfnamefont{J.}~\bibnamefont{Piekarewicz}},
  \bibinfo{journal}{Phys. Rev.} \textbf{\bibinfo{volume}{C90}},
  \bibinfo{pages}{044305} (\bibinfo{year}{2014}), \eprint{1408.4159}.

\bibitem[{\citenamefont{Jiang et~al.}(2012)\citenamefont{Jiang, Li, and
  Chen}}]{Jiang2012}
\bibinfo{author}{\bibfnamefont{W.-Z.} \bibnamefont{Jiang}},
  \bibinfo{author}{\bibfnamefont{B.-A.} \bibnamefont{Li}}, \bibnamefont{and}
  \bibinfo{author}{\bibfnamefont{L.-W.} \bibnamefont{Chen}},
  \bibinfo{journal}{The Astrophysical Journal} \textbf{\bibinfo{volume}{756}},
  \bibinfo{pages}{56} (\bibinfo{year}{2012}),
  \urlprefix\url{https://doi.org/10.1088/0004-637x/756/1/56}.

\bibitem[{\citenamefont{Takatsuka et~al.}(2002)\citenamefont{Takatsuka,
  Nishizaki, Yamamoto, and Tamagaki}}]{Takatsuka2002}
\bibinfo{author}{\bibfnamefont{T.}~\bibnamefont{Takatsuka}},
  \bibinfo{author}{\bibfnamefont{S.}~\bibnamefont{Nishizaki}},
  \bibinfo{author}{\bibfnamefont{Y.}~\bibnamefont{Yamamoto}}, \bibnamefont{and}
  \bibinfo{author}{\bibfnamefont{R.}~\bibnamefont{Tamagaki}},
  \bibinfo{journal}{Progress of Theoretical Physics Supplement}
  \textbf{\bibinfo{volume}{146}}, \bibinfo{pages}{279} (\bibinfo{year}{2002}),
  ISSN \bibinfo{issn}{0375-9687},
  \eprint{https://academic.oup.com/ptps/article-pdf/doi/10.1143/PTPS.146.279/5268098/146-279.pdf},
  \urlprefix\url{https://doi.org/10.1143/PTPS.146.279}.

\bibitem[{\citenamefont{Baldo et~al.}(2000)\citenamefont{Baldo, Burgio, and
  Schulze}}]{Baldo2000}
\bibinfo{author}{\bibfnamefont{M.}~\bibnamefont{Baldo}},
  \bibinfo{author}{\bibfnamefont{G.~F.} \bibnamefont{Burgio}},
  \bibnamefont{and} \bibinfo{author}{\bibfnamefont{H.~J.}
  \bibnamefont{Schulze}}, \bibinfo{journal}{Phys. Rev. C}
  \textbf{\bibinfo{volume}{61}}, \bibinfo{pages}{055801}
  (\bibinfo{year}{2000}), \eprint{nucl-th/9912066}.

\bibitem[{\citenamefont{Vidana et~al.}(2000)\citenamefont{Vidana, Polls, Ramos,
  Engvik, and Hjorth-Jensen}}]{Vidana2000}
\bibinfo{author}{\bibfnamefont{I.}~\bibnamefont{Vidana}},
  \bibinfo{author}{\bibfnamefont{A.}~\bibnamefont{Polls}},
  \bibinfo{author}{\bibfnamefont{A.}~\bibnamefont{Ramos}},
  \bibinfo{author}{\bibfnamefont{L.}~\bibnamefont{Engvik}}, \bibnamefont{and}
  \bibinfo{author}{\bibfnamefont{M.}~\bibnamefont{Hjorth-Jensen}},
  \bibinfo{journal}{Phys. Rev. C} \textbf{\bibinfo{volume}{62}},
  \bibinfo{pages}{035801} (\bibinfo{year}{2000}), \eprint{nucl-th/0004031}.

\bibitem[{\citenamefont{Vidana et~al.}(2011)\citenamefont{Vidana, Logoteta,
  Providencia, Polls, and Bombaci}}]{Vidana2010}
\bibinfo{author}{\bibfnamefont{I.}~\bibnamefont{Vidana}},
  \bibinfo{author}{\bibfnamefont{D.}~\bibnamefont{Logoteta}},
  \bibinfo{author}{\bibfnamefont{C.}~\bibnamefont{Providencia}},
  \bibinfo{author}{\bibfnamefont{A.}~\bibnamefont{Polls}}, \bibnamefont{and}
  \bibinfo{author}{\bibfnamefont{I.}~\bibnamefont{Bombaci}},
  \bibinfo{journal}{EPL} \textbf{\bibinfo{volume}{94}}, \bibinfo{pages}{11002}
  (\bibinfo{year}{2011}), \eprint{1006.5660}.

\bibitem[{\citenamefont{Logoteta}(2013)}]{Logoteta2013}
\bibinfo{author}{\bibfnamefont{D.}~\bibnamefont{Logoteta}}, Ph.D. thesis,
  \bibinfo{school}{University of Coimbra} (\bibinfo{year}{2013}).

\bibitem[{\citenamefont{Lonardoni et~al.}(2015)\citenamefont{Lonardoni, Lovato,
  Gandolfi, and Pederiva}}]{Lonardoni2014}
\bibinfo{author}{\bibfnamefont{D.}~\bibnamefont{Lonardoni}},
  \bibinfo{author}{\bibfnamefont{A.}~\bibnamefont{Lovato}},
  \bibinfo{author}{\bibfnamefont{S.}~\bibnamefont{Gandolfi}}, \bibnamefont{and}
  \bibinfo{author}{\bibfnamefont{F.}~\bibnamefont{Pederiva}},
  \bibinfo{journal}{Phys. Rev. Lett.} \textbf{\bibinfo{volume}{114}},
  \bibinfo{pages}{092301} (\bibinfo{year}{2015}), \eprint{1407.4448}.

\bibitem[{\citenamefont{Beznogov and Yakovlev}(2015)}]{Beznogov_MNRAS_2015}
\bibinfo{author}{\bibfnamefont{M.~V.} \bibnamefont{Beznogov}} \bibnamefont{and}
  \bibinfo{author}{\bibfnamefont{D.~G.} \bibnamefont{Yakovlev}},
  \bibinfo{journal}{Mon. Not. Roy. Astron. Soc.}
  \textbf{\bibinfo{volume}{452}}, \bibinfo{pages}{540} (\bibinfo{year}{2015}),
  \eprint{1507.04206}.

\bibitem[{\citenamefont{Lalazissis et~al.}(1997)\citenamefont{Lalazissis,
  Konig, and Ring}}]{nl3}
\bibinfo{author}{\bibfnamefont{G.~A.} \bibnamefont{Lalazissis}},
  \bibinfo{author}{\bibfnamefont{J.}~\bibnamefont{Konig}}, \bibnamefont{and}
  \bibinfo{author}{\bibfnamefont{P.}~\bibnamefont{Ring}},
  \bibinfo{journal}{Phys. Rev.} \textbf{\bibinfo{volume}{C55}},
  \bibinfo{pages}{540} (\bibinfo{year}{1997}), \eprint{nucl-th/9607039}.

\bibitem[{\citenamefont{Pais and Providência}(2016)}]{Pais16}
\bibinfo{author}{\bibfnamefont{H.}~\bibnamefont{Pais}} \bibnamefont{and}
  \bibinfo{author}{\bibfnamefont{C.}~\bibnamefont{Providência}},
  \bibinfo{journal}{Phys. Rev.} \textbf{\bibinfo{volume}{C94}},
  \bibinfo{pages}{015808} (\bibinfo{year}{2016}), \eprint{1607.05899}.

\bibitem[{\citenamefont{Horowitz and Piekarewicz}(2001)}]{Horowitz01}
\bibinfo{author}{\bibfnamefont{C.~J.} \bibnamefont{Horowitz}} \bibnamefont{and}
  \bibinfo{author}{\bibfnamefont{J.}~\bibnamefont{Piekarewicz}},
  \bibinfo{journal}{Phys. Rev. Lett.} \textbf{\bibinfo{volume}{86}},
  \bibinfo{pages}{5647} (\bibinfo{year}{2001}), \eprint{astro-ph/0010227}.

\bibitem[{\citenamefont{Sugahara and Toki}(1994)}]{tm1}
\bibinfo{author}{\bibfnamefont{Y.}~\bibnamefont{Sugahara}} \bibnamefont{and}
  \bibinfo{author}{\bibfnamefont{H.}~\bibnamefont{Toki}},
  \bibinfo{journal}{Nucl. Phys.} \textbf{\bibinfo{volume}{A579}},
  \bibinfo{pages}{557} (\bibinfo{year}{1994}).

\bibitem[{\citenamefont{Bao and Shen}(2014)}]{Bao2014}
\bibinfo{author}{\bibfnamefont{S.~S.} \bibnamefont{Bao}} \bibnamefont{and}
  \bibinfo{author}{\bibfnamefont{H.}~\bibnamefont{Shen}},
  \bibinfo{journal}{Phys. Rev.} \textbf{\bibinfo{volume}{C89}},
  \bibinfo{pages}{045807} (\bibinfo{year}{2014}), \eprint{1405.3837}.

\bibitem[{\citenamefont{Typel et~al.}(2010)\citenamefont{Typel, R{\"o}pke,
  Kl{\"a}hn, Blaschke, and Wolter}}]{typel10}
\bibinfo{author}{\bibfnamefont{S.}~\bibnamefont{Typel}},
  \bibinfo{author}{\bibfnamefont{G.}~\bibnamefont{R{\"o}pke}},
  \bibinfo{author}{\bibfnamefont{T.}~\bibnamefont{Kl{\"a}hn}},
  \bibinfo{author}{\bibfnamefont{D.}~\bibnamefont{Blaschke}}, \bibnamefont{and}
  \bibinfo{author}{\bibfnamefont{H.}~\bibnamefont{Wolter}},
  \bibinfo{journal}{Phys.Rev.} \textbf{\bibinfo{volume}{C81}},
  \bibinfo{pages}{015803} (\bibinfo{year}{2010}).

\bibitem[{\citenamefont{Lalazissis et~al.}(2005)\citenamefont{Lalazissis,
  Nik\ifmmode \check{s}\else \v{s}\fi{}i\ifmmode~\acute{c}\else \'{c}\fi{},
  Vretenar, and Ring}}]{ddme2}
\bibinfo{author}{\bibfnamefont{G.~A.} \bibnamefont{Lalazissis}},
  \bibinfo{author}{\bibfnamefont{T.}~\bibnamefont{Nik\ifmmode \check{s}\else
  \v{s}\fi{}i\ifmmode~\acute{c}\else \'{c}\fi{}}},
  \bibinfo{author}{\bibfnamefont{D.}~\bibnamefont{Vretenar}}, \bibnamefont{and}
  \bibinfo{author}{\bibfnamefont{P.}~\bibnamefont{Ring}},
  \bibinfo{journal}{Phys. Rev. C} \textbf{\bibinfo{volume}{71}},
  \bibinfo{pages}{024312} (\bibinfo{year}{2005}),
  \urlprefix\url{https://link.aps.org/doi/10.1103/PhysRevC.71.024312}.

\bibitem[{\citenamefont{Dutra et~al.}(2014)\citenamefont{Dutra, Lourenço,
  Avancini, Carlson, Delfino et~al.}}]{Dutra2014}
\bibinfo{author}{\bibfnamefont{M.}~\bibnamefont{Dutra}},
  \bibinfo{author}{\bibfnamefont{O.}~\bibnamefont{Lourenço}},
  \bibinfo{author}{\bibfnamefont{S.}~\bibnamefont{Avancini}},
  \bibinfo{author}{\bibfnamefont{B.}~\bibnamefont{Carlson}},
  \bibinfo{author}{\bibfnamefont{A.}~\bibnamefont{Delfino}},
  \bibnamefont{et~al.}, \bibinfo{journal}{Phys.Rev.}
  \textbf{\bibinfo{volume}{C90}}, \bibinfo{pages}{055203}
  (\bibinfo{year}{2014}).

\bibitem[{\citenamefont{Fortin et~al.}(2018)\citenamefont{Fortin, Oertel, and
  Providência}}]{Fortin18}
\bibinfo{author}{\bibfnamefont{M.}~\bibnamefont{Fortin}},
  \bibinfo{author}{\bibfnamefont{M.}~\bibnamefont{Oertel}}, \bibnamefont{and}
  \bibinfo{author}{\bibfnamefont{C.}~\bibnamefont{Providência}},
  \bibinfo{journal}{Publ. Astron. Soc. Austral.} \textbf{\bibinfo{volume}{35}},
  \bibinfo{pages}{44} (\bibinfo{year}{2018}), \eprint{1711.09427}.

\bibitem[{\citenamefont{Khaustov et~al.}(2000)}]{khaustov00}
\bibinfo{author}{\bibfnamefont{P.}~\bibnamefont{Khaustov}} \bibnamefont{et~al.}
  (\bibinfo{collaboration}{AGS E885}), \bibinfo{journal}{Phys. Rev.}
  \textbf{\bibinfo{volume}{C61}}, \bibinfo{pages}{054603}
  (\bibinfo{year}{2000}), \eprint{nucl-ex/9912007}.

\bibitem[{\citenamefont{{Nakazawa} et~al.}(2015)\citenamefont{{Nakazawa},
  {Endo}, {Fukunaga}, {Hoshino}, {Hwang}, {Imai}, {Ito}, {Itonaga}, {Kand a},
  {Kawasaki} et~al.}}]{kiso}
\bibinfo{author}{\bibfnamefont{K.}~\bibnamefont{{Nakazawa}}},
  \bibinfo{author}{\bibfnamefont{Y.}~\bibnamefont{{Endo}}},
  \bibinfo{author}{\bibfnamefont{S.}~\bibnamefont{{Fukunaga}}},
  \bibinfo{author}{\bibfnamefont{K.}~\bibnamefont{{Hoshino}}},
  \bibinfo{author}{\bibfnamefont{S.~H.} \bibnamefont{{Hwang}}},
  \bibinfo{author}{\bibfnamefont{K.}~\bibnamefont{{Imai}}},
  \bibinfo{author}{\bibfnamefont{H.}~\bibnamefont{{Ito}}},
  \bibinfo{author}{\bibfnamefont{K.}~\bibnamefont{{Itonaga}}},
  \bibinfo{author}{\bibfnamefont{T.}~\bibnamefont{{Kand a}}},
  \bibinfo{author}{\bibfnamefont{M.}~\bibnamefont{{Kawasaki}}},
  \bibnamefont{et~al.}, \bibinfo{journal}{Progress of Theoretical and
  Experimental Physics} \textbf{\bibinfo{volume}{2015}}, \bibinfo{eid}{033D02}
  (\bibinfo{year}{2015}).

\bibitem[{\citenamefont{Sun et~al.}(2016)\citenamefont{Sun, Hiyama, Sagawa,
  Schulze, and Meng}}]{Sun2016}
\bibinfo{author}{\bibfnamefont{T.~T.} \bibnamefont{Sun}},
  \bibinfo{author}{\bibfnamefont{E.}~\bibnamefont{Hiyama}},
  \bibinfo{author}{\bibfnamefont{H.}~\bibnamefont{Sagawa}},
  \bibinfo{author}{\bibfnamefont{H.~J.} \bibnamefont{Schulze}},
  \bibnamefont{and} \bibinfo{author}{\bibfnamefont{J.}~\bibnamefont{Meng}},
  \bibinfo{journal}{Phys. Rev.} \textbf{\bibinfo{volume}{C94}},
  \bibinfo{pages}{064319} (\bibinfo{year}{2016}), \eprint{1611.03661}.

\bibitem[{\citenamefont{Gal et~al.}(2016)\citenamefont{Gal, Hungerford, and
  Millener}}]{Gal2016}
\bibinfo{author}{\bibfnamefont{A.}~\bibnamefont{Gal}},
  \bibinfo{author}{\bibfnamefont{E.~V.} \bibnamefont{Hungerford}},
  \bibnamefont{and} \bibinfo{author}{\bibfnamefont{D.~J.}
  \bibnamefont{Millener}}, \bibinfo{journal}{Rev. Mod. Phys.}
  \textbf{\bibinfo{volume}{88}}, \bibinfo{pages}{035004}
  (\bibinfo{year}{2016}),
  \urlprefix\url{https://link.aps.org/doi/10.1103/RevModPhys.88.035004}.

\bibitem[{\citenamefont{Lattimer and Lim}(2013)}]{Lattimer13}
\bibinfo{author}{\bibfnamefont{J.~M.} \bibnamefont{Lattimer}} \bibnamefont{and}
  \bibinfo{author}{\bibfnamefont{Y.}~\bibnamefont{Lim}},
  \bibinfo{journal}{Astrophys. J.} \textbf{\bibinfo{volume}{771}},
  \bibinfo{pages}{51} (\bibinfo{year}{2013}), \eprint{1203.4286}.

\bibitem[{\citenamefont{Oertel et~al.}(2017)\citenamefont{Oertel, Hempel,
  Kl\"ahn, and Typel}}]{Oertel17}
\bibinfo{author}{\bibfnamefont{M.}~\bibnamefont{Oertel}},
  \bibinfo{author}{\bibfnamefont{M.}~\bibnamefont{Hempel}},
  \bibinfo{author}{\bibfnamefont{T.}~\bibnamefont{Kl\"ahn}}, \bibnamefont{and}
  \bibinfo{author}{\bibfnamefont{S.}~\bibnamefont{Typel}},
  \bibinfo{journal}{Rev. Mod. Phys.} \textbf{\bibinfo{volume}{89}},
  \bibinfo{pages}{015007} (\bibinfo{year}{2017}),
  \urlprefix\url{https://link.aps.org/doi/10.1103/RevModPhys.89.015007}.

\bibitem[{\citenamefont{Zimmerman et~al.}(2020)\citenamefont{Zimmerman, Carson,
  Schumacher, Steiner, and Yagi}}]{Zimmerman_2020}
\bibinfo{author}{\bibfnamefont{J.}~\bibnamefont{Zimmerman}},
  \bibinfo{author}{\bibfnamefont{Z.}~\bibnamefont{Carson}},
  \bibinfo{author}{\bibfnamefont{K.}~\bibnamefont{Schumacher}},
  \bibinfo{author}{\bibfnamefont{A.~W.} \bibnamefont{Steiner}},
  \bibnamefont{and} \bibinfo{author}{\bibfnamefont{K.}~\bibnamefont{Yagi}}
  (\bibinfo{year}{2020}), \eprint{2002.03210}.

\bibitem[{\citenamefont{Mondal et~al.}(2017)\citenamefont{Mondal, Agrawal, De,
  Samaddar, Centelles, and Vi\~nas}}]{Mondal_PRC_2017}
\bibinfo{author}{\bibfnamefont{C.}~\bibnamefont{Mondal}},
  \bibinfo{author}{\bibfnamefont{B.~K.} \bibnamefont{Agrawal}},
  \bibinfo{author}{\bibfnamefont{J.~N.} \bibnamefont{De}},
  \bibinfo{author}{\bibfnamefont{S.~K.} \bibnamefont{Samaddar}},
  \bibinfo{author}{\bibfnamefont{M.}~\bibnamefont{Centelles}},
  \bibnamefont{and} \bibinfo{author}{\bibfnamefont{X.}~\bibnamefont{Vi\~nas}},
  \bibinfo{journal}{Phys. Rev. C} \textbf{\bibinfo{volume}{96}},
  \bibinfo{pages}{021302} (\bibinfo{year}{2017}),
  \urlprefix\url{https://link.aps.org/doi/10.1103/PhysRevC.96.021302}.

\bibitem[{\citenamefont{Malik et~al.}(2018)\citenamefont{Malik, Alam, Fortin,
  Provid\^encia, Agrawal, Jha, Kumar, and Patra}}]{Malik_PRC_2018}
\bibinfo{author}{\bibfnamefont{T.}~\bibnamefont{Malik}},
  \bibinfo{author}{\bibfnamefont{N.}~\bibnamefont{Alam}},
  \bibinfo{author}{\bibfnamefont{M.}~\bibnamefont{Fortin}},
  \bibinfo{author}{\bibfnamefont{C.}~\bibnamefont{Provid\^encia}},
  \bibinfo{author}{\bibfnamefont{B.~K.} \bibnamefont{Agrawal}},
  \bibinfo{author}{\bibfnamefont{T.~K.} \bibnamefont{Jha}},
  \bibinfo{author}{\bibfnamefont{B.}~\bibnamefont{Kumar}}, \bibnamefont{and}
  \bibinfo{author}{\bibfnamefont{S.~K.} \bibnamefont{Patra}},
  \bibinfo{journal}{Phys. Rev. C} \textbf{\bibinfo{volume}{98}},
  \bibinfo{pages}{035804} (\bibinfo{year}{2018}),
  \urlprefix\url{https://link.aps.org/doi/10.1103/PhysRevC.98.035804}.

\bibitem[{\citenamefont{Zhang and Li}(2019)}]{Zhang_EPJA_2018}
\bibinfo{author}{\bibfnamefont{N.-B.} \bibnamefont{Zhang}} \bibnamefont{and}
  \bibinfo{author}{\bibfnamefont{B.-A.} \bibnamefont{Li}},
  \bibinfo{journal}{Eur. Phys. J. A} \textbf{\bibinfo{volume}{55}},
  \bibinfo{pages}{39} (\bibinfo{year}{2019}), \eprint{1807.07698}.

\bibitem[{\citenamefont{Newton and Crocombe}(2020)}]{Newton_2020}
\bibinfo{author}{\bibfnamefont{W.~G.} \bibnamefont{Newton}} \bibnamefont{and}
  \bibinfo{author}{\bibfnamefont{G.}~\bibnamefont{Crocombe}}
  (\bibinfo{year}{2020}), \eprint{2008.00042}.

\bibitem[{\citenamefont{Raaijmakers et~al.}(2020)}]{Raaijmakers_2019}
\bibinfo{author}{\bibfnamefont{G.}~\bibnamefont{Raaijmakers}}
  \bibnamefont{et~al.}, \bibinfo{journal}{Astrophys. J. Lett.}
  \textbf{\bibinfo{volume}{893}}, \bibinfo{pages}{L21} (\bibinfo{year}{2020}),
  \eprint{1912.11031}.

\bibitem[{\citenamefont{Miller et~al.}(2019)}]{Miller_2019}
\bibinfo{author}{\bibfnamefont{M.~C.} \bibnamefont{Miller}}
  \bibnamefont{et~al.}, \bibinfo{journal}{Astrophys. J. Lett.}
  \textbf{\bibinfo{volume}{887}}, \bibinfo{pages}{L24} (\bibinfo{year}{2019}),
  \eprint{1912.05705}.

\bibitem[{\citenamefont{Riley et~al.}(2019)}]{Riley_2019}
\bibinfo{author}{\bibfnamefont{T.~E.} \bibnamefont{Riley}}
  \bibnamefont{et~al.}, \bibinfo{journal}{Astrophys. J. Lett.}
  \textbf{\bibinfo{volume}{887}}, \bibinfo{pages}{L21} (\bibinfo{year}{2019}),
  \eprint{1912.05702}.

\bibitem[{\citenamefont{Abbott et~al.}(2017)}]{GW170817}
\bibinfo{author}{\bibfnamefont{B.~P.} \bibnamefont{Abbott}}
  \bibnamefont{et~al.} (\bibinfo{collaboration}{the Virgo, The LIGO
  Scientific}), \bibinfo{journal}{Phys. Rev. Lett.}
  \textbf{\bibinfo{volume}{119}}, \bibinfo{eid}{161101} (\bibinfo{year}{2017}).

\bibitem[{\citenamefont{{Antoniadis} et~al.}(2013)\citenamefont{{Antoniadis},
  {Freire}, {Wex}, {Tauris}, {Lynch}, {van Kerkwijk}, {Kramer}, {Bassa},
  {Dhillon}, {Driebe} et~al.}}]{Antoniadis2013}
\bibinfo{author}{\bibfnamefont{J.}~\bibnamefont{{Antoniadis}}},
  \bibinfo{author}{\bibfnamefont{P.~C.~C.} \bibnamefont{{Freire}}},
  \bibinfo{author}{\bibfnamefont{N.}~\bibnamefont{{Wex}}},
  \bibinfo{author}{\bibfnamefont{T.~M.} \bibnamefont{{Tauris}}},
  \bibinfo{author}{\bibfnamefont{R.~S.} \bibnamefont{{Lynch}}},
  \bibinfo{author}{\bibfnamefont{M.~H.} \bibnamefont{{van Kerkwijk}}},
  \bibinfo{author}{\bibfnamefont{M.}~\bibnamefont{{Kramer}}},
  \bibinfo{author}{\bibfnamefont{C.}~\bibnamefont{{Bassa}}},
  \bibinfo{author}{\bibfnamefont{V.~S.} \bibnamefont{{Dhillon}}},
  \bibinfo{author}{\bibfnamefont{T.}~\bibnamefont{{Driebe}}},
  \bibnamefont{et~al.}, \bibinfo{journal}{Science}
  \textbf{\bibinfo{volume}{340}}, \bibinfo{pages}{448} (\bibinfo{year}{2013}).

\bibitem[{\citenamefont{Haensel and Zdunik}(1990)}]{HZ90}
\bibinfo{author}{\bibfnamefont{P.}~\bibnamefont{Haensel}} \bibnamefont{and}
  \bibinfo{author}{\bibfnamefont{J.~L.} \bibnamefont{Zdunik}},
  \bibinfo{journal}{Astron. Astrophys.} \textbf{\bibinfo{volume}{227}},
  \bibinfo{pages}{431} (\bibinfo{year}{1990}).

\bibitem[{\citenamefont{{Haensel} and {Zdunik}}(2008)}]{HZ08}
\bibinfo{author}{\bibfnamefont{P.}~\bibnamefont{{Haensel}}} \bibnamefont{and}
  \bibinfo{author}{\bibfnamefont{J.~L.} \bibnamefont{{Zdunik}}},
  \bibinfo{journal}{Astron. Astrophys.} \textbf{\bibinfo{volume}{480}},
  \bibinfo{pages}{459} (\bibinfo{year}{2008}), \eprint{0708.3996}.

\bibitem[{\citenamefont{{Potekhin} et~al.}(2019)\citenamefont{{Potekhin},
  {Chugunov}, and {Chabrier}}}]{Potekhin2019}
\bibinfo{author}{\bibfnamefont{A.~Y.} \bibnamefont{{Potekhin}}},
  \bibinfo{author}{\bibfnamefont{A.~I.} \bibnamefont{{Chugunov}}},
  \bibnamefont{and}
  \bibinfo{author}{\bibfnamefont{G.}~\bibnamefont{{Chabrier}}},
  \bibinfo{journal}{A\&A} \textbf{\bibinfo{volume}{629}}, \bibinfo{eid}{A88}
  (\bibinfo{year}{2019}), \eprint{1907.08299}.

\bibitem[{\citenamefont{{Levenfish} and {Yakovlev}}(1994)}]{Levenfish_1994}
\bibinfo{author}{\bibfnamefont{K.~P.} \bibnamefont{{Levenfish}}}
  \bibnamefont{and} \bibinfo{author}{\bibfnamefont{D.~G.}
  \bibnamefont{{Yakovlev}}}, \bibinfo{journal}{Astronomy Letters-A Journal of
  Astronomy and Space Astrophysics} \textbf{\bibinfo{volume}{20}},
  \bibinfo{pages}{43} (\bibinfo{year}{1994}).

\bibitem[{\citenamefont{{Yakovlev} and {Levenfish}}(1995)}]{Yakovlev_AA_1995}
\bibinfo{author}{\bibfnamefont{D.~G.} \bibnamefont{{Yakovlev}}}
  \bibnamefont{and} \bibinfo{author}{\bibfnamefont{K.~P.}
  \bibnamefont{{Levenfish}}}, \bibinfo{journal}{Astronomy and Astrophysics}
  \textbf{\bibinfo{volume}{297}}, \bibinfo{pages}{717} (\bibinfo{year}{1995}).

\bibitem[{\citenamefont{Leinson and Perez}(2006)}]{Leison_PLB_2006}
\bibinfo{author}{\bibfnamefont{L.}~\bibnamefont{Leinson}} \bibnamefont{and}
  \bibinfo{author}{\bibfnamefont{A.}~\bibnamefont{Perez}},
  \bibinfo{journal}{Physics Letters B} \textbf{\bibinfo{volume}{638}},
  \bibinfo{pages}{114 } (\bibinfo{year}{2006}), ISSN \bibinfo{issn}{0370-2693},
  \urlprefix\url{http://www.sciencedirect.com/science/article/pii/S0370269306005995}.

\bibitem[{\citenamefont{{Leinson}}(2016)}]{Leinson2016}
\bibinfo{author}{\bibfnamefont{L.~B.} \bibnamefont{{Leinson}}},
  \bibinfo{journal}{arXiv e-prints} \bibinfo{eid}{arXiv:1611.03794}
  (\bibinfo{year}{2016}), \eprint{1611.03794}.

\bibitem[{\citenamefont{{Fortin}
  et~al.}(2018{\natexlab{a}})\citenamefont{{Fortin}, {Taranto}, {Burgio},
  {Haensel}, {Schulze}, and {Zdunik}}}]{Fortin2018}
\bibinfo{author}{\bibfnamefont{M.}~\bibnamefont{{Fortin}}},
  \bibinfo{author}{\bibfnamefont{G.}~\bibnamefont{{Taranto}}},
  \bibinfo{author}{\bibfnamefont{G.~F.} \bibnamefont{{Burgio}}},
  \bibinfo{author}{\bibfnamefont{P.}~\bibnamefont{{Haensel}}},
  \bibinfo{author}{\bibfnamefont{H.~J.} \bibnamefont{{Schulze}}},
  \bibnamefont{and} \bibinfo{author}{\bibfnamefont{J.~L.}
  \bibnamefont{{Zdunik}}}, \bibinfo{journal}{MNRAS}
  \textbf{\bibinfo{volume}{475}}, \bibinfo{pages}{5010}
  (\bibinfo{year}{2018}{\natexlab{a}}), \eprint{1709.04855}.

\bibitem[{\citenamefont{Sedrakian and Clark}(2019)}]{Sedrakian_EPJA_2019}
\bibinfo{author}{\bibfnamefont{A.}~\bibnamefont{Sedrakian}} \bibnamefont{and}
  \bibinfo{author}{\bibfnamefont{J.~W.} \bibnamefont{Clark}},
  \bibinfo{journal}{Eur. Phys. J.} \textbf{\bibinfo{volume}{A55}},
  \bibinfo{pages}{167} (\bibinfo{year}{2019}), \eprint{1802.00017}.

\bibitem[{\citenamefont{Ding et~al.}(2016)\citenamefont{Ding, Rios, Dussan,
  Dickhoff, Witte, Carbone, and Polls}}]{Ding_PRC_2016}
\bibinfo{author}{\bibfnamefont{D.}~\bibnamefont{Ding}},
  \bibinfo{author}{\bibfnamefont{A.}~\bibnamefont{Rios}},
  \bibinfo{author}{\bibfnamefont{H.}~\bibnamefont{Dussan}},
  \bibinfo{author}{\bibfnamefont{W.~H.} \bibnamefont{Dickhoff}},
  \bibinfo{author}{\bibfnamefont{S.~J.} \bibnamefont{Witte}},
  \bibinfo{author}{\bibfnamefont{A.}~\bibnamefont{Carbone}}, \bibnamefont{and}
  \bibinfo{author}{\bibfnamefont{A.}~\bibnamefont{Polls}},
  \bibinfo{journal}{Phys. Rev. C} \textbf{\bibinfo{volume}{94}},
  \bibinfo{pages}{025802} (\bibinfo{year}{2016}),
  \urlprefix\url{https://link.aps.org/doi/10.1103/PhysRevC.94.025802}.

\bibitem[{\citenamefont{Page et~al.}(2004)\citenamefont{Page, Lattimer,
  Prakash, and Steiner}}]{Page_2004}
\bibinfo{author}{\bibfnamefont{D.}~\bibnamefont{Page}},
  \bibinfo{author}{\bibfnamefont{J.~M.} \bibnamefont{Lattimer}},
  \bibinfo{author}{\bibfnamefont{M.}~\bibnamefont{Prakash}}, \bibnamefont{and}
  \bibinfo{author}{\bibfnamefont{A.~W.} \bibnamefont{Steiner}},
  \bibinfo{journal}{ApJ Supplement Series} \textbf{\bibinfo{volume}{155}},
  \bibinfo{pages}{623} (\bibinfo{year}{2004}),
  \urlprefix\url{http://stacks.iop.org/0067-0049/155/i=2/a=623}.

\bibitem[{\citenamefont{Entem and Machleidt}(2003)}]{N3LO}
\bibinfo{author}{\bibfnamefont{D.~R.} \bibnamefont{Entem}} \bibnamefont{and}
  \bibinfo{author}{\bibfnamefont{R.}~\bibnamefont{Machleidt}},
  \bibinfo{journal}{Phys. Rev. C} \textbf{\bibinfo{volume}{68}},
  \bibinfo{pages}{041001} (\bibinfo{year}{2003}),
  \urlprefix\url{https://link.aps.org/doi/10.1103/PhysRevC.68.041001}.

\bibitem[{\citenamefont{Schwenk et~al.}(2003)\citenamefont{Schwenk, Friman, and
  Brown}}]{SFB_2003}
\bibinfo{author}{\bibfnamefont{A.}~\bibnamefont{Schwenk}},
  \bibinfo{author}{\bibfnamefont{B.}~\bibnamefont{Friman}}, \bibnamefont{and}
  \bibinfo{author}{\bibfnamefont{G.~E.} \bibnamefont{Brown}},
  \bibinfo{journal}{Nucl. Phys. A} \textbf{\bibinfo{volume}{713}},
  \bibinfo{pages}{191 } (\bibinfo{year}{2003}), ISSN \bibinfo{issn}{0375-9474},
  \urlprefix\url{http://www.sciencedirect.com/science/article/pii/S0375947402012903}.

\bibitem[{\citenamefont{Baldo et~al.}(1992)\citenamefont{Baldo, Cugnon,
  Lejeune, and Lombardo}}]{Baldo_NPA1992}
\bibinfo{author}{\bibfnamefont{M.}~\bibnamefont{Baldo}},
  \bibinfo{author}{\bibfnamefont{J.}~\bibnamefont{Cugnon}},
  \bibinfo{author}{\bibfnamefont{A.}~\bibnamefont{Lejeune}}, \bibnamefont{and}
  \bibinfo{author}{\bibfnamefont{U.}~\bibnamefont{Lombardo}},
  \bibinfo{journal}{Nucl. Phys. A} \textbf{\bibinfo{volume}{536}},
  \bibinfo{pages}{349 } (\bibinfo{year}{1992}), ISSN \bibinfo{issn}{0375-9474},
  \urlprefix\url{http://www.sciencedirect.com/science/article/pii/037594749290387Y}.

\bibitem[{\citenamefont{Chen et~al.}(1993)\citenamefont{Chen, Clark, Dave, and
  Khodel}}]{Chen_NPA1993}
\bibinfo{author}{\bibfnamefont{J.}~\bibnamefont{Chen}},
  \bibinfo{author}{\bibfnamefont{J.}~\bibnamefont{Clark}},
  \bibinfo{author}{\bibfnamefont{R.}~\bibnamefont{Dave}}, \bibnamefont{and}
  \bibinfo{author}{\bibfnamefont{V.}~\bibnamefont{Khodel}},
  \bibinfo{journal}{Nucl. Phys. A} \textbf{\bibinfo{volume}{555}},
  \bibinfo{pages}{59 } (\bibinfo{year}{1993}), ISSN \bibinfo{issn}{0375-9474},
  \urlprefix\url{http://www.sciencedirect.com/science/article/pii/037594749390314N}.

\bibitem[{\citenamefont{Lagaris and Pandharipande}(1981)}]{Argonne_v14}
\bibinfo{author}{\bibfnamefont{I.}~\bibnamefont{Lagaris}} \bibnamefont{and}
  \bibinfo{author}{\bibfnamefont{V.}~\bibnamefont{Pandharipande}},
  \bibinfo{journal}{Nuclear Physics A} \textbf{\bibinfo{volume}{359}},
  \bibinfo{pages}{331 } (\bibinfo{year}{1981}), ISSN \bibinfo{issn}{0375-9474},
  \urlprefix\url{http://www.sciencedirect.com/science/article/pii/0375947481902402}.

\bibitem[{\citenamefont{Chao et~al.}(1972)\citenamefont{Chao, Clark, and
  Yang}}]{Chao_NPA_1972}
\bibinfo{author}{\bibfnamefont{N.-C.} \bibnamefont{Chao}},
  \bibinfo{author}{\bibfnamefont{J.}~\bibnamefont{Clark}}, \bibnamefont{and}
  \bibinfo{author}{\bibfnamefont{C.-H.} \bibnamefont{Yang}},
  \bibinfo{journal}{Nuclear Physics A} \textbf{\bibinfo{volume}{179}},
  \bibinfo{pages}{320 } (\bibinfo{year}{1972}), ISSN \bibinfo{issn}{0375-9474},
  \urlprefix\url{http://www.sciencedirect.com/science/article/pii/0375947472903739}.

\bibitem[{\citenamefont{{Fortin}
  et~al.}(2018{\natexlab{b}})\citenamefont{{Fortin}, {Taranto}, {Burgio},
  {Haensel}, {Schulze}, and {Zdunik}}}]{Taranto}
\bibinfo{author}{\bibfnamefont{M.}~\bibnamefont{{Fortin}}},
  \bibinfo{author}{\bibfnamefont{G.}~\bibnamefont{{Taranto}}},
  \bibinfo{author}{\bibfnamefont{G.~F.} \bibnamefont{{Burgio}}},
  \bibinfo{author}{\bibfnamefont{P.}~\bibnamefont{{Haensel}}},
  \bibinfo{author}{\bibfnamefont{H.-J.} \bibnamefont{{Schulze}}},
  \bibnamefont{and} \bibinfo{author}{\bibfnamefont{J.~L.}
  \bibnamefont{{Zdunik}}}, \bibinfo{journal}{Monthly Notices of the Royal
  Astronomical Society} \textbf{\bibinfo{volume}{475}}, \bibinfo{pages}{5010}
  (\bibinfo{year}{2018}{\natexlab{b}}), \eprint{1709.04855}.

\bibitem[{\citenamefont{Beznogov et~al.}(2018)\citenamefont{Beznogov, Rrapaj,
  Page, and Reddy}}]{Beznogov_PRC_2018}
\bibinfo{author}{\bibfnamefont{M.~V.} \bibnamefont{Beznogov}},
  \bibinfo{author}{\bibfnamefont{E.}~\bibnamefont{Rrapaj}},
  \bibinfo{author}{\bibfnamefont{D.}~\bibnamefont{Page}}, \bibnamefont{and}
  \bibinfo{author}{\bibfnamefont{S.}~\bibnamefont{Reddy}},
  \bibinfo{journal}{Phys. Rev. C} \textbf{\bibinfo{volume}{98}},
  \bibinfo{pages}{035802} (\bibinfo{year}{2018}), \eprint{1806.07991}.

\bibitem[{\citenamefont{{Potekhin} et~al.}(2003)\citenamefont{{Potekhin},
  {Yakovlev}, {Chabrier}, and {Gnedin}}}]{PY03}
\bibinfo{author}{\bibfnamefont{A.~Y.} \bibnamefont{{Potekhin}}},
  \bibinfo{author}{\bibfnamefont{D.~G.} \bibnamefont{{Yakovlev}}},
  \bibinfo{author}{\bibfnamefont{G.}~\bibnamefont{{Chabrier}}},
  \bibnamefont{and} \bibinfo{author}{\bibfnamefont{O.~Y.}
  \bibnamefont{{Gnedin}}}, \bibinfo{journal}{ApJ}
  \textbf{\bibinfo{volume}{594}}, \bibinfo{pages}{404} (\bibinfo{year}{2003}),
  \eprint{astro-ph/0305256}.

\bibitem[{\citenamefont{{Heinke} et~al.}(2009)\citenamefont{{Heinke}, {Jonker},
  {Wijnands}, {Deloye}, and {Taam}}}]{Heinke09}
\bibinfo{author}{\bibfnamefont{C.~O.} \bibnamefont{{Heinke}}},
  \bibinfo{author}{\bibfnamefont{P.~G.} \bibnamefont{{Jonker}}},
  \bibinfo{author}{\bibfnamefont{R.}~\bibnamefont{{Wijnands}}},
  \bibinfo{author}{\bibfnamefont{C.~J.} \bibnamefont{{Deloye}}},
  \bibnamefont{and} \bibinfo{author}{\bibfnamefont{R.~E.}
  \bibnamefont{{Taam}}}, \bibinfo{journal}{The Astrophysical Journal}
  \textbf{\bibinfo{volume}{691}}, \bibinfo{pages}{1035} (\bibinfo{year}{2009}),
  \eprint{0810.0497}.

\bibitem[{\citenamefont{{Coriat} et~al.}(2012)\citenamefont{{Coriat}, {Fender},
  and {Dubus}}}]{Coriat_MNRAS_2012}
\bibinfo{author}{\bibfnamefont{M.}~\bibnamefont{{Coriat}}},
  \bibinfo{author}{\bibfnamefont{R.~P.} \bibnamefont{{Fender}}},
  \bibnamefont{and} \bibinfo{author}{\bibfnamefont{G.}~\bibnamefont{{Dubus}}},
  \bibinfo{journal}{MNRAS} \textbf{\bibinfo{volume}{424}},
  \bibinfo{pages}{1991} (\bibinfo{year}{2012}), \eprint{1205.5038}.

\end{thebibliography}

\end{document}